\magnification=\magstep1
\input epsf

\input amstex
\documentstyle{amsppt}




\input eplain
%
%
%






\def\setRuledStrut{\relax}

\catcode`@=11                                   
\catcode`\|=12                                  
\catcode`\&=4                                   

\newcount\ncols         \ncols=\z@              
\newcount\nrows         \nrows=\z@              
\newcount\curcol        \curcol=\z@             
     
\newdimen\thinsize      \thinsize=0.6pt         
\newdimen\thicksize     \thicksize=1.5pt        
\newdimen\framesize     \framesize=1.5pt        
\newdimen\tablewidth    \tablewidth=-\maxdimen  
\newdimen\parasize      \parasize=4in           

\newif\iftableinfo      \tableinfofalse         
\newif\ifcentertables   \centertablestrue       
%
%
     
\let\plaincr=\cr                        
\let\plainspan=\span                    
\let\plaintab=&                         
\let\lparen=(                           
\let\NX=\noexpand                       

     
\def\ruledtable{\relax                          
    \@BeginRuledTable                     
        \@RuledTable}


\def\@BeginRuledTable{
   \ncols=0\nrows=0                             
   \begingroup                                  
    \offinterlineskip                           
    \def~{\phantom{0}}
    \def\span{\plainspan\omit\relax\colcount\plainspan}
    \let\cr=\crrule                             
    \let\CR=\crthick                            
    \let\nr=\crnorule                           
    \let\|=\Vb                                  
    \def\hfill{\hskip0pt plus1fill\hbox{}}
%
%
    \ifx\tablestrut\undefined\relax             
    \else\let\tstrut=\tablestrut\fi             
    \catcode`\|=13 \catcode`\&=13\relax         
    \TableActive                                
    \curcol=1                                   
%
%
    \ifdim\tablewidth>-\maxdimen\relax          %
      \edef\@Halign{\NX\halign to \NX\tablewidth\NX\bgroup\TablePreamble}%
      \tabskip=0pt plus 1fil                    
    \else                                       %
      \edef\@Halign{\NX\halign\NX\bgroup\TablePreamble}%
      \tabskip=0pt                              
    \fi                                         %
%
%
    \ifcentertables                             
       \ifhmode\vskip 0pt\fi                    
       \line\bgroup\hss                         
    \else\hbox\bgroup                           
    \fi}


\long\def\@RuledTable#1\endruledtable{
   \vrule width\framesize                       
     \vbox{\@Halign                             
       \framerule                               
       #1\killspace                             
       \tstrut                                  
       \linecount                               
       \plaincr\framerule                       
     \egroup}
   \vrule width\framesize                       
   \ifcentertables\hss\fi\egroup                
  \endgroup                                     
  \global\tablewidth=-\maxdimen                 
  \iftableinfo                                  
      \immediate\write16{[Nrows=\the\nrows, Ncols=\the\ncols]}%
   \fi}
     

\def\TablePreamble{
   \TableItem{####}
   \plaintab\plaintab                   
   \TableItem{####}
   \plaincr}


\def\@TableItem#1{
   \hfil\tablespace                     
   #1\killspace
   \tablespace\hfil                     
    }%

\def\@tableright#1{
   \hfil\tablespace\relax               
   #1\killspace
   \tablespace\relax}

\def\@tableleft#1{
   \tablespace\relax                    
   #1\killspace
   \tablespace\hfil}

\let\TableItem=\@TableItem              
     
\def\RightJustifyTables{\let\TableItem=\@tableright}
\def\LeftJustifyTables{\let\TableItem=\@tableleft}
\def\NoJustifyTables{\let\TableItem=\@TableItem}


\def\LooseTables{\let\tablespace=\quad}
\def\TightTables{\let\tablespace=\space}
\LooseTables                                    


\def\TrailingSpaces{\let\killspace=\relax}      
\def\NoTrailingSpaces{\let\killspace=\unskip}   
\TrailingSpaces                                 

%


\def\setRuledStrut{
   \dimen@=\baselineskip                        
   \advance\dimen@ by-\normalbaselineskip       
   \ifdim\dimen@<.5ex \dimen@=.5ex\fi           
   \setbox0=\hbox{\lparen}
   \dimen1=\dimen@ \advance\dimen1 by \ht0      
   \dimen2=\dimen@ \advance\dimen2 by \dp0      
   \def\tstrut{\vrule height\dimen1 depth\dimen2 width\z@}%
   }%

\def\tstrut{\vrule height 3.1ex depth 1.2ex width 0pt}


\def\bigitem#1{
   \dimen@=\baselineskip                        
   \advance\dimen@ by-\normalbaselineskip       
   \ifdim\dimen@<.5ex \dimen@=.5ex\fi           
   \setbox0=\hbox{#1}
   \dimen1=\dimen@ \advance\dimen1 by \ht0      
   \dimen2=\dimen@ \advance\dimen2 by \dp0      
   \vrule height\dimen1 depth\dimen2 width\z@   
   \copy0}

     
%

     
\def\nextcolumn#1{
   \plaintab\omit#1\relax\colcount              
   \plaintab}
     
\def\tab{
   \nextcolumn{\relax}}


\def\vb{
   \nextcolumn{\vrule width\thinsize}}

\def\Vb{
   \nextcolumn{\vrule width\thicksize}}


     
{\catcode`\|=13 \let|0
 \catcode`\&=13 \let&0
 \gdef\TableActive{\let|=\vb \let&=\tab}%
}


\def\crrule{\killspace                  
   \tstrut                              
   \linecount                           
   \plaincr\tablerule                   
  }%

\def\crthick{\killspace                 
   \tstrut                              
   \linecount                           
   \plaincr\thickrule                   
  }%
     
\def\crnorule{\killspace                
   \tstrut                              
   \linecount                           
   \plaincr                             
   }%
   

     
\def\tablerule{\noalign{\hrule height\thinsize depth 0pt}}%
\def\thickrule{\noalign{\hrule height\thicksize depth 0pt}}%
\def\framerule{\noalign{\hrule height\framesize depth 0pt}}


%
%
%
     

\def\linecount{
   \global\advance\nrows by1
   \ifnum\ncols>0
      \ifnum\curcol=\ncols\relax\else           
      \immediate\write16
      {\NX\ruledtable warning: Ncols=\the\curcol\space for Nrow=\the\nrows}%
      \fi\fi                                    
   \global\ncols=\curcol                        
   \global\curcol=1}                            

\def\colcount{\relax                            %
   \global\advance\curcol by 1\relax}


%

%

\def\begintable{\relax                          
    \@BeginRuledTable                           
    \@begintable}

\long\def\@begintable#1\endtable{
   \@RuledTable#1\endruledtable}



\catcode`@=12                                   


\let\ref=\eplainref  

\TagsOnRight
\parskip 1ex

\hsize=6.5truein
\vsize=9truein

\loadbold
\loadeurm

\font\bbf=cmbx12
\font\bbigbf=cmbx12 scaled\magstep1
\font\bi=cmbxti10	
\font\bigbf=cmbx12
\font\bigbi=cmbxti10 scaled\magstep1	

\font\sc=cmcsc10
\font\smallrm=cmr9

\font\smallbf=cmbx9
\font\smallbi=cmbxti10 at 8pt
\font\sf=cmss10

\font\smalltt=cmtt8
\font\ssf=cmss8

\document

\define\de{\define}
\de\AAA{{\tx{${\scr A}$}}}
\de\AAa{{\tx{${\Bbb A}$}}}
\de\ABBB{{\tx{$A^\BBB$}}}
\de\ABN{{\tx{$A^{B,N}$}}}
\de\AN{{\tx{$A^N$}}}
\de\Ab{{\bf A}}
\de\Abi{{\bi A}}
\de\Abar{{\tx{$\bar A$}}}
\de\Abars{\tx{$\Abar_s$}}
\de\Abaru{\tx{$\Abar^u$}}
\de\AinKK{\tx{$A \in \KK$}}
\de\AinAlgSig{\tx{$A \in \AlgSig$}}
\de\AinNStdAlgSig{\tx{$A \in \NStdAlgSig$}}
\de\AinStdAlgSig{\tx{$A \in \StdAlgSig$}}
\de\Akvec{\tx{$A[\kvec]$}}
\de\Akxvec{\tx{$A[\kxvec]$}}
\de\Aklxvec{\tx{$A[\kvec,\lxvec]$}}
\de\Alg{\tx{\bi Alg}}
\de\AlgSig{\Alg\,(\Sig)}
\de\Alvec{\tx{$A[\lvec\,]$}}
\de\Alxvec{\tx{$A[\lxvec]$}}
\de\Aps{\tx{\sf Ap}}
\de\As{\tx{$A_s$}}
\de\Asbar{\tx{$\Abar_s$}}
\de\Asx{\tx{$A_s^*$}}
\de\AssLang{\tx{\bi Ass}\Lang}
\de\Assumption{\pr{Assumption}\sl}
\de\Assumptionn#1{\pr{Assumption #1}\sl}
\de\Assumptions{\pr{Assumptions}\sl}
\de\Asuu{\tx{$A^\uu_s$}}
\de\At{\tx{\bi At}}
\de\AtSig{\At(\Sig)}
\de\AtSt{\tx{\bi AtSt}}
\de\AtStSig{\tx{$\AtSt(\Sig)$}}
\de\Atil{{\tx{$\til A$}}}
\de\Au{\tx{$A^u$}}
\de\Aul{{\tx{$\ul A$}}}
\de\Auu{\tx{$A^\uu$}}
\de\AuuN{\tx{$A^{\uu,N}$}}
\de\Av{\tx{$A^v$}}
\de\Aw{\tx{$A^w$}}
\de\Ax{{\tx{$A^*$}}}
\de\Axs{\tx{$A^*_s$}}
\de\Axx{(\Ax)\str}
\de\Bbi{\tx{\bi B}}
\de\BB{{\tx{${\Bbb B}$}}}
\de\BBB{{\tx{${\scr B}$}}}
\de\BBbar{\tx{$\ol \BB$}}
\de\BBuu{\tx{${\Bbb B}^\uu$}}
\de\Bb{\tx{\bf B}}
\de\Bool{\tx{\bi Bool}}
\de\BoolSig{\tx{$\Bool(\Sig)$}}
\de\Bs{\tx{\sf B}}
\de\Bss{{\tx{\ssf B}}}
\de\Btil{\tx{$\til B$}}
\de\Bul{\tx{$\ul B$}}
\de\CC{\tx{${\Bbb C}$}}
\de\CCn{\tx{$\CC^n$}}
\de\CCC{\tx{$\scr C$}}
\de\CCCL{\tx{${\CCC}^<$}}
\de\CR{\tx{\bi CR}}
\de\CRSig{\tx{$\CR(\Sig)$}}
\de\CT{{\tx{\bi CT}}}
\de\CTSig{\CT(\Sig)}
\de\CTsmall{{\tx{\smallbi CT}}}
\de\Cbar{\tx{$\ol C$}}
\de\Cf{{\it Cf\.}}
\de\Chead#1{\bn\cbb{#1}\bn}
\de\Comp{\tx{\bi Comp}}
\de\CompA{\tx{$\Comp^A$}}
\de\CompLengthA{\tx{$\Comp\Length^A$}}
\de\CompSeqA#1{\tx{$\Comp\Seq^A(#1)$}}
\de\Compu{\tx{\bi Compu}}
\de\Con{\tx{\bi Con}}
\de\Cond{\tx{\bi Cond}}
\de\CondSig{\tx{$\Cond(\Sig)$}}
\de\Constrn#1{\pr{Construction \ #1}\rm}
\de\Cont{\tx{\bi Cont}}
\de\ContA{\tx{$\Cont(A)$}}
\de\Conventionn#1{\pr{Convention \ #1}\sl}
\de\Cor{\pr{Corollary}\sl}
\de\Corn#1{\prn{Corollary #1}\sl}
\de\Cors{\pr{Corollaries}\sl}
\de\Cs{\tx{\sf C}}
\de\Cul{\tx{$\ul C$}}
\de\DD{{\tx{${\Bbb D}$}}}
\de\DDD{{\tx{$\scr D$}}}
\de\Dbar{\tx{$\ol{D}$}}
\de\Dbars{\tx{$\ol{D}_s$}}
\de\Def{\pr{Definition}\rm}
\de\Defn#1{\prn{Definition #1}\rm}
\de\Defs{\pr{Definitions}\rm}
\de\Defsn#1{\prn{Definitions #1}\rm}
\de\Del{\tx{$\Delta$}}
\de\Discussion{\pr{Discussion}\rm}
\de\Discussionn#1{\pr{Discussion #1}\rm}
\de\Ds{\tx{\sf D}}
\de\Dss{{\tx{\ssf D}}}
\de\Ebar{\tx{$\ol E$}}
\de\Eg{{\it E.g.}}
\de\ElemInd{\tx{\sf ElemInd}}
\de\ElemIndSig{\tx{$\ElemInd(\Sig)$}}
\de\EqSort{\tx{\bi EqSort\/}}
\de\EqSortSig{\EqSort(\Sig)}
\de\Eqb{\tx{\bi Eq}}
\de\Eqs{\tx{\sf Eq}}
\de\Equiv{\tx{$\pmb{Equiv}$}}
\de\Etil{\tx{$\til{E}$}}
\de\Example{\pr{Example}}
\de\Examplen#1{\prn{Example {#1}}}
\de\Examples{\pr{Examples}}
\de\Examplesn#1{\prn{Examples {#1}}}
\de\Exercise{\pr{Exercise}}
\de\Exercisen#1{\prn{Exercise {#1}}}
\de\Exercises{\pr{Exercises}}
\de\Exercisesn#1{\prn{Exercises {#1}}}
\de\F{\tx{\bi F}}
\de\FA{\tx{$F^A$}}
\de\FAuu{\tx{$F^{A,\uu}$}}
\de\FFF{\tx{$\scr F$}}
\de\FFt{\tx{\tt F}}
\de\FinFuncSig{\tx{$F \in \FuncSig$}}
\de\First{\tx{\bi First}}
\de\For{\tx{\bi For}}
\de\ForA{\tx{$\For(A)$}}
\de\ForN{\tx{$\For^N$}}
\de\ForNA{\tx{$\ForN(A)$}}
\de\ForNSig{\tx{$\ForN(\Sig)$}}
\de\ForSig{\tx{$\For(\Sig)$}}
\de\ForSigN{\tx{$\For(\SigN)$}}
\de\ForSigx{\tx{$\For(\Sigx)$}}
\de\Form{\tx{\bi Form}}
\de\FormSig{\tx{$\Form(\Sig)$}}
\de\Foro{\tx{$\For_0$}}
\de\Forolam{\tx{$\For_0^{\lam}$}}
\de\Forx{\tx{$\For^\starb$}}
\de\ForxA{\tx{$\Forx(A)$}}
\de\ForxSig{\tx{$\Forx(\Sig)$}}
\de\Fs{{\tx{\smallbi F}}}
\de\Ftil{\tx{$\til{F}$}}
\de\Func{\tx{\bi Func}}
\de\FuncSig{\Func\,(\Sig)}
\de\FuncSigp{\tx{$\Func(\Sigp)$}}
\de\FuncSigus{\tx{$\FuncSig_\utos$}}
\de\GG{\tx{${\Bbb G}$}}
\de\GGG{\tx{$\scr G$}}
\de\Gtt{\tx{\tt G}}
\de\Gam{\tx{$\Gamma$}}
\de\Gbar{\tx{$\ol G$}}
\de\Gs{{\tx{\sf G}}}
\de\Halt{\tx{\bi Halt}}
\de\HaltTests{\tx{\sf HaltTest}}
\de\HaltA#1{\tx{$\Halt^A\,(#1)$}}
\de\Hbar{\tx{$\bar H$}}
\de\Hh{\tx{\bi H}}
\de\Hs{{\tx{\sf H}}}
\de\Hss{{\tx{\ssf H}}}
\de\Htt{\tx{\tt H}}
\de\Ib{\tx{{\bf I}}}
\de\II{{\tx{${\Bbb I}$}}}
\de\III{\tx{${\scr I}$}}
\de\IIIN{{\tx{$\III^N$}}}
\de\IL{\tx{\bi IL}}
\de\IO{\tx{\bi IO}}
\de\Ibar{{\tx{$\bar I$}}}
\de\Ie{{\it I.e.}}
\de\Ind{\tx{\sf Ind}}
\de\IndSig{\tx{$\Ind(\Sig)$}}
\de\Indi{\tx{$\Ind_i$}}
\de\IndiSig{\Indi(\Sig)}
\de\Init{\tx{\bi Init}}
\de\Is{\tx{\sf I}}
\de\Iss{{\tx{\ssf I}}}
\de\JJJ{\tx{${\scr J}$}}
\de\Jbar{{\tx{$\bar J$}}}
\de\KK{{\tx{${\Bbb K}$}}}
\de\KKbar{\tx{$\ol{\KK}$}}
\de\KKN{\tx{${\Bbb K}^N$}}
\de\KKuu{\tx{${\Bbb K}^\uu$}}
\de\KKx{\tx{${\Bbb K}^*$}}
\de\LL{\tx{${\Bbb L}$}}
\de\LLL{\tx{${\scr L}$}}
\de\Lam{\tx{$\Lambda$}}
\de\Lang{\tx{\bi Lang}}
\de\LangSig{\tx{$\Lang(\Sig)$}}
\de\LangSigx{\tx{$\Lang(\Sigx)$}}
\de\Langx{\tx{$\Lang^*$}}
\de\LangxSig{\tx{$\Langx(\Sig)$}}
\de\Lemma{\pr{Lemma}\sl}
\de\Lemman#1{\prn{Lemma #1}\sl}
\de\Lemmas{\pr{Lemmas}\sl}
\de\Lemmasn#1{\prn{Lemmas #1}\sl}
\de\Length{\tx{\bi Length}}
\de\Lgths{\tx{\sf Lgth}}
\de\Lss{{\tx{\ssf L}}}
\de\MMM{\tx{${\scr M}$}}
\de\Min{\tx{\bi Min}}
\de\MinNStdAlg{\Min\NStdAlg}
\de\Mod{\tx{\bi Mod}}
\de\NN{{\tx{${\Bbb N}$}}}
\de\NNN{{\tx{${\scr N}$}}}
\de\NNNB{{\tx{${\NNN^B}$}}}
\de\NNNo{{\tx{$\NNN_0$}}}
\de\NNu{\tx{${\Bbb N}^\uu$}}
\de\NStdAlg{\tx{\bi NStdAlg\/}}
\de\NStdAlgSig{\tx{$\NStdAlg\,(\Sig)$}}
\de\Nbd{\tx{\bi Nbd}}
\de\Nbi{{\tx{\bi N}}}
\de\Nbismall{{\tx{\smallbi N}}}
\de\Newlengths{\tx{\sf Newlength}}
\de\Notation{\pr{Notation}\rm}
\de\Notationn#1{\pr{Notation \ #1}\rm}
\de\Note{\pr{Note}\rm}
\de\Noten#1{\pr{Note #1}\rm}
\de\Notes{\pr{Notes}\rm}
\de\Notesn#1{\pr{Notes #1}\rm}
\de\Ns{\tx{\sf N}}
\de\Nss{{\tx{\ssf N}}}
\de\Nnu{\tx{$\Nn^\uu$}}
\de\Nulls{\tx{\sf Null}}
\de\Nullvec{\tx{$\vec{\Nulls}$}}
\de\Om{{\tx{$\Omega$}}}
\de\Pp{\tx{\bk P}}
\de\PA{\tx{$P^A$}}
\de\PE{\tx{\bi PE}}
\de\PL#1{\tx{$\ProgLang_{#1}$}}
\de\PPP{\tx{$\scr P$}}
\de\PR{\tx{PR}}
\de\PRA{\tx{$\PR(A)$}}
\de\PRAutos{\tx{$\PRA_\utos$}}
\de\PRSig{\tx{$\PR(\Sig)$}}
\de\PRSigb{\tx{$\PRb(\Sigb)$}}
\de\PRSigutos{\tx{$\PRSig_\utos$}}
\de\PRb{\tx{\bf PR}}
\de\PRmin{\tx{$\PR_-$}}
\de\PRo{\tx{$\PR_0$}}
\de\PRolam{\tx{$\PR_0^{\lam}$}}
\de\PRutos{\tx{$\PR_\utos$}}
\de\PRx{\PR\str}
\de\PRxA{\tx{$\PRx(A)$}}
\de\PRxSig{\tx{\PRx(\Sig)}}
\de\PRxSigb{\tx{\PRxb(\Sigb)}}
\de\PRxb{\PRb\strb}
\de\PTE{\tx{\bi PTE}}
\de\PTEA{\tx{$\PTE^A$}}
\de\PTerm{\tx{\bi PTerm}}
\de\PTermSig{\tx{$\PTerm(\Sig)$}}
\de\Pbi{\tx{\bi P}}
\de\Pf{\n{\bf Proof:\ \,}}
\de\Ph{\tx{$\Phi$}}
\de\PreStr{\tx{\bi PreStr}}
\de\PreStrSig{\PreStr\,(\Sig)}
\de\Problem{\pr{Problem}}
\de\Problemn#1{\pr{Problem {#1}}}
\de\Problems{\pr{Problems}}
\de\Problemsn#1{\pr{Problems {#1}}}
\de\Propsn#1{\pr{Propositions {#1}}}
\de\Proc{\tx{\bi Proc}}
\de\ProcN{\tx{$\Proc^N$}}
\de\ProcNSig{\ProcN(\Sig)}
\de\ProcSig{\Proc(\Sig)}
\de\ProcSigN{\Proc(\SigN)}
\de\ProcSiguv{\tx{$\Proc(\Sig)_\utov$}}
\de\ProcSigx{\tx{$\Proc(\Sigx)$}}
\de\Procuv{\tx{$\Proc_\utov$}}
\de\ProcuvSig{\tx{$\Procuv(\Sig)$}}
\de\Procx{\Proc\str}
\de\ProcxSig{\Procx(\Sig)}
\de\ProcxSiguv{\tx{$\ProcxSig_\utov$}}
\de\Procxuv{\tx{$\Proc^*_\utov$}}
\de\Prod{\tx{\bi Prod}}
\de\ProdType{\tx{\bi ProdType}}
\de\ProdTypeSig{\tx{$\ProdType(\Sig)$}}
\de\Prog{\tx{\bi Prog}}
\de\ProgLang{\Prog\Lang}
\de\ProgTerm{\Prog\Term}
\de\ProgTermSig{\tx{$\ProgTerm(\Sig)$}}
\de\Prop{\pr{Proposition}\sl}
\de\Propn#1{\prn{Proposition #1}\sl}
\de\Propp{\tx{\bi Prop}}
\de\Props{\pr{Propositions}\sl}
\de\Ps{\tx{$\Psi$}}
\de\Q{{\bf Q}}
\de\QQ{\tx{${\Bbb Q}$}}
\de\Qs{\tx{\sf Q}}
\de\Question{\pr{Question}\sl}
\de\RR{{\tx{${\Bbb R}$}}}
\de\RRR{{\tx{$\scr R$}}}
\de\RRRB{\tx{$\RRR^B$}}
\de\RRRL{\tx{${\RRR}^<$}}
\de\RRRN{\tx{$\RRR^N$}}
\de\RRn{\tx{$\RR^n$}}
\de\RRx{\tx{$\RR\str$}}
\de\RRq{\tx{$\RR^q$}}
\de\Ref{\tx{\bf Ref:\/}}
\de\Rem{\tx{\bi Rem}}
\de\RemA{\tx{$\Rem^A$}}
\de\RemSeqA#1{\tx{$\Rem\Seq^A(#1)$}}
\de\RemSet#1{\tx{\bi RemSet\,$(#1)$}}
\de\Remark{\pr{Remark}\rm}
\de\Remarkn#1{\prn{Remark #1}\rm}
\de\Remarks{\pr{Remarks}\rm}
\de\Remarksn#1{\prn{Remarks #1}\rm}
\de\Rep{\tx{\bi Rep}}
\de\Rest{\tx{\bi Rest}}
\de\Rs{\tx{\sf R}}
\de\Rss{{\tx{\ssf R}}}
\de\SA{\tx{$\bb{S}^A$}}
\de\SS{{\tx{${\Bbb S}$}}}
\de\SSS{{\tx{${\scr S}$}}}
\de\Sarrow{\tx{$S^{\to}$}}
\de\Sat{\tx{$S_{\tx{\ssf at}}$}}
\de\Sbar{\tx{$\ol \Sbi$}}
\de\Sbi{\tx{\bi S}}
\de\Seq{\tx{\bi Seq}}
\de\Shead#1#2{\bn{\bigbf {#1}\ \ \ {#2}}\sn}
\de\Sheads#1#2{\bn{\bigbf {#1}\ \ \ {#2}}}
\de\Sig{{\tx{$\varSigma$}}}
\de\SigA{\tx{$\Sig(A)$}}
\de\SigBBB{\tx{$\Sig^\BBB$}}
\de\SigBN{\tx{$\Sig^{B,N}$}}
\de\SigN{\tx{$\Sig^N$}}
\de\SigT{\tx{$(\Sig,T)$}}
\de\Sigb{\tx{$\bs\varSigma$}}
\de\Sigbar{\tx{$\ol \Sig$}}
\de\Sigo{\tx{$\varSigma_0$}}
\de\Sigox{\tx{$\varSigma_0^*$}}
\de\Sigone{\tx{$\Sigb_1$}}
\de\SigoneInd{\tx{$\Sigb_{\pmb1}$-$\pmb{Ind}\,$}}
\de\Sigonex{\tx{$\Sigb_1^*$}}
\de\SigonexInd{\tx{$\Sigb_{\pmb1}^{\pmb*}$-$\pmb{Ind}\,$}}
\de\SigonexIndl{\tx{$\Sigb_{\pmb1}^{\pmb*}$-$\pmb{Ind_\ell}\,$}}
\de\SigonexIndc{\tx{$\Sigb_{\pmb1}^{\pmb*}$-$\pmb{Ind_c}\,$}}
\de\SigonexIndi{\tx{$\Sigb_{\pmb1}^{\pmb*}$-$\pmb{Ind_i}\,$}}
\de\Sigp{{\tx{$\Sig'$}}}
\de\SigpTp{\tx{$(\Sig',T')$}}
\de\Sigpp{{\tx{$\Sig''$}}}
\de\SigppTpp{\tx{$(\Sig'',T'')$}}
\de\Sigstk{\tx{$\Sig^\stkss$}}
\de\Siguu{\tx{$\varSigma^\uu$}}
\de\SiguuN{\tx{$\varSigma^{\uu,N}$}}
\de\Sigx{\tx{$\varSigma^*$}}
\de\Snap{\tx{\bi Snap}}
\de\SnapA{\tx{$\Snap^A$}}
\de\SnapSeqA#1{\tx{$\Snap\Seq^A(#1)$}}
\de\Sort{\tx{\bi Sort\/}}
\de\SortSig{\Sort(\Sig)}
\de\SortSigsmall{\Sortsmall\,(\Sig)}
\de\SortSigp{\tx{$\Sort(\Sigp)$}}
\de\SortSigx{\tx{$\Sort(\Sigx)$}}
\de\Sortsmall{\tx{\smallbi Sort}}
\de\Spec{\tx{\sf Spec}}
\de\Ss{\tx{\sf S}}
\de\Sss{{\tx{\ssf S}}}
\de\State{\tx{\bi State}}
\de\StateA{\tx{$\State(A)$}}
\de\Step{\tx{\bi Step}}
\de\Stmt{\tx{\bi Stmt}}
\de\StmtSig{\tx{$\Stmt(\Sig)$}}
\de\StmtStepA{\tx{$\Stmt\Step^A$}}
\de\Std{\tx{\bi Std\/}}
\de\StdAlg{\Std\Alg}
\de\StdAlgSig{\tx{$\StdAlg\,(\Sig)$}}
\de\StdAlgSigx{\tx{$\StdAlg\,(\Sigx)$}}
\de\StdMod{\Std\Mod}
\de\Str{\tx{\bi Str}}
\de\StrSig{\Str\,(\Sig)}
\de\SubalgA#1{\tx{${\bi \ Subalg}_A(#1)$}}
\de\Tb{\tx{\bf T}}
\de\Tbi{\tx{\bi T}}
\de\TComp{\tx{\bi TComp}}
\de\TE{\tx{\bi TE}}
\de\TEA{\tx{$\TE^A$}}
\de\TOL{\TagsOnLeft}
\de\TOR{\TagsOnRight}
\de\TSig{\Tbi(\Sig)}
\de\TT{\tx{${\Bbb T}$}}
\de\TTT{\tx{$\scr T$}}
\de\Tab{\tx{\bi Tab}}
\de\TabSig{\Tab(\Sig)}
\de\Tbar{\tx{$\ol T$}}
\de\Term{\tx{\bi Term}}
\de\TermSig{\tx{$\Term(\Sig)$}}
\de\TermTup{\Term\Tup}
\de\TermTupSig{\tx{$\TermTup(\Sig)$}}
\de\Terminology{\pr{Terminology}\rm}
\de\Th{\tx{\bi Th}}
\de\Thesisn#1{\pr{Thesis \ #1}\sl}
\de\Thm{\pr{Theorem}\sl}
\de\Thmn#1{\prn{Theorem #1}\sl}
\de\Thms{\pr{Theorems}\sl}
\de\Tm{\tx{\bi Tm}}
\de\TmSig{\tx{$\Tm(\Sig)$}}
\de\Tms{\tx{$\Tm_s$}}
\de\TmsSig{\tx{$\Tms(\Sig)$}}
\de\Ts{\tx{$\Tbi_s$}}
\de\TsSig{\tx{$\Ts(\Sig)$}}
\de\Tup{\tx{\bi Tup}}
\de\Ub{\tx{\bf U}}
\de\Ubi{\tx{\bi U}}
\de\Univ{\tx{\bi Univ}}
\de\Univs{\tx{\sf Univ}}
\de\Univuv{\tx{$\Univs_\utov$}}
\de\UnivuvA{\tx{$\Univs_\utov^A$}}
\de\Univuvx{\tx{$\Univs_\utov^*$}}
\de\UnivuvxA{\tx{$\Univs_\utov^{*,A}$}}
\de\Univus{\tx{$\Univ_\utos$}}
\de\UnivusA{\tx{$\Univ_\utos^A$}}
\de\Unspecs{\tx{\sf Unspec}}
\de\Updates{\tx{\sf Update}}
\de\UpdateD{\tx{$\Updates^D$}}
\de\Us{\tx{\sf U}}
\de\VVV{\tx{$\scr V$}}
\de\Val{\tx{\bi Val}}
\de\Var{\tx{\bi Var}}
\de\VarSig{\tx{$\Var(\Sig)$}}
\de\VarTup{\Var\Tup}
\de\VarTupSig{\tx{$\VarTup(\Sig)$}}
\de\Vars{\tx{$\Var_s$}}
\de\VarsSig{\tx{$\Vars(\Sig)$}}
\de\Vbb{\tx{\bf V}}
\de\WWW{\tx{${\scr W}$}}
\de\Wb{\tx{\bf W}}
\de\While{\tx{\bi While}}
\de\WhileA{\tx{$\While(A)$}}
\de\WhileN{\tx{$\While^\Nbismall$}}
\de\WhileNA{\tx{$\WhileN(A)$}}
\de\WhileNSig{\tx{$\WhileN(\Sig)$}}
\de\WhileSig{\tx{$\While(\Sig)$}}
\de\WhileSigN{\tx{$\While(\SigN)$}}
\de\WhileSigx{\tx{$\While(\Sigx)$}}
\de\Whilebig{\tx{\bigbi While}}
\de\WhilebigSig{\tx{$\Whilebig(\Sigb)$}}
\de\WhilebigN{\tx{$\Whilebig^\Nbi$}}
\de\Whilebigx{\tx{$\Whilebig^\starb$}}
\de\Whilex{\tx{$\While^\starb$}}
\de\WhilexA{\Whilex(A)}
\de\WhilexSig{\Whilex(\Sig)}
\de\WhilexSigN{\Whilex(\SigN)}
\de\Whilexbig{\tx{$\Whilebig^{\dz{\starb}}$}}
\de\XXX{{\tx{${\scr X}$}}}
\de\ZZ{\tx{${\Bbb Z}$}}
\de\ZZZ{{\tx{${\scr Z}$}}}
\de\Zbi{\tx{\bi Z}}
\de\Zs{\tx{\sf Z}}
\de\Zss{{\tx{\ssf Z}}}
\de\Zz{\tx{\sf Z}}
\de\abi{\tx{\bi a}}
\de\aas{\tx{\sf a}}
\de\abar{\tx{$\bar a$}}
\de\adt{{\bi adt}}
\de\aee{\tx{\bi ae}}
\de\ainAu{\tx{$a \in \Au$}}
\de\ainAv{\tx{$a \in \Av$}}
\de\ainAs{\tx{$a \in \As$}}
\de\al{{\tx{$\alpha$}}}
\de\alA{\tx{$\alpha^A$}}
\de\alb{\tx{$\bs\alpha$}}
\de\albar{{\tx{$\ol \alpha$}}}
\de\algebras{\tx{\sf algebra}}
\de\alhat{{\tx{$\hat{\alpha}$}}}
\de\aliass{\tx{\sf alias}}
\de\aliasss{{\tx{\ssf alias}}}
\de\all{\forall}
\de\alvec{\tx{$\vec{\alpha}$}}
\de\ands{\tx{\sf and}}
\de\ang#1{\tx{$\langle #1 \rangle$}}
\de\angg#1{\tx{$\langg #1 \rangg$}}
\de\arity#1{\tx{\rm arity$(#1)$}}
\de\arr{\arrow <0.1in> [0.2,0.5]}
\de\att{{\tx{\tt a}}}
\de\atts{{\tx{\smalltt a}}}
\de\attx{\att\str}
\de\auxs{\tx{\sf aux}}
\de\avec{\tx{$\vec{a}$}}
\de\ax{\tx{$a^*$}}
\de\bA{\tx{$\bb{b}^A$}}
\de\bB{\bk{B}}
\de\bFor{\tx{\bigbi For}}
\de\bForN{\tx{$\bFor^\Nbi$}}
\de\bForx{\tx{$\bFor^\starb$}}
\de\bProc{\tx{\bigbi Proc}}
\de\bb#1{\tx{$\lbb #1 \rbb$}}
\de\bbar{[\ns]}
\de\bbi{\tx{\bi b}}
\de\bbs{\tx{\sf b}}
\de\bdot{{\bs\cdot}}
\de\be{{\tx{$\beta$}}}
\de\beA{\tx{$\beta^A$}}
\de\beb{\tx{$\bs\beta$}}
\de\begins{\tx{\sf begin}}
\de\bigcaplim#1{\underset#1\to\bigcap}
\de\bigcon{\bigwedge}
\de\bigdis{\bigvee}
\de\bigskipn{\bigskip\nin}
\de\biu{\tx{\bi u}}
\de\bk{\boldkey}
\de\bn{\bigskip\nin}
\de\bools{\tx{\sf bool}}
\de\boolss{{\tx{\ssf bool}}}
\de\br{\ |\ }
\de\bs{\boldsymbol}
\de\btt{{\tx{\tt b}}}
\de\btts{{\tx{\smalltt b}}}
\de\bttx{\btt\str}
\de\bu{$\bullet$}
\de\bul{\item{\bu}}
\de\bull{\itemm{\bu}}
\de\bulll{\itemmm{\bu}}
\de\bx{\tx{$b^*$}}
\de\capt#1#2{\ce{{\sc #1}. \ \ #2}}
\de\card#1{\tx{\bi card\,$(#1)$}}
\de\carrierss{\tx{\sf carriers}}
\de\cart#1{\tx{\bi cart\, $(#1)$}}
\de\cb#1{\ce{\bf #1}}
\de\cbb#1{\ce{\bbf #1}}
\de\cbbb#1{\ce{\bbigbf #1}}
\de\cbi#1{\ce{\bi #1}}
\de\cbbi#1{\ce{\bigbi #1}}
\de\ccc{{\tx{\bi c}}}
\de\ce{\centerline}
\de\cf{{\it cf\.}}
\de\chead#1{\bn\cb{#1}\mn}
\de\chooses{\tx{\sf choose}}
\de\cnr#1{\tx{$\ulc #1 \urc$}}
\de\comlabs{\tx{\com\bf-\labs}}
\de\comlam{\tx{$\com^\lam$}}
\de\compb{\tx{\bi comp}}
\de\compl#1{\tx{\bi compl\,$(#1)$}}
\de\complexs{\tx{\sf complex}}
\de\complexss{{\tx{\ssf complex}}}
\de\comps{\tx{\sf comp}}
\de\con{\land}
\de\concon{{\bigcon\nss\nss\nss\bigcon}}
\de\conconlim#1{\underset#1\to\concon}
\de\conconlims#1#2{\overset{#2}\to{\conconlim{#1}}}
\de\cons{\tx{\sf con}}
\de\conss{{\tx{\ssf con}}}
\de\constantss{\tx{\sf constants}}
\de\consts{\tx{\sf const}}
\de\constss{{\tx{\ssf const}}}
\de\cpxs{\tx{\sf complex}}
\de\ccs{\tx{\sf c}}
\de\css{{\tx{\ssf c}}}
\de\ctt{{\tx{\tt c}}}
\de\ctts{{\tx{\smalltt c}}}
\de\cttx{\ctt\str}
\de\curl#1{\tx{$\{#1\}$}}
\de\curly#1{\tx{$\{\,#1\,\}$}}
\de\cvals{\tx{\sf cval}}
\de\cvec{\tx{$\vec{c}$}}
\de\cx{\tx{$c^*$}}
\de\da{\downarrow}
\de\dash{\item{---}}
\de\dashh{\itemm{---}}
\de\dashhh{\itemmm{---}}
\de\datas{{\tx{\sf data}}}
\de\datass{{\tx{\ssf data}}}
\de\dcs{\tx{\sf dc}}
\de\ddis{\lor\!\!\!\!\!\lor}
\de\defaults{\tx{\sf default}}
\de\degg#1{\tx{\bi deg$(#1)$}}
\de\del{{\tx{$\delta$}}}
\de\delb{\tx{$\bs\delta$}}
\de\delhat{\hat{\delta}}
\de\depth#1{\tx{\bi depth$(#1)$}}
\de\dis{\lor}
\de\disdis{{\bigdis\nss\nss\nss\bigdis}}
\de\disdislim#1{\underset#1\to\disdis}
\de\disdislims#1#2{\overset{#2}\to{\disdislim{#1}}}
\de\displayquote{\narrower\narrower\smallrm\nin}
\de\displaytext{\narrower\narrower\sl\nin}
\de\diss{\tx{\sf dis}}
\de\divN{\tx{$\divs_{\natss}$}}
\de\divNR{\tx{$\divs_\natss^\RRR$}}
\de\divs{\tx{\sf div}}
\de\dos{\tx{\sf do}}
\de\dom#1{\tx{\bi dom$(#1)$}}
\de\down{\tx{$^\vee$}}
\de\ds{{\tx{\sf d}}}
\de\dt{\tx{\bi dt}}
\de\dtt{{\tx{\tt d}}}
\de\dz{\dsize}
\de\ebar{{\tx{$\bar e$}}}
\de\eg{{\it e.g.}}
\de\ehat{tx{$\hat e$}}
\de\el{\tx{$\ell$}}
\de\elses{\tx{\sf else}}
\de\endpf{\qed\smskip}
\de\endpr{\rm\smskip}
\de\ends{\tx{\sf end}}
\de\enum{\tx{\bi enum}}
\de\eps{{\tx{$\epsilon$}}}
\de\epsb{{\tx{$\bs\epsilon$}}}
\de\eqbool{\tx{$\eqs_{\boolss}$}}
\de\eqdata{\tx{$\eqs_\datass$}}
\de\eqdf{=_{df}}
\de\eqint{\tx{$\eqs_{\intss}$}}
\de\eqintvl{\tx{$\eqs_{\intvlss}$}}
\de\eqnat{\tx{$\eqs_{\natss}$}}
\de\eqnatN{\tx{$\eqs_{\natss}^\NNN$}}
\de\eqreal{\tx{$\eqs_{\realss}$}}
\de\eqrealR{\tx{$\eqs_{\realss}^\RRR$}}
\de\eqs{\tx{\sf eq}}
\de\eqss{\tx{$\eqs_s$}}
\de\equal{\ &= \ }
\de\es{\tx{\sf e}}
\de\etal{{\it et al.}}
\de\ett{{\tx{\tt e}}}
\de\eval{\tx{\bi eval}}
\de\evals{\tx{\sf eval}}
\de\evec{\tx{$\vec{e}$}}
\de\ex{\exists}
\de\expands{\tx{\sf expand}}
\de\fA{\tx{$f^A$}}
\de\falses{\tx{\sf false}}
\de\fbi{\tx{\bi f}}
\de\fff{\tx{{\sf f}\!{\sf f}}}
\de\ffrom{\tx{$\Leftarrow$}}
\de\fcheck{\tx{$\check f$}}
\de\fhat{\tx{$\hat f$}}
\de\first{\tx{\bi first}}
\de\fis{\tx{\sf fi}}
\de\floor#1{\tx{$\llc #1 \lrc$}}
\de\fn{\footnote}
\de\fors{\tx{\sf for}}
\de\free{\tx{\bi free}}
\de\from{\leftarrow}
\de\fs{{\tx{\sf f}}}
\de\fss{{\tx{\ssf f}}}
\de\ftil{\tx{$\til f$}}
\de\funcs{\tx{\sf func}}
\de\functionss{\tx{\sf functions}}
\de\fuu{\tx{$f^\uu$}}
\de\fvec{\tx{$\vec{f}$}}
\de\gam{{\tx{$\gamma$}}}
\de\gamA{\tx{$\gam^A$}}
\de\gamb{\tx{$\bs\gamma$}}
\de\gamx{\tx{$\gam^*$}}
\de\ghat{\tx{$\hat g$}}
\de\gtil{\tx{$\til g$}}
\de\gn{\tx{\bi gn}}
\de\gotos{\tx{\sf goto}}
\de\graph{\tx{\bi graph}}
\de\gs{{\tx{\sf g}}}
\de\gtt{{\tx{\tt g}}}
\de\gvec{\tx{$\vec{g}$}}
\de\halt{\tx{\bi halt}}
\de\hs{{\tx{\sf h}}}
\de\htil{\tx{$\til h$}}
\de\htt{{\tx{\tt h}}}
\de\ibar{{\tx{$\bar{\imath}$}}}
\de\id{\tx{\bi id}}
\de\ident{\equiv}
\de\idi{\tx{$\id_i$}}
\de\ie{{\it i.e.}}
\de\ifbool{\tx{$\ifs_{\boolss}$}}
\de\ifdata{\tx{$\ifs_{\datass}$}}
\de\ifff{\ \,\llongtofrom\ }
\de\ifffdf{\ \ \tx{$\llongtofrom_{df}$}\ }
\de\ifint{\tx{$\ifs_{\intss}$}}
\de\ifintvl{\tx{$\ifs_{\intvlss}$}}
\de\ifnat{\tx{$\ifs_{\natss}$}}
\de\ifnatN{\tx{$\ifs_{\natss}^\NNN$}}
\de\ifreal{\tx{$\ifs_{\realss}$}}
\de\ifrealR{\tx{$\ifs_\realss^\RRR$}}
\de\ifs{\tx{\sf if}}
\de\ifss{\tx{$\ifs_s$}}
\de\ifstk{\tx{$\ifs_{\stkss}$}}
\de\ift#1{&\tx{if \ {#1}}}
\de\ii{{\tx{\bi i}}}
\de\iii{{\tx{\sf i}}}
\de\iinI{{\tx{$i \in I$}}}
\de\ims{\tx{\sf im}}
\de\imp{\to}
\de\impp{\ \,\llongto\ }
\de\imps{\tx{\sf imp}}
\de\imports{\tx{\sf import}}
\de\inbull{\itemitem{\bu}}
\de\indash{\itemitem{---}}
\de\indentt{\indent\quad}
\de\ins{\tx{\sf in}}
\de\ints{{\tx{\sf int}}}
\de\intss{{\tx{\ssf int}}}
\de\intvls{{\tx{\sf intvl}}}
\de\intvlss{{\tx{\ssf intvl}}}
\de\io{{\tx{$\iota$}}}
\de\is{\tx{\sf i}}
\de\isom{\cong}
\de\itt{{\tx{\tt i}}}
\de\ivec{\tx{$\vec{\imath}$}}
\de\ivecstrut{\vec{\mathstrut i}}
\de\jbar{{\tx{$\bar{\jmath}$}}}
\de\js{\tx{\sf j}}
\de\jvec{\tx{$\vec{\jmath}$}}
\de\jvecstrut{\vec{\mathstrut j}}
\de\kap{{\tx{$\kappa$}}}
\de\kapb{\tx{$\bs\kappa$}}
\de\kbar{\tx{$\bar k$}}
\de\kinNN{\tx{$k\in\NN$}}
\de\ktt{{\tx{\tt k}}}
\de\kvec{\tx{$\vec{k}$}}
\de\kvecstrut{\vec{\mathstrut k}}
\de\kxvec{\tx{$\kvec^*$}}
\de\labs{\lamb\tx{\bi abs}}
\de\lam{{\tx{$\lambda$}}}
\de\lamPR{\lam\PR}
\de\lamPRmin{\lam\PRmin}
\de\lamPRx{\lam\PRx}
\de\lamb{\tx{$\bs\lambda$}}
\de\lammuPR{\lam\muu\PR}
\de\lammuPRmin{\lam\muu\PRmin}
\de\lammuPRx{\lam\muu\PRx}
\de\langg{\langle\!|}
\de\lb{\linebreak}
\de\lbar{\tx{$\ol l$}}
\de\lbb{[\![}
\de\leasts{\tx{\sf least}}
\de\lele#1#2#3{{\tx{$#1 \le #2 \le #3$}}}
\de\lel#1#2#3{{\tx{$#1 \le #2 < #3$}}}
\de\lgth#1{\tx{\bi lgth$(#1)$}}
\de\llc{\llcorner}
\de\lle#1#2#3{{\tx{$#1 < #2 \le #3$}}}
\de\llongfrom{\tx{$\Longleftarrow$}}
\de\llongto{\tx{$\Longrightarrow$}}
\de\llongtofrom{\tx{$\Longleftrightarrow$}}
\de\loccit{{\it loc\. cit.}}
\de\locs{\tx{\sf loc}}
\de\longfrom{\longleftarrow}
\de\longimp{\longrightarrow}
\de\longto{\longrightarrow}
\de\longtofrom{\longleftrightarrow}
\de\loops{\tx{\sf loop}}
\de\lowerbox#1#2{{\lower#1pt\hbox{$#2$}}}
\de\lrc{\lrcorner}
\de\lss{\tx{\sf less}}
\de\lsint{\tx{$\lss_\intss$}}
\de\lsintvl{\tx{$\lss_\intvlss$}}
\de\lsnat{\tx{$\lss_\natss$}}
\de\lsnatN{\tx{$\lss_\natss^\NNN$}}
\de\lsreal{\tx{$\lss_\realss$}}
\de\lsrealR{\tx{$\lss_\realss^\RRR$}}
\de\lvec{\tx{$\vec{l}$}}
\de\lxvec{\tx{$\vec{l}^*$}}
\de\lxx{\lower.2ex}
\de\lxxx{\lower.3ex}
\de\lxxxx{\lower.4ex}
\de\macros{\tx{\sf macro}}
\de\macross{{\tx{\ssf macro}}}
\de\magn#1{\magnification=\magstep#1}
\de\mapdownl#1{\mapdown\lft{\dz{#1}}}
\de\mapdownlr#1#2{\mapdown\lft{\dz{#1}}\rt{\dz{#2}}}
\de\mapdownr#1{\mapdown\rt{\dz{#1}}}
\de\maprightd#1{\mapright_{\dz{#1}}}
\de\maprightdlower#1#2{\mapright_{\lower#1pt\hbox{$\dz{#2}$}}}
\de\maprightu#1{\mapright^{\dz{#1}}}
\de\maprightud#1#2{\mapright^{\dz{#1}}_{\dz{#2}}}
\de\maprighturaise#1#2{\mapright^{\raise#1pt\hbox{$\dz{#2}$}}}
\de\maprighturaised#1#2#3{\mapright^{\raise#1pt\hbox{$\dz{#2}$}}_{\dz{#3}}}
\de\maprightudlower#1#2#3{\mapright^{\dz{#1}}_{\lower#2pt\hbox{$\dz{#3}$}}}
\de\mapupl#1{\mapup\lft{\dz{#1}}}
\de\mapuplr#1#2{\mapup\lft{\dz{#1}}\rt{\dz{#2}}}
\de\mapupr#1{\mapup\rt{\dz{#1}}}
\de\mbar{\tx{$\bar m$}}
\de\medskipn{\medskip\nin}
\de\mg#1{\tx{\bi mg$(#1)$}}
\de\mi{\tx{$^-$}}
\de\mins{\tx{\sf min}}
\de\mn{\medskip\nin}
\de\mtt{{\tx{\tt m}}}
\de\muCR{\mub\CR}
\de\muCRA{\tx{$\muCR(A)$}}
\de\muCRSig{\tx{$\muCR(\Sig)$}}
\de\muPR{\muu\PR}
\de\muPRA{\tx{$\muPR(A)$}}
\de\muPRSig{\tx{$\muPR(\Sig)$}}
\de\muPRSigb{\tx{$\muPRb(\Sigb)$}}
\de\muPRb{\mub\PRb}
\de\muPRmin{\muu\PRmin}
\de\muPRo{\muu\PRo}
\de\muPRolam{\muu\PRolam}
\de\muPRx{\muPR\str}
\de\muPRxb{\mub\PRxb}
\de\muPRxA{\tx{$\muPRx(A)$}}
\de\muPRxSig{\tx{$\muPRx(\Sig)$}}
\de\muPRxSigutos{\tx{$\muPRxSig_\utos$}}
\de\muPRxSigb{\tx{$\muPRxb(\Sigb)$}}
\de\muPRxutos{\tx{$\muPRx_\utos$}}
\de\mub{\tx{$\bs\mu$}}
\de\muu{{\tx{$\mu$}}}
\de\mvec{\tx{$\vec{m}$}}
\de\mxvec{\tx{$\mvec^*$}}
\de\n{\noindent}
\de\names{\tx{\sf name}}
\de\namess{{\tx{\ssf name}}}
\de\nats{\tx{\sf nat}}
\de\natss{{\tx{\ssf nat}}}
\de\nbar{\tx{$\bar n$}}
\de\negs{\tx{\sf neg}}
\de\news{\tx{\sf new}}
\de\newss{{\tx{\ssf new}}}
\de\nil{\tx{$\emptyset$}}
\de\nin{\noindent}
\de\ninNN{\tx{$n\in\NN$}}
\de\nl{\newline}
\de\nneg{\neg\!\!\!\!\!\neg}
\de\notover{\tx{\bi notover}}
\de\notimplies{\tx{$\ \not\nsss\llongto$}}
\de\notimplied{\tx{$\ \not\nsss\llongfrom$}}
\de\nots{\tx{\sf not}}
\de\npn{\NoPageNumbers}
\de\ns{\negthinspace}
\de\nss{\negthickspace}
\de\nsss{\nss\nss}
\de\nssss{\nsss\nsss}
\de\ntt{{\tx{\tt n}}}
\de\nub{\tx{$\bs\nu$}}
\de\nx{\tx{$n^*$}}
\de\ol{\overline}
\de\opcit{{\it op\. cit\.}}
\de\ods{\tx{\sf od}}
\de\om{{\tx{$\omega$}}}
\de\omb{{\tx{$\bs\omega$}}}
\de\onto{\twoheadrightarrow}
\de\ors{\tx{\sf or}}
\de\outs{\tx{\sf out}}
\de\ow{&\tx{otherwise}}
\de\pa{{\it p.a.}}
\de\pairs{\tx{\sf pair}}
\de\pe{\tx{\bi pe}}
\de\ph{{\tx{$\varphi$}}}
\de\pie{\tx{$\pi$}}
\de\pmapsto{{\,\overset\ {\cdot} \to\mapsto\,}}
\de\pr#1{\mn{\bf #1}. \,}
\de\prb#1{\bn{\bbf #1}.\sn}
\de\prc#1{\sn{\sc #1}.\ \ }
\de\prn#1{\mn{\bf #1}.\,}
\de\prims{\tx{\sf prim}}
\de\procs{\tx{\sf proc}}
\de\procss{{\tx{\ssf proc}}}
\de\proj{\tx{\bi proj}}
\de\projs{\tx{\sf proj}}
\de\ps{{\tx{$\psi$}}}
\de\pte{\tx{\bi pte}}
\de\pto{{\,\overset\ {\cdot} \to\longto\,}}
\de\ptt{{\tx{\tt p}}}
\de\pttx{\ptt\str}
\de\px{\tx{$p^*$}}
\de\qqquad{\qquad \qquad}
\de\qqqquad{\qqquad \qqquad}
\de\qfors{`\,\fors'}
\de\qforxs{`$\,\fors^*$'}
\de\qtt{{\tx{\tt q}}}
\de\qwhiles{`\,\whiles'}
\de\qwhilex{`$\,\whiles^*$'}
\de\rR{\tx{$\bk R$}}
\de\raisebox#1#2{{\raise#1pt\hbox{$#2$}}}
\de\ran#1{\tx{\bi ran$(#1)$}}
\de\rand{\ \tx{\rm and}\ }
\de\rangg{|\!\rangle}
\de\rats{\tx{\sf rat}}
\de\ratss{{\tx{\ssf rat}}}
\de\rbb{]\!]}
\de\reals{\tx{\sf real}}
\de\realss{{\tx{\ssf real}}}
\de\rec{\tx{\bi rec}}
\de\redSig{\,|\,_{\Sig}}
\de\reffs{\tx{\sf ref}}
\de\refss{{\tx{\ssf ref}}}
\de\rel#1{\tx{\rm (rel $#1$)}}
\de\rem{\tx{\bi rem}}
\de\res{\tx{\sf re}}
\de\rest{\restriction}
\de\restt{\tx{\bi rest}}
\de\rh{\tx{$\rho$}}
\de\rtt{{\tx{\tt r}}}
\de\rvec{\tx{$\vec{r}$}}
\de\rxx{\raise.2ex}
\de\rxxx{\raise.3ex}
\de\rxxxx{\raise.4ex}
\de\sbar{{\tx{$\bar s$}}}
\de\scr{\Cal}
\de\se{\tx{\bi se}}
\de\seq{\tx{\bi seq}}
\de\seqt{\longmapsto}
\de\setss{{\tx{\ssf set}}}
\de\shead#1#2{\mn{\bf {#1}\ \ \ {#2}}\sn}
\de\sheadb#1#2{\mn{\bf {#1}\ \ \ {#2}}\nl}
\de\sheadrun#1#2{\mn{\bf {#1}\ \ \ {#2}.\ \ }}
\de\sheads#1#2{\mn{\bf {#1}\ \ \ {#2}}}
\de\sig{{\tx{$\sigma$}}}
\de\signatures{\tx{\sf signature}}
\de\sigo{\tx{$\sig_0$}}
\de\sigp{\tx{$\sig'$}}
\de\sinSortSig{{\tx{$s \in \SortSig$}}}
\de\sinSortSigx{\tx{$s \in \SortSigx$}}
\de\skips{\tx{\sf skip}}
\de\smskip{\smallskip}
\de\smskipn{\smallskip\nin}
\de\sn{\smallskip\nin}
\de\snap{\tx{\bi snap}}
\de\sort#1{\tx{\bi sort\,{$(#1)$}}}
\de\sortss{\tx{\sf sorts}}
\de\sq{\simeq}
\de\sqsseq{\sqsubseteq}
\de\srchs{\tx{\sf srch}}
\de\ssA{{\ssz A}}
\de\ssB{{\ssz B}}
\de\ssbe{{\ssz \beta}}
\de\sset{\subset}
\de\sseq{\subseteq}
\de\sshead#1#2{{\bf {#1}\ \ \ {#2}}\sn\!}
\de\ssheadrun#1#2{{\bf {#1}\ \ \ {#2}.\ \ }}
\de\ssheads#1#2{{\bf {#1}\ \ \ {#2}}}
\de\sss{\tx{\sf s}}
\de\sz{\ssize}
\de\ssz{\sssize}
\de\starb{{\tx{$\bk *$}}}
\de\state{\tx{\bi state}}
\de\stks{{\tx{\sf stk}}}
\de\stkss{{\tx{\ssf stk}}}
\de\str{\tx{$^*$}}
\de\stt{{\tx{\tt s}}}
\de\strb{\tx{$^{\bk*}$}}
\de\subex{\tx{\bi subex}}
\de\subst{\tx{\bi subst}}
\de\sucs{\tx{\sf suc}}
\de\suu{{\tx{$s^\uu$}}}
\de\sx{{\tx{$s^*$}}}
\de\tA{\tx{$\bb{t}^A$}}
\de\ta{\tx{$\tau$}}
\de\tagx{\tag{$*$}}
\de\tagxx{\tag{$**$}}
\de\tagxxx{\tag{${*}{*}{*}$}}
\de\tagxxxx{\tag{${*}{*}{*}{*}$}}
\de\tagxxxxx{\tag{${*}{*}{*}{*}{*}$}}
\de\tagxxxxxx{\tag{${*}{*}{*}{*}{*}{*}$}}
\de\tbar{\tx{$\bar t$}}
\de\te{\tx{\bi te}}
\de\term{\tx{\bi term}}
\de\that{\tx{$\hat t$}}
\de\thens{\tx{\sf then}}
\de\tif#1{&\tx{if \ {#1}}}
\de\til{\tilde}
\de\tiltil#1{\tilde{\tilde#1}}
\de\tinT{\tx{$t\in\TT$}}
\de\th{\tx{$\theta$}}
\de\tofrom{\leftrightarrow}
\de\tos{\tx{\sf to}}
\de\toto{\rightrightarrows}
\de\trues{\tx{\sf true}}
\de\tstile{\vdash}
\de\tto{\tx{$\Rightarrow$}}
\de\ttofrom{\tx{$\Leftrightarrow$}}
\de\ttstile{\models}
\de\ttt{\tx{{\sf t}\!{\sf t}}}
\de\tup#1#2#3{\tx{$#1_{#2},\dots,#1_{#3}$}}
\de\tuptimes#1#2#3{\tx{$#1_{#2} \times \dots \times #1_{#3}$}}
\de\tvec{\tx{$\vec{t}$}}
\de\tvecstrut{\vec{\mathstrut t}}
\de\twiddle{{\tx{$\sim$}}}
\de\twiddlek{\tx{$\sim_k$}}
\de\tx{\text}
\de\txvec{\tx{$\vec{t^*}$}}
\de\type#1{\tx{\bi type$(#1)$}}
\de\ua{{\tx{$\uparrow$}}}
\de\ub#1{\tx{\underbar{#1}}}
\de\ubi{{\tx{\bi u}}}
\de\ubrivec{\tx{$u | \ivec$}}
\de\uinProdTypeSig{\tx{$u \in \ProdTypeSig$}}
\de\ul{\underline}
\de\ulc{\ulcorner}
\de\up{\tx{$^\wedge$}}
\de\urc{\urcorner}
\de\unspecs{\tx{\sf unspec}}
\de\unspecN{\tx{$\unspecs_\natss$}}
\de\unspecB{\tx{$\unspecs_\boolss$}}
\de\ups{{\tx{$\upsilon$}}}
\de\utos{{\tx{$u \to s$}}}
\de\utov{{\tx{$u \to v$}}}
\de\uu{{\tx{\ssf u}}}
\de\uuu{\tx{{\sf u}\!\i}}
\de\uuus{\tx{$\uuu_s$}}
\de\uvec{\tx{$\vec{u}$}}
\de\ux{{\tx{$u^*$}}}
\de\val{\tx{\bi val}}
\de\vals{\tx{\sf val}}
\de\valss{{\tx{\ssf val}}}
\de\var#1{\tx{\bi var$(#1)$}}
\de\vars{\tx{\sf var}}
\de\vects{\tx{\sf vect}}
\de\vinProdTypeSig{\tx{$v \in \ProdTypeSig$}}
\de\vvec{\tx{$\vec{v}$}}
\de\vvecstrut{\vec{\mathstrut v}}
\de\vx{{\tx{$v^*$}}}
\de\vxvec{\tx{$\vec{v}\,^*$}}
\de\whiles{\tx{\sf while}}
\de\wrt{w.r.t\.}
\de\wvec{\tx{$\vec{w}$}}
\de\wx{{\tx{$w^*$}}}
\de\x{\tx{\bi x}}
\de\xbar{\tx{$\bar x$}}
\de\xib{\tx{$\bs\xi$}}
\de\xii{\tx{$\xi$}}
\de\xinAbaru{\tx{$x \in \Abaru$}}
\de\xinAu{\tx{$x \in \Au$}}
\de\xinAw{\tx{$x \in \Aw$}}
\de\xtt{{\tx{\tt x}}}
\de\xttx{\tx{$\xtt^*$}}
\de\xtts{{\tx{\smalltt x}}}
\de\xttvec{\tx{$\vec{\xtt}$}}
\de\xvec{\tx{$\vec{x}$}}
\de\xx{\tx{$x^*$}}
\de\xxvec{\tx{$\xvec^*$}}
\de\xxx{\tx{${*}{*}{*}$}}
\de\xxxx{\tx{${*}{*}{*}{*}$}}
\de\xxxxx{\tx{${*}{*}{*}{*}{*}$}}
\de\xxxxxx{\tx{${*}{*}{*}{*}{*}{*}$}}
\de\yinAs{\tx{$y \in \As$}}
\de\yinAv{\tx{$y \in \Av$}}
\de\yvec{\tx{$\vec{y}$}}
\de\yx{\tx{$y^*$}}
\de\yxvec{\tx{$\yvec\,^*$}}
\de\ytt{{\tx{\tt y}}}
\de\yttx{\ytt\str}
\de\ztt{{\tx{\tt z}}}
\de\zttx{\ztt\str}
\de\zvec{\tx{$\vec{z}$}}
\de\zx{\tx{$z^*$}}
\de\zxvec{\tx{$\zvec\,^*$}}


\newcount\mmm
\de\itemn#1{\mmm=#1\par\indentn\mmm \advance\mmm by 1
      \hangindent\mmm \parindent \textindent
 }

\newcount\nnn

\global\de\indentn#1{
\nnn=#1
\loop \ifnum\nnn>0 
      \indent
      \advance\nnn by -1
\repeat
}

\de\itemm{\itemn1}
\de\itemmm{\itemn2}

\newcount\ppp

\global\de\quadn#1{
\n=#1
\loop \ifnum\ppp>0
        \quad
        \advance\ppp by -1
\repeat
}

\de\AEE{\tx{\bi AE}}
\de\AExA{\tx{$\AEE\,_\xtts^A$}}
\de\Abe{\tx{$(A,\be)$}}
\de\Abes{{\tx{$A,\ssbe$}}}
\de\AbsComp{\tx{\bi AbsComp}}
\de\AbsCompA{\tx{$\AbsComp(A)$}}
\de\Ai{\tx{$A_i$}}
\de\Ao{$\tx{\rm A}_0$}
\de\Asprod{\tx{$A_{s_1}\times\dots\times A_{s_n}$}}
\de\Ass{\tx{$A_{s_1},\dots,, A_{s_n}$}}
\de\Asua{\tx{$\As^\ua$}}
\de\AtStx{\tx{$\AtSt\,_\xtts$}}
\de\Aua{\tx{$A^\ua$}}
\de\Avua{\tx{$\Av^\ua$}}
\de\BBBbar{\tx{$\ol \BBB$}}
\de\Bbals{\tx{$\Bb_{\al,s}$}}
\de\Bbalu{\tx{$\Bb_{\al,u}$}}
\de\Bbe{\tx{$(B,\be)$}}
\de\Bbs{\tx{$\Bb_s$}}
\de\Bbu{\tx{$\Bb_u$}}
\de\Bsua{\tx{$B_s^\ua$}}
\de\Bu{\tx{$B^u$}}
\de\Bua{\tx{$B^\ua$}}
\de\Bv{\tx{$B^v$}}
\de\Bvua{\tx{$\Bv^\ua$}}
\de\CS{\tx{\bf CS}}
\de\CSo{\tx{$\CS_0$}}
\de\CSn{\tx{$\CS^n$}}
\de\CSon{\tx{$\CS_0^n$}}
\de\CalLQQe{\tx{$\Cals(\LQQe)$}}
\de\CalQQ{\tx{$\Cals(\QQ)$}}
\de\CalX{\tx{$\Cals(X)$}}
\de\CalXalbar{\tx{$(\CalX, \,\albar)$}}
\de\CalXs{\tx{$\CalX_s$}}
\de\CalXx{\tx{$\CalX^*$}}
\de\Calbar{\tx{$\Cs_\albar$}}
\de\Cals{\tx{$\Cs_\al$}}
\de\CompI{\tx{$\Comp_1$}}
\de\CompIA{\tx{$\Comp_1^A$}}
\de\CompStep{\Comp\Step}
\de\CompStepA{\tx{$\CompStep^A$}}
\de\CompStepxA{\tx{$\CompStep\,_\xtts^A$}}
\de\CompTree{\Comp\Tree}
\de\CompTreeA{\tx{$\CompTree^A$}}
\de\CompTreeASsig{\tx{$\CompTreeA(S,\sig)$}}
\de\CompTreeStage{\tx{\bi CompTreeStage}}
\de\CompTreeStageA{\tx{$\CompTreeStage^A$}}
\de\CompTreeStageB{\tx{$\CompTreeStage^B$}}
\de\CompTreeStagexA{\tx{$\CompTreeStage\,_\xtts^A$}}
\de\CompTreexA{\tx{$\CompTree\,_\xtts^A$}}
\de\CompalA{\tx{$\Comp_\al(A)$}}
\de\ConcRep{\tx{\bi ConcRep}}
\de\ConcRepA{\tx{$\ConcRep(A)$}}
\de\Cua{\tx{$C^\ua$}}
\de\Dal{\tx{$D_\al$}}
\de\Dals{\tx{$D_{\al,s}$}}
\de\Dalu{\tx{$D_\al^u$}}
\de\Dalbar{\tx{$D_\albar$}}
\de\Dalbars{\tx{$D_{\albar,s}$}}
\de\Du{\tx{$D^u$}}
\de\Dx{{\tx{$D^*$}}}
\de\Dxal{{\tx{$D^*_\al$}}}
\de\Dxalbar{\tx{$D^*_\albar$}}
\de\Dxalbars{\tx{$D^*_{\albar,s}$}}
\de\Dxbar{\tx{$\ol{D^*}$}}
\de\FAPCS{\tx{\bi FAPCS}}
\de\Findroots{\tx{\sf Findroots}}
\de\IIIp{\tx{$\III_p$}}
\de\IIItN{\tx{$\III_t^N$}}
\de\Iq{\tx{$I^q$}}
\de\LQQe{\tx{$\tx{L}(\QQ,\es)$}}
\de\LU{\tx{\bf LU}}
\de\LUF{\tx{$\LU_F$}}
\de\LUs{\tx{\smallbf LU}}
\de\LUFs{\tx{$\LUs_F$}}
\de\Lp{\tx{$L^p$}}
\de\MC{\tx{\bf MC}}
\de\MCF{\tx{$\MC_F$}}
\de\MCs{\tx{\smallbf MC}}
\de\MCFs{\tx{$\MCs_F$}}
\de\NNNN{\tx{$\NN^\NN$}}
\de\NNtoAs{\tx{$[\NN\to \!A_s]$}}
\de\NNua{\tx{$\NN^\ua$}}
\de\Nbials{\tx{$\Nbi_{\al,s}$}}
\de\Nbialu{\tx{$\Nbi_{\al,u}$}}
\de\Nbis{\tx{$\Nbi_s$}}
\de\Nbiu{\tx{$\Nbi_u$}}
\de\Omal{\tx{$\Om_\al$}}
\de\Omali{\tx{$\Om_{\al,i}$}}
\de\Omalbar{\tx{$\Om_\albar$}}
\de\Omalbars{\tx{$\Om_{\albar,s}$}}
\de\Omalbarreal{\tx{$\Om_{\albar,\realss}$}}
\de\Omalbaru{\tx{$\Om_\albar^u$}}
\de\Omalbarvua{\tx{$\Omalbarv^\ua$}}
\de\Omalbarv{\tx{$\Om_\albar^v$}}
\de\Omalbarw{\tx{$\Om_\albar^w$}}
\de\Omalreal{\tx{$\Om_{\al,\realss}$}}
\de\Omals{\tx{$\Om_{\al,s}$}}
\de\Omalu{\tx{$\Om_\al^u$}}
\de\Ombe{\tx{$\Om_\be$}}
\de\Ombes{\tx{$\Om_{\be,s}$}}
\de\Ombeu{\tx{$\Om_\be^u$}}
\de\Ombev{\tx{$\Om_\be^v$}}
\de\Ombew{\tx{$\Om_\be^w$}}
\de\Oms{\tx{$\Om_s$}}
\de\Omxs{\tx{$\Om^*_s$}}
\de\PAN{\tx{$P^{A^N}$}}
\de\PAn{\tx{$P^A_n$}}
\de\PAua{\tx{$\PA^\ua$}}
\de\PEabc{\tx{$\PE_\abctts$}}
\de\PEabcA{\tx{$\PE\,_\abctts^A$}}
\de\PPPom{\tx{$\PPP_\om$}}
\de\PPPomp{\tx{$\PPP_\om^+$}}
\de\PPPfin{\tx{$\PPP_\fin$}}
\de\PPPfinp{\tx{$\PPP_\fin^+$}}
\de\PTerms{\tx{$\PTerm_s$}}
\de\PTermx{\tx{$\PTerm_{\,\xtts}$}}
\de\PTermxs{\tx{$\PTerm_{\,\xtts,s}$}}
\de\PfA{\tx{$P_f^A$}}
\de\Pfs{\tx{$P_\fs$}}
\de\Procabc{\tx{$\Proc\,_\abctts$}}
\de\PsqcupA{\tx{$P_\sqcup^A$}}
\de\QA{\tx{$Q^A$}}
\de\QN{\tx{$\QQ^\NN$}}
\de\RRRLN{\tx{$\RRR^{<,N}$}}
\de\RRRi{\tx{$\RRR_1$}}
\de\RRRo{\tx{$\RRR_0$}}
\de\RRRp{\tx{$\RRR_p$}}
\de\RRRtN{\tx{$\RRR_t^N$}}
\de\Rat{\tx{\bi Rat\/}}
\de\RestA{\tx{$\Rest\,^A$}}
\de\RestB{\tx{$\Rest\,^B$}}
\de\RestxA{\tx{$\Rest\,_\xtts^A$}}
\de\SE{\tx{\bi SE}}
\de\SExA{\tx{$\SE\,_\xtts^A$}}
\de\Sf{\tx{$S_\fs$}}
\de\SigRRR{\tx{$\Sig(\RRR)$}}
\de\Sigf{\tx{$\Sig_\fs$}}
\de\Sinit{\tx{$S_{init}$}}
\de\StateAcupxua{\tx{$\StateA\cup\{*,\ua\}$}}
\de\StateAua{\tx{$\StateA^\ua$}}
\de\StateB{\tx{$\State(B)$}} 
\de\StatesA{\tx{$\State_s(A)$}}
\de\Statex{\tx{$\State\,_\xtts$}}
\de\StatexA{\tx{$\State\,_\xtts(A)$}}
\de\StatexAua{\tx{$\State\,_\xtts(A)^\ua$}}
\de\Stmtx{\tx{$\Stmt_{\,\xtts}$}}
\de\TExsA{\tx{$\TE\,_{\xtts,s}^A$}}
\de\PTExsA{\tx{$\PTE\,_{\xtts,s}^A$}}
\de\Terms{\tx{$\Term_s$}}
\de\Termx{\tx{$\Term_{\,\xtts}$}}
\de\Termxs{\tx{$\Term_{\,\xtts,s}$}}
\de\Tree{\tx{\bi Tree}}
\de\WhileCC{\tx{\bi WhileCC\/}}
\de\WhileCCSig{\WhileCC(\Sig)}
\de\WhileCCbig{\tx{\bigbi WhileCC}}
\de\WhileCCx{\tx{$\WhileCC^\starb$}}
\de\WhileCCxSig{\WhileCCx(\Sig)}
\de\WhileCCxSigN{\WhileCCx(\SigN)}
\de\WhileCCxbig{\tx{$\WhileCCbig^\starb$}}
\de\XXXo{\tx{$\XXX_0$}}
\de\Xal{\tx{$(X,\al)$}}
\de\Xals{\tx{$X_{\al,s}$}}
\de\Xalu{\tx{$X_\al^u$}}
\de\Xalbar{\tx{$X_\albar$}}
\de\Xalbars{\tx{$X_{\albar,s}$}}
\de\Xnorm{\tx{$(X,\,\normfn)$}}
\de\Xu{\tx{$X^u$}}
\de\Xs{\tx{$X_s$}}
\de\Xsal{\tx{$(X_s,\al)$}}
\de\Xx{{\tx{$X^*$}}}
\de\Xxal{{\tx{$X^*_\al$}}}
\de\Xxalbar{\tx{$X^*_\albar$}}
\de\Xxalbars{\tx{$X^*_{\albar,s}$}}
\de\Xxbar{\tx{$\ol{X^*}$}}
\de\Xua{\tx{$X^\ua$}}
\de\Yua{\tx{$Y^\ua$}}
\de\Zua{\tx{$Z^\ua$}}
\de\abctt{{\tx{$\att,\btt,\ctt$}}}
\de\abctts{{\tx{$\atts,\btts,\ctts$}}}
\de\aexAbe{\tx{$\aee\,_\xtts^\Abes$}}
\de\alalbar{\tx{$(\al,\albar)$}}
\de\albars{{\tx{$\albar_s$}}}
\de\albarreal{{\tx{$\albar_\realss$}}}
\de\albaru{{\tx{$\albar^u$}}}
\de\albarv{{\tx{$\albar^v$}}}
\de\albarx{{\tx{$\ol\alx$}}}
\de\albarxs{{\tx{$\ol \alxs$}}}
\de\alreal{{\tx{$\alpha_\realss$}}}
\de\als{{\tx{$\alpha_s$}}}
\de\alu{{\tx{$\alpha^u$}}}
\de\alx{{\tx{$\alpha^*$}}}
\de\alxs{{\tx{$\alpha_s^*$}}}
\de\bes{{\tx{$\beta_s$}}}
\de\beu{{\tx{$\beta^u$}}}
\de\bev{{\tx{$\beta^v$}}}
\de\bigomp{{}_{\dz\om}^{\dz+}}
\de\bimaprighturaise#1#2{\bimapright^{\raise#1pt\hbox{$\dz{#2}$}}}
\de\cP{\cnr{P}}
\de\cS{\cnr{S}}
\de\choicesttx{\tx{\tt choices\str}}
\de\choosezb{\tx{$\chooses \ \ztt:b$}}
\de\choosezt{\tx{$\chooses \ \ztt:t$}}
\de\comp{\tx{\rm comp}}
\de\complength{\tx{\bi complength}}
\de\complengthxAbe{\tx{$\complength\,_\xtts^\Abes$}}
\de\compseq{\tx{\bi compseq}}
\de\compseqxAbe{\tx{$\compseq\,_{\xtts}^\Abes$}}
\de\compstage{\tx{\bi compstage}}
\de\compstagexA{\tx{$\compstage\,_{\xtts}^A$}}
\de\compstep{\tx{\bi compstep}}
\de\compstepxAbe{\tx{$\compstep\,_{\xtts}^\Abes$}}
\de\comptree{\tx{\bi comptree}}
\de\comptreexA{\tx{$\comptree\,_\xtts^A$}}
\de\delbA{\tx{$\delb_A$}}
\de\delbAu{\tx{$\delb_A^u$}}
\de\delbAv{\tx{$\delb_A^v$}}
\de\delbAw{\tx{$\delb_A^w$}}
\de\delbs{\tx{$\delb^s$}}
\de\delbsA{\tx{$\delb^s_A$}}
\de\delbu{\tx{$\delb^u$}}
\de\delbuA{\tx{$\delb^u_A$}}
\de\domal#1{\tx{$\tx{\bi dom}_\albar(#1)$}}
\de\dsH{\tx{$\ds_\Hss$}}
\de\dss{\tx{$\ds_s$}}
\de\dsu{\tx{$\ds_u$}}
\de\dsx{\tx{$\ds^*$}}
\de\dsxs{\tx{$\ds^*_s$}}
\de\ebarn{\tx{$\ebar[n]$}}
\de\econk{\tx{$e_{\conss[k]}$}}
\de\eqp{\tx{$\eqs_p$}}
\de\enumSig{\tx{$\enum_\Sig$}}
\de\enumC{\tx{$\enums_C$}}
\de\enumX{\tx{$\enums_X$}}
\de\enums{\tx{\sf enum}}
\de\evalG{\tx{$\eval_G$}}
\de\expmins{\tx{\sf expmin}}
\de\fin{{\tx{\rm fin}}}
\de\fmin{\tx{$f^-$}}
\de\foralls{\tx{\sf forall}}
\de\fua{\tx{$f^\ua$}}
\de\gua{\tx{$g^\ua$}}
\de\haltt{\tx{\tt halt}}
\de\hyphenCCb{\tx{\bi -CC}}
\de\invR{\tx{$\invs^\RRR$}}
\de\invs{\tx{\sf inv}}
\de\isints{\tx{\sf isint}}
\de\iI{\tx{$\is_I$}}
\de\iN{\tx{$\is_N$}}
\de\kapn{\tx{$\kap[n]$}}
\de\kbarn{\tx{$\kbar[n]$}}
\de\lp{\tx{$\ell^p$}}
\de\lsp{\tx{$\lss_p$}}
\de\norm#1{\tx{$\|#1\|$}}
\de\normfn{\tx{$\|\cdot\|$}}
\de\peabcAbe{\tx{$\pe\,_\abctts^\Abes$}}
\de\piv{\tx{\rm piv}}
\de\pivo{\tx{$\piv_0$}}
\de\pivom{\tx{$\piv_\om$}}
\de\pteboolsAbe{\tx{$\pte\,_{\boolss,s}^\Abes$}}
\de\ptexsAbe{\tx{$\pte\,_{\xtts,s}^\Abes$}}
\de\qchooses{\tx{`\chooses'}}
\de\restxAbe{\tx{$\restt\,_\xtts^\Abes$}}
\de\scalars{\tx{\sf scalar}}
\de\sexAbe{\tx{$\se\,_\xtts^\Abes$}}
\de\sigG{\tx{$\sig_G$}}
\de\sigs{\tx{$\sig_s$}}
\de\sigzn{\sig\zn}
\de\spaces{\tx{\sf space}}
\de\sprod{\tx{${s_1}\times\dots\times {s_n}$}}
\de\statex{\tx{$\state\,_\xtts$}}
\de\statexA{\tx{$\state\,_\xtts(A)$}}
\de\taun{\tx{$\tau[n]$}}
\de\texboolA{\tx{$\te\,_{\boolss,s}^A$}}
\de\tom{\to_\om}
\de\tomp{\to_\om^+}
\de\totop{\toto^+}
\de\truncs{\tx{\sf truncs}}
\de\vart{\tx{\bi vart}}
\de\vartxbe{\tx{$\vart\,_\xtts^\ssbe$}}
\de\vectors{\tx{\sf vector}}
\de\xo{\tx{$x_0$}}
\de\zn{\{\ztt/n\}}

\redefine\Ds{\tx{$D_s$}}


\cbb{Abstract versus Concrete Computation on Metric Partial Algebras}
\bigskip
\bigskip
\ce{{\bf J.V. Tucker}
}
\sn
\ce{\it Department of Computer Science,}
\ce{\it University of Wales, Swansea  SA2 8PP, Wales}
\ce{\tt J.V.Tucker@swansea.ac.uk}
\mn
\ce{{\bf J.I. Zucker}\footnotemark"*"}
\footnotetext"*\ "{
\n The research of the second author was supported 
by a grant from the Natural Sciences
and Engineering Research Council (Canada)
and by a Visiting Fellowship from the Engineering
and Physical Sciences Research Council (U.K.)
}
\nl
\ce{\it Department of Computing and Software,}
\ce{\it McMaster University, Hamilton, Ont\.  L8S 4L7, Canada}
\ce{\tt zucker@mcmaster.ca}
\bn
\bn

\cbb{Abstract}
\mn
{\narrower\smallrm
Data types containing infinite data, such as the real
numbers, functions, bit streams and waveforms, are
modelled by topological many-sorted algebras. In the
theory of computation on topological algebras there is a
considerable gap between so-called abstract and concrete
models of computation. We prove theorems that bridge the
gap in the case of metric algebras with partial operations.

With an abstract model of computation on an algebra, the
computations are invariant under isomorphisms and do not
depend on any representation of the algebra. Examples of
such models are the \qwhiles\ programming language and the BCSS model. 
With a concrete model of computation, the computations
depend on the choice of a representation of the algebra
and are not invariant under isomorphisms. Usually, the
representations are made from the set \NN\ of natural numbers,
and computability is reduced to classical computability
on \NN. Examples of such models are computability via
effective metric spaces, effective domain representations, and type
two enumerability.

The theory of abstract models is stable:
there are many models of computation, and conditions under
which they are equivalent are largely known. The theory
of concrete models is not yet stable, though it seems to be
converging: several interesting models are known to be
equivalent over special types of topological algebra. We
investigate the problem of comparing the two types of
models and, hence, establishing a unified and stable theory of
computation for topological algebras.

First, we show that to compute functions on topological
algebras using an abstract model, it is necessary
that one must use algebras with partial operations and
computable functions that are continuous and multivalued.
This multivaluedness is needed even to compute
single-valued functions, and so {\it abstract models must
be nondeterministic even to compute deterministic
problems\/}. Then we choose the \qwhiles-array programming language
as an abstract model for computing on any data type,
and extend it with a nondeterministic assignment
of ``countable choice". This is the new \WhileCCx\ model.
Finally, we introduce the notion of approximable multivalued
computation on metric algebras.  As a concrete model,
we choose effective metric spaces. 
Among a number of results we prove the following.

For any metric algebra $A$ with an effective representation,
any function \WhileCCx\ approximable over $A$ is computable
in the effective representation of the metric algebra
$A$. Conversely, we show that, under certain reasonable
conditions on the effective metric algebra $A$, any
function that is effective is also \WhileCCx\ approximable. We
give an equivalence theorem, and examples 
of algebras where equivalence holds.

}

\bn
{\bf Keywords:\/}
{\smallrm
data types, abstract models of computation, concrete models of computation,
partial algebra, \qwhiles\ language, countable choice, nondeterminism,
multivalued functions, metric algebras, topological algebras,
approximation by \qwhiles\ programs, effective metric spaces, 
effective Banach spaces}

\Shead0{Introduction}
The theory of data in computer science is based on many sorted algebras
and homomorphisms.  The theory originates in the 1960s, and has
developed a wealth of theoretical concepts, methods and techniques for
the specification, construction, and verification of software and
hardware systems. It is a significant achievement in computer science
and has exerted a profound influence on programming 
\cite{wirsing,adj78,meseguer-goguen}. 
However, given the
absolutely fundamental nature of its subject matter --- data --- there are
many fascinating and significant open problems.  An important general
problem is:

{\displaytext
To develop a comprehensive theory of specification, computation and
reasoning with infinite data.

}
By infinite data we mean real numbers, spaces of functions, streams of
bits or reals, waveforms, multidimensional graphics objects, video, and
analogue and digital interfaces. The application areas are obvious:
scientific modelling and simulation, embedded systems, graphics and
multimedia communications. 

Data types containing infinite data are modelled by topological 
many-sorted algebras. In this paper we consider computability theory
on topological algebras and investigate the problem

{\displaytext
To compare and integrate high-level, representation independent,
abstract models of computation with low-level, representation dependent,
concrete models of computation in topological algebras.

}
Computability theory lies at the technical heart of 
theories of both specification
and reasoning about such systems. There are many disparate ways of defining
computable functions on topological algebras and some have (different)
significant mathematical theories. 
In the case of real numbers one can contrast the approaches in books such as
\cite{aberth80,aberth01,pourel-richards,weih:book,bcss}.

Generally speaking, the models
of computation for an algebra can be divided into two kinds: the
{\it abstract\/} and {\it concrete\/}. 

With an {\it abstract model of computation\/} for an algebra the programs 
do not depend on any representation of
the algebra and are
invariant under isomorphisms.
Abstract models originated in the late 1950s in formalising
flowcharts, and include program schemes and
many general models of recursion.
Examples of such models are the \,\While\ \,
programming language over any algebra
and the 
Blum-Cucker-Shub-Smale model \cite{bss,bcss}
over the rings of real or complex numbers.
The theory of abstract models is stable: there are
many models of computation and the conditions under which they are
equivalent are largely known \cite{tz:book,tz:hb}.
For example, \qwhiles\ programs, flow charts, register machines,
Kleene schemes, etc., are equivalent on {\it any\/} algebra;
the BCSS models are simply instances obtained
by choosing the algebra appropriate to the 
ring or ordered ring.

With a {\it concrete model of computation\/} for an algebra the programs 
and computations are not
invariant under isomorphisms, but depend on the choice
of a representation of the algebra. Usually, the representations are
made from the set \NN\ of natural numbers, and computability on an algebra
is reduced to classical computability on \NN. 
Concrete models originated in the 1940s, in formalising the computable
functions on real numbers.
Examples of general models are
computability via 
\bull
effective metric spaces \cite{moscho64},
\bull
computable sequence structures \cite{pourel-richards},
\bull
domain representations 
\cite{stolt-jvt88,stolt-jvt95,edalat95:icomp,edalat97}, and
\bull
type two enumerability \cite{weih:book}. 

\n
The theory of concrete models
is not stable though it seems to be converging: several basic models are
known to be equivalent in special cases (see, \eg, \cite{stolt-jvt99:tcs}
where the four general approaches above are shown to be
equivalent). 

In the theory of computation on algebras, abstract models are implemented
by concrete models. Thus, the gap between the models is the gap between
high level programming abstractions 
and low level implementations, and can be explored in
terms of the following concepts:

\bul
{\it Soundness of abstract model\/}: 
The functions computable in the abstract model
are also computable in the concrete model.
\bul
{\it Adequacy of abstract model\/}: 
The functions computable in the concrete model
are computable in the abstract model. 
\bul
{\it Completeness of abstract model\/}: 
Functions are computable in the abstract model
if, and only if, they are computable in the concrete model.

\n
However, there is a considerable gap between abstract and concrete
models of computation, especially over topological data types. For
example, the popular abstract model in \cite{bcss} is {\it not\/} sound 
for the main
concrete models because of its assumptions about 
the total computability of relations such as equality.
Equality on the real numbers is not everywhere continuous,
but in all the concrete models
computable functions are continuous
(\cf\ Ceitin's Theorem \cite{moscho64}).
The connection between abstract and concrete models of
computation on the real numbers is examined in \cite{tz:top}
where approximation by \qwhiles\ programs
over a {\it particular\/} algebra
was shown to be equivalent to the standard
concrete model of GL computability over the unit interval.

First attempts at bridging the gap for all topological algebras
in general have been made in 
\cite{brattka96,brattka:thesis},
using a generalisation of recursion schemes (abstract
computability) and Weihrauch's type two enumerability  (concrete
computability). Here we investigate further the problems in comparing
the two classes of models and in establishing a unified and stable theory
of computation on topological algebras.  We prove new theorems that bridge
the gap in the case of computations on metric algebras with partial
operations. 
	
By reflecting on a series of examples,
we show that to compute functions on topological algebras, it is
necessary that one must consider 

\itemm{($i$)}
algebras with partial operations,
\itemm{($ii$)}
computable functions that are both continuous and multivalued, and
\itemm{($iii$)}
approximations by abstract programs.

\n
In particular, {\it multivalued functions are needed, even to compute 
single-valued functions\/}. Thus, to prove an equivalence between abstract and
concrete models we must 
include a nondeterministic construct to define
multivalued functions, and in this way
use nondeterministic abstract  models even to
compute deterministic problems. 
We find that

{\displaytext
imperative and other abstract programming
models must be nondeterministic to express even simple programs on 
topological data types.

}
We choose the \,\While\ \,programming language as an abstract model for
computing on any data type, and extend it with the {\it nondeterministic
assignment of countable choice\/} having the form:
$$
\xtt\ ::= \ \chooses\ \ztt: b(\ztt, \xtt, \ytt)
$$
where \ztt\ is a natural number variable and $b$ is a Boolean-valued
operation. This new model is called \,\WhileCCx\ \,computability
(`{\it CC\/}' for ``countable choice", 
`\str' for array variables.) \,In particular,
we introduce a notion of {\it approximable multivalued computation\/}, and
formulate and prove the continuity of their semantics. 
We thus have the partial multivalued functions
approximable by a \WhileCCx\ program on $A$.

As a concrete model, we choose {\it effective metric spaces\/}; 
this is known to
be equivalent with several other concrete models. 
In computation with effective metric spaces $A$ we pick an enumeration
\al\ of a subspace $X$ of $A$, and construct the subspace
\,\CalX\ \,of \al-computable elements of $A$, enumerated by \albar.
We thus have the partial functions computable on \,\CalX\
\,in the representation \albar.

We then prove two theorems that can be summarised
(a little loosely) as follows.

\n
{\bi Soundness Theorem\/}: \sl Let $A$ be any  metric partial  algebra with an
effective representation \al. 
Suppose \,\CalX\ \,is a subalgebra of $A$, effective under \albar.
Then  any function $F$ on $A$
that is
\WhileCCx\ \,approximable over $A$ is computable on \,\CalX\ \,in \albar.
\endpr

The soundness theorem is technically involved but quite general,
and gives new insight into the semantics of
imperative programs applied to topological data types.
The converse theorem is more restricted in its data types:

\n
{\bi Adequacy Theorem\/}: 
\sl Let $A$ be any  metric partial  algebra A with an
effective representation \al.  Suppose the representation \al\ is \,\WhileCCx\
\,computable and dense.  Then  any function \,$F\: A \to A$  
\,that is computable on \,\CalX\ \,in \albar\
and effectively locally uniformly continuous in \al\
is  \,\WhileCCx\ \,approximable over $A$. 
\endpr

These are combined into a {\bi Completeness Theorem\/}.

The proper statements of these theorems are given as
Theorems A, B and C (in Sections 6, 7 and 8).
Some interesting applications to algebras of real numbers and to
Banach spaces are studied.

Here is the structure of the paper. We begin, in Section 1, by
explaining the role of partiality, continuity and multivaluedness
in computation, using simple examples on the real numbers. 
In Section 2 we describe topological
and metric partial algebras and their extensions. In Section 3 we
introduce the \,\WhileCCx\ \,language, give it an algebraic semantics,
and define approximable \,\WhileCCx\ \,computability.
We will see that the \WhileCCx\ \,language has a complex
semantics.  However on total algebras it defines precisely the 
\Whilex\ \,computable functions.
Section 4 is devoted to examples. In Section 5 we prove the
continuity of these \,\WhileCCx\ \,computable multivalued functions. In Section
6 we introduce our concrete model, effective metric spaces,
and prove a Soundness Theorem (Theorem $A_0$) for the special case
of surjective enumerations
of countable (not necessarliy metric) algebras.
In Section 7
we define the subspace (or subalgebra) 
of elements computable in a metric algebra,
and then prove the more general Soundness Theorem 
(Theorem A) and, in Section 8, the Adequacy Theorem (Theorem B).
These are combined into a Completeness Theorem (Theorem C)
in Section 9.
Concluding remarks are made in Section 10.

This work is part of a research programme  --- starting in \cite{tz:book}
and most recently surveyed in \cite{tz:hb} ---
on the theory of computability on algebras, and its application to
specifiability and verifiability in different areas of computer science
and mathematics. Specifically, it has developed from our studies of real
and complex number computation in \cite{tz:sanantonio,tz:top,tz:hb},
stream algebras in \cite{tz:prague,tz:markt}
and metric algebras in \cite{tz:spec}.

We thank Vasco Brattka and Kristian Stewart for
invaluable discussions.

\bn

\itemm{\bbf1\ \ }{\bbf Nondeterminism, many-valuedness, non-extensionality,
\nl
continuity and partiality: \,Some real number examples}
\sn
When one considers the relation between abstract and concrete models,
a number of intriguing problems appear.
We explain them by considering a series of examples.
Then we formulate our strategy for solving these problems.

Our chosen abstract and concrete models are introduced later (in
Sections 3 and 5, respectively), so we must explain the problems of computing
on the real number data type in rather general terms. First, we sketch
the abstract and concrete forms of the real number data type. The
picture for topological algebras in general will be clear from the
examples.
 
\newpage

\itemm{\bf 1.1}
{\bf Abstract versus concrete data types of real numbers;
\,Continuity; 
\nl
Partiality}
\sheadrun{1.1.1}{Abstract and concrete data types of reals}
To compute on the set \RR\  of real numbers with an
abstract model of computation, we have only to select an
algebra $A$ in which \RR\ is a carrier set. Abstract
computability on an algebra $A$ is a computability {\it relative
to\/} $A$: a function is computable over $A$ if it can
be programmed from the operations of $A$ using the programming constructs
of the abstract model.  Clearly, there are infinitely many choices
of operations with which to make an algebra $A$, and hence
there are infinitely many choices of classes of abstractly computable
functions. All the classes of abstractly computable functions  on \RR\
have decent mathematical theories, resembling the theory of
the computable functions on the natural numbers --- thanks to
the general theory of computable functions on many sorted
algebras \cite{tz:hb}.

In contrast, to compute on \RR\ with a concrete model
of computation, we choose an appropriate concrete
representation $R$, and map
$$
\al\: R\ \to\ \RR
$$
where $R$ is an algebra made from the set \NN\ of natural numbers. 
For example, the map will be based on the fact that the reals
can be built from the rationals, and hence the
naturals, in a variety of equivalent ways (such
as Cauchy sequences, decimal expansions, etc.). The
computability of functions on the reals is
investigated using the theory of computable functions on
\NN, applied to \RR\ via \al.

To compare this computation theory with abstract models, we
choose an algebra $A$ in which \RR\ is a carrier set
and, in particular, the operations of $A$ are computable with
respect to the representation \al. For example, multiplication by 3 is not
computable in the decimal representation, but the field
operations on \RR\ are computable in the Cauchy sequence
representation.

We assume that our concrete model is the subspace \,\CS\
\,of Baire space \,\NNNN\
\,consisting of codings of fast Cauchy sequences of rationals,
\ie, sequences \,$(k_n)$ \,of naturals such that
for all $n$ and all $m>n$, \,$|r_{k_m} - r_{k_n}| < 2^{-n}$,
where \,$r_0,r_1,r_2,\dots$
\,is some standard enumeration of the rationals.
The representing function
$$
\al\: \CS\ \to \ \RR
$$
is continuous and onto.

\sheadrun{1.1.2}
{Continuity}
Computations with real numbers involve infinite data. The
topology of \RR\ defines a process of
approximation for infinite data; the functions on the
data that are continuous in the topology are exactly the
functions that can be approximated to any desired degree
of precision.

For abstract models we assume the algebra $A$ that contains
\RR\ is a topological  algebra, \ie, one in which
the basic operations are continuous in its topologies.  We expect
further that all the computable functions will be continuous. 
The class
of functions that can be abstractly computed exactly
can be quite limited! 
With abstract models, 
approximate computations also turn out to be necessary \cite{tz:top}.

In the concrete models, moreover, it follows from 
Ceitin's Theorem \cite{moscho64} that if a function
is computable then it is continuous.

Thus, in both abstract and concrete approaches, an analysis
of basic concepts shows that 
computability implies continuity.
 
\sheadrun{1.1.3}
{Partiality}
In computing with an abstract model on $A$ we assume 
$A$ has some boolean-valued functions to test data. For
example, in computing on \RR\ we need to
use the functions  
$$
=_R\: \RR^2 \ \to  \ \BB\ \qquad\tx{and}\qquad  <_R\: \RR^2 \ \to\  \BB
$$
where \,$\BB = \curl{\ttt,\fff}$
\,is the set of booleans.

Use of these functions presents a problem, since
total continuous boolean-valued functions on the reals 
must be constant. 
This is because the only continuous functions from a connected space
to a discrete space are the constant functions.
Furthermore,
in \cite{tz:top} it was shown that on connected total topological algebras,
the \qwhiles\ and \qwhiles-array computable functions
are precisely the functions explicitly definable by terms over the algebra.

To study the full range of real number computations, we
must therefore redefine these tests as {\it partial\/} 
boolean-valued functions.
Computation with partial algebras has interesting effects
on the theory of computable functions,
as indicated in \cite{tz:top}.

On the basis of these preliminary remarks on the data type of reals, we 
turn to the examples.

\shead{1.2}{Examples of nondeterminism and many-valuedness}
We now look at three examples of computing functions on \RR.
\Examplen{1.2.1: \,Pivot function} \ Define the function
$$ \piv\:\RRn \ \pto\ \curly{1,\dots,n} $$ by $$
\piv(\tup{x}1n) \ = \ \cases \tx{some}\ i: \ x_i\ne0
\tif{such an $i$ exists}\\ \ua \ow \endcases \tag1 $$
Computation of this pivot is a vital step in the Gaussian
elimination algorithm for inverting matrices.

Note that (depending on the precise semantics
for the phrase ``some $i$" in (1))
\,\piv\ \,is {\it nondeterministic\/} or (alternatively)
{\it many-valued\/} on 
\,$\dom{\piv} = \RRn\backslash\curl0$.
\ Further:
\sn
($a$) \,There is no {\it single-valued\/} function
which satisfies the definition (1) and is 
{\it continuous\/} on \,\RRn.
For such a function,
\,being continuous and integer-valued,
would have to be constant 
on its domain \,$\RRn\backslash\curl0$,
\,with constant value (say) $j\in\curly{1,\dots,n}$.
But its value on the $x_j$-axis would have to be different from $j$,
leading to a contradiction.

\sn
($b$) \,However there {\it is\/} a computable (and hence continuous!)
single-valued function
$$
\pivo\: \CSn\ \pto\ \curly{1,\dots,n}
\tag2
$$
with a simple algorithm.
Note however that \,\pivo\ \,is {\it not extensional\/} on 
\,\CSn\  \,(\ie, not well defined on \RRn),
or (equivalently) the map (2) cannot be factored through
\,\RRn:
$$
\commdiag{
\CSn \cr
\mapdown\lft{\dz\al} &\arrow(2,-1)\lft\bdot\rt{\raisebox3{\dz\pivo}}\cr
\RRn &\mapright^\bdot_{\dz\tx{?}}&\ \curly{1,\dots,n}
}
$$
In effect, we can regain continuity (for a single-valued function),
by foregoing extensionality.

\sn
($c$)
\,Alternatively, we can maintain continuity {\it and\/} extensionality
by giving up single-valuedness.
For the many-valued function
$$
\pivom\: \RRn\ \to \ \PPPom(\curl{1,\dots,n})
$$
(where \,$\PPPom(\dots)$
\,denotes the set of countable subsets of \,\dots)
defined by: \,for all $k\in\curl{1,\dots,n}$
$$
k\,\in\,\pivom(\tup{x}1n)\ \ifff\ x_k\ne0,
$$
is {\it extensional\/} and {\it continuous\/},
where a function
$$
f\: A\ \to\ \PPPom(B)
$$
is defined to be continuous iff for all open $Y\sseq B$,
$$
f^{-1}[Y]\ := \ \curly{x\in A\br f(x)\cap Y \ne\nil}
$$
is open in $A$.
(We will consider continuity of many-valued functions
systematically in Section 5.)

\Remarksn{1.2.2}
($i$) The many-valued function \,\pivom\ \,is ``tracked"
(in a sense to be elucidated in Section 6)
by (any implementation of) \,\pivo.
\sn
($ii$) We could only recover continuity of the \,\piv\ 
\,function by giving up either extensionality
(as in ($b$)) or single-valuedness (as in ($c$)).
\sn
($ii$) Note however that the complete algorithm for inverting matrices
\,is {\it continuous\/} and {\it deterministic\/}
(hence {\it single-valued\/})
and {\it extensional\/},
even though it contains \,\pivo\ \,as an essential 
component!
\endpr

\Examplen{1.2.3: \,``Choose" a rational arbitrarily near a real}
\ Define a function
$$
F\: \RR\times\NN\ \to\ \NN
$$
by
$$
F(x,n) \ =\ \tx{``some"}\ k: \ \ds(x,r_k)\,<\,2^{-n}
\tag3
$$
where (as before) \,$r_0,r_1,r_2,\dots$
\,is some standard enumeration of the rationals.
Note again (as in Example 1.1): 
\sn
($a$)
\,There is no {\it single-valued, continuous\/} function $F$
satisfying (3).  This is because such a function, being continuous
with discrete range space, would
have to be constant in the first argument.
\sn
($b$) \,But there {\it is\/} a single-valued computable (and continuous)
function
$$
F_0\: \CS\times\NN\ \to\ \NN
$$
trivially -- just define
$$
F_0(\xi,n) \ = \ \xi_n.
$$
This is, again, {\it non-extensional\/} on \RR.

\sn
($c$)
\,Further, there is a {\it many-valued, continuous, extensional\/}
function satisfying (1):
$$
F_\om\: \RR\times\NN\ \to\ \PPPom(\NN)
$$
where
$$
F_\om(x,n) \ = \ \curly{k\br \ds(x,r_k)<2^{-n}}.
$$

\Examplen{1.2.4: \,Finding the root of a function}
\ This example is adapted from \cite{weih:book}.
Consider the function $f_a$ shown in Figure 1,
where $a$ is a parameter which can assume any real value.

\midinsert
\epsfxsize = 4.5in
\ce{\epsfbox{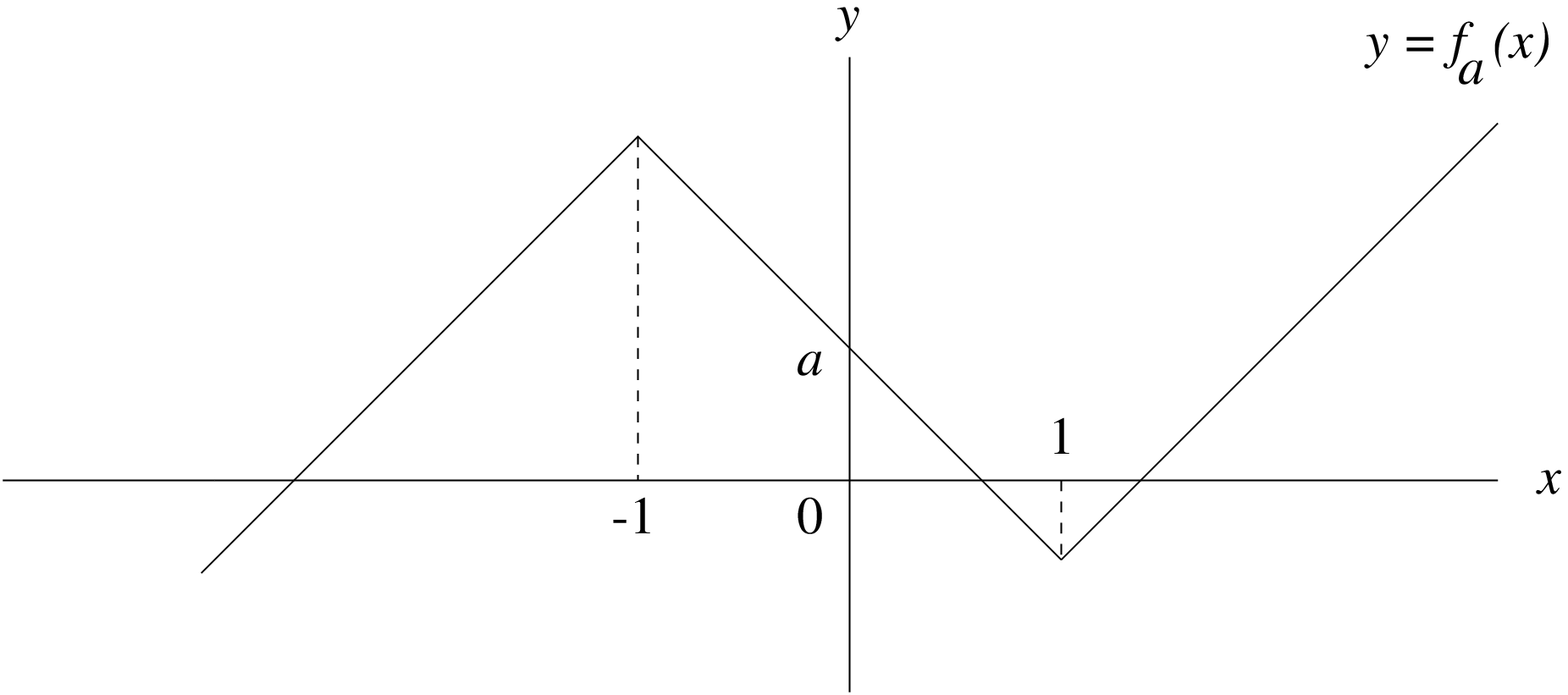}}
\mn
\ce{\sc Figure 1}
\sn
\endinsert

\sn
It is defined by
$$
f_a(x) \ = \
\cases
x+a+2 \ift{$x\le-1$}\\
a-x \ift{$-1\le x\le1$}\\
x+a-2 \ift{$1\le x$}.
\endcases
$$
This function has either 1 or 3 roots, depending 
on the size of $a$.
For $a<-1$, \,$f_a$ has a single (large positive) root;
\,for $a>1$, \,$f_a$ has a single (large negative) root;
\,and for
$-1<a<1$, \,$f_a$ has three roots,
two of which become equal when $a = \pm1$.

Let $g$ be the (many-valued) function, such that
$g(a)$ gives all the non-repeated roots of $f_a$.
This is shown in Figure 2.

\midinsert
\epsfxsize = 2in
\ce{\epsfbox{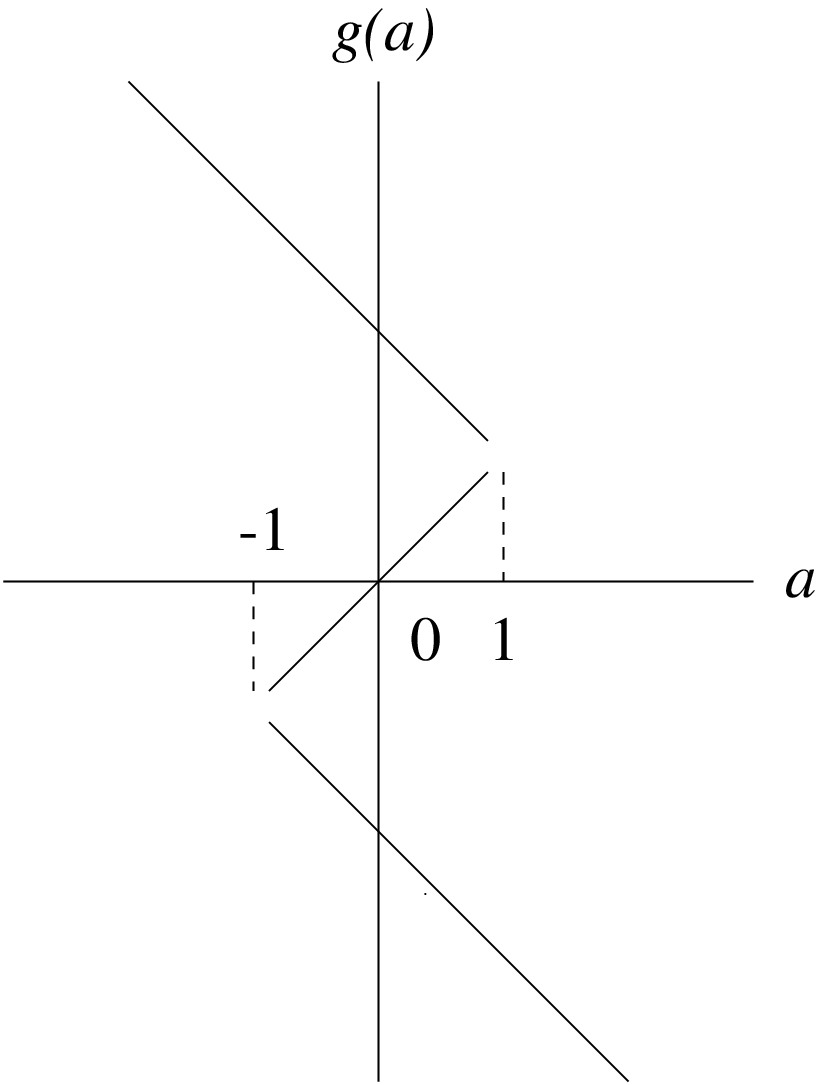}}
\mn
\ce{\sc Figure 2}
\sn
\endinsert

Again, we have the same situation as in the previous examples:
\sn
($a$) \,We cannot choose a (single) root of $f_a$ continuously
as a function of $a$.
\sn
($b$) \,However, one can easily choose and compute 
a root of $f_a$ continuously as a function
of a {\it Cauchy sequence representation\/} of $a$,
\ie, non-extensionally in $a$.
\sn
($c$) Finally, $g(a)$, as a {\it many-valued\/} function of $a$,
is continuous.
(Note that in order to have continuity,
we must exclude the 
repeated roots of $f_a$, at \,$a=\pm1$.)

\Remarkn{1.2.5}
Other examples of a similar nature abound, 
and can be treated similarly;  for example,
the problem of finding, for a given real number $x$,
an integer $n>x$.
\endpr

\newpage

\shead{1.3}{Solutions for the abstract model}
In the above three examples we have given:
\itemm{($i$)}
a number of single-valued functions 
\,$f\: \RR^n \to  \RR$ \,that we want to compute;
\itemm{($ii$)}
arguments that they are not continuous;
\itemm{($iii$)}
a prima facie case that they cannot be abstractly computed on the
abstract data type $A$ containing \RR\ because they are not continuous;
\itemm{($iv$)}
a prima facie case that they can be computed in the concrete data
type \CS;
\itemm{($v$)}
arguments that they are selection functions for many-valued
functions on \RR\ that are continuous.

\n
At the level of {\it concrete models\/} of computation,
there is not really a problem with the
issues raised by the above examples,
since concrete models work only by 
computations on {\it representations\/}
of the reals
(say by Cauchy sequences),
as described fully in Sections 5 and 7.

The real problem arises with 
the construction of {\it abstract models\/}
of computation on the reals
which should model the phenomena illustrated by these examples,
and should, moreover, correspond, in some sense,
to the concrete models.
Thus we have the question:

{\displaytext
Can such continuous many-valued functions be computed on the abstract
data type $A$ containing \RR\  using new abstract models of computation? If
they can, are the concrete and abstract models then equivalent?

}

The rest of this paper deals with these issues.
We answer the above question more generally, over many-sorted
partial metric algebras $A$.

The solution presented in this paper is to extend the \Whilex\ \,programming
language over $A$ \cite{tz:hb} with 
a nondeterministic ``countable choice" programming construct,
so that in the rules of program term formation,
$$
\chooses \ \ztt:b
$$
is a new term of type \,\nats,
\,where \,\ztt\ \,is a variable of type \,\nats
\,and $b$ is a term of type \,\bools.
We will revisit the examples after giving the language definition
in Section 3.

Alternatively, one could use other abstract models; for example,
one can modify the \,\muPRx\
\,function schemes \cite[\S8.1]{tz:hb}
by replacing the constructive least number
(\muu) operator
$$
f(x) \ \sq \ \mu z\in\NN[g(x,z) = \ttt],
$$
where $g$ is a boolean-valued function,
by a nondeterministic
choice operator:
$$
f(x) \ \sq \ \chooses \ z\in\NN[g(x,z) = \ttt].
$$

\mn
Given suitable semantics,
these two approaches turn out to be equivalent.

In \cite{brattka:thesis} a more elaborate set of recursive schemes
over many-sorted algebras,
with many-valued operations, was presented.

\Shead{2}{Topological partial algebras and continuity}
We define some basic notions concerning 
topological and metric many-sorted partial algebras.
We begin with some basic ideas and examples.
\shead{2.1}{Basic algebraic definitions}
A {\it signature} \Sig\ (for a many-sorted partial algebra)
is a pair consisting of
($i$) a finite set \SortSig\ of {\it sorts},
and ($ii$) a finite set \FuncSig\ of {\it (basic)
function symbols},
each symbol $F$ having a {\it type}  \ $\tuptimes{s}{1}{m}\to s$,
where $s_1 ,\ldots, s_m, s \in \SortSig$;
in that case we write
\ $F:\ \tuptimes{s}{1}{m}\to s$.
\ (The case $m = 0$ corresponds to {\it constant symbols}.)

A 
\Sig-{\it product type\/}
has the form
\ $u = \tuptimes{s}{1}{m}$ \ ($m\ge0$),
where \,\tup{s}{1}{m} \,are \Sig-sorts.
We use the notation
\ $u,v,w,\dots$ \ for \Sig-product types.

A partial \Sig-{\it algebra} $A$ has, for each sort $s$ of \Sig,
a non-empty {\it carrier set} \,\As\
\,of sort $s$,
and for each \Sig-function symbol
\ $F:\utos$,
\ a partial function
\ $\FA : \Au \pto \As$,
\,where, for the \Sig-product type \ $u  = \tuptimes{s}{1}{m}$,
\ we write
\ $\Au \ \eqdf \ A_{s_1} \times\dots\times A_{s_m}$.
(The notation \,$f: X\pto Y$ \,refers in general to
a partial function from $X$ to $Y$.)

The algebra $A$ is {\it total} if 
\,\FA\ is total for each \Sig-function symbol $F$.
Without such a totality assumption,
$A$ is called {\it partial}.

In this paper we deal mainly with partial algebras.
{\it The default assumption is that ``algebra"
refers to partial algebra.}
We will, nevertheless, for the sake of emphasis,
often speak explicitly of ``partial algebras".

Given an algebra $A$, we write \,$\Sig(A)$ \,for its signature.

\Examplesn{2.1.1}
The following algebras will be used repeatedly as examples 
in this paper.
All but one are total.
\sn
($a$)
\,The algebra of {\it booleans\/}
has the carrier \ $\BB = \{\ttt,\,\fff\}$
\ of sort \,\bools. 
The signature $\Sig(\BBB)$
and algebra \BBB\ respectively can be displayed as follows:
$$
\boxed{
\matrix \format\l&\quad\l\\
\signatures &\Sig(\BBB)\\
\sortss &\bools\\
\functionss&\trues, \falses:\ \ \to\bools,\\
&\ands, \ors:\bools^2\to\bools\\
&\nots: \bools\to\bools\\
\ends&
\endmatrix
}
\quad
\tx{and}
\quad
\boxed{
\matrix \format\l&\quad\l\\
\algebras &\BBB\\
\carrierss &\BB\\
\functionss&\ttt, \fff:\ \ \to\BB,\\
&\ands^\BBB, \ors^\BBB:\BB^2\to\BB\\
&\nots^\BBB: \BB\to\BB\\
\ends&
\endmatrix
}
$$
Usually the signature can essentially be inferred from the algebra;
indeed we will not define the signature where no confusion will arise.
Further, for notational simplicity, we will not always distinguish between
function names in the signature (\trues, etc.)
and their intended interpretations ($\trues^\BBB = \ttt$, etc.)
\sn
($b$) \,The algebra \,\NNNo\ \,of naturals
has a carrier \,\NN\ \,of sort \,\nats,
\,together with the zero constant and successor function:
$$
\boxed{
\matrix \format\l&\quad\l\\
\algebras &\NNNo \\
\carrierss &\NN\\
\functionss&0:\ \ \to\NN,\\
&\Ss:\NN\to\NN\\
\ends&
\endmatrix
}
$$

($c$) \,The ring \,\RRRo\ \,of reals
has a carrier \RR\ of sort \reals:
$$
\boxed{
\matrix \format\l&\quad\l\\
\algebras &\RRRo \\
\carrierss &\RR\\
\functionss&0,1:\ \ \to\RR,\\
&+,\times:\RR^2\to\RR,\\
&-:\RR\to\RR\\
\ends&
\endmatrix
}
$$
\endpr

($d$) The field \,\RRRi\ \,of reals is
formed by adding the multiplicative inverse to the ring \RRRo:
$$
\boxed{
\matrix \format\l&\quad\l\\
\algebras &\RRRi \\
\imports &\RRRo\\
\functionss &\invR:\RR\to\RR\\
\ends&
\endmatrix
}
$$
where \,
$$
\invR(x) \ = \ 
\cases 
1/x \ift{$x\ne 0$}\\
\ua \ow.
\endcases.
$$
This is an example of a  partial algebra.
More examples of partial algebras wil be given later.

Throughout this work we make the following assumption about 
the signatures \Sig.

\Assumptionn{2.1.2 \ (Instantiation Assumption)}
{\sl For every sort $s$ of \Sig, there is a closed term of that sort,
called the default term \,\delbs\ \,of that sort.}
\endpr

This guarantees the presence of {\it default values} \,\delbsA\ \, 
in a \Sig-algebra $A$ at all sorts $s$, 
and {\it default tuples} \,\delbuA\ \,at all product types $u$.

\Defn{2.1.3 \,(Expansions and reducts)}
Let \Sig\ and \Sigp\ be signatures with $\Sig \subset \Sig'$.
\sn
\ ($a$) If $A'$ is a \Sigp-algebra, \,then the \Sig-{\it reduct of} $A'$,
\ $A' \redSig$,
\ is the algebra
of signature \Sig, consisting of the carriers of $A'$
named by the sorts of \Sig\ and equipped with the functions
of $A'$ named by the function symbols of \Sig.
\sn
\ ($b$) If \,$A$ is a \,\Sig-algebra and $A'$ is a \Sigp-algebra,
\,then $A'$ is a {\it \Sigp-expansion} of $A$ iff $A$ is the
\Sig-reduct of $A'$.
\endpr

\shead{2.2}{Adding booleans: \,Standard signatures and algebras}
The algebra \BBB\ of booleans (Example 2.1.1($a$))
plays an essential role in computation, as we will see.
This motivates the following definition.

\Defn{2.2.1 \ (Standard signature)}
A signature \Sig\ is {\it standard} if
\itemitem{($i$)} 
it contains the signature of booleans, \ie,
\,$\Sig(\BBB) \ \sseq \Sig$, \ and
\itemitem{($ii$)}
The function symbols of \Sig\ include
a {\it conditional\/}
$$
\ifs_s :\bools\times s^2 \to s
$$
for all sorts $s$ of \Sig\ other than \bools.
\endpr

Now given a standard signature \Sig,
a sort of \Sig\
is called an {\it equality sort\/}
if \Sig\ includes an
{\it equality  operator}
$$
\eqs_s: s^2 \to \bools.
$$

\Defn{2.2.2 \ (Standard algebra)}
Given a standard signature \Sig,
a \Sig-algebra $A$ is a {\it standard} if
\itemm{($i$)}
it is an expansion of \BBB;
\itemm{($ii$)}
the conditional operator on each sort $s$
has its standard interpretation in $A$;
\ie, \,for \,$b\in\BB$ \,and \,$x,y \in \As$,
$$
\ifs^A_s(b,x,y) \ = \
\cases
   x \tif{$b = \ttt$}\\
   y \tif{$b = \fff$};
\endcases
$$
\itemm{($iii$)}
the equality operator
\,$\eqs_s$ \,is interpreted as a {\it partial identity\/} on each
equality sort $s$, \ie, for any two elements of \As,
if they are identical, then
the operator at these arguments returns 
\,\ttt\ \,if it returns anything; and
if they are not identical,
it returns \,\fff\
\,if anything.
More specifically, there are three possible cases.
First, the case
$$
\eqs^A_s(x,y) \ = \ 
\cases
  \ttt \ift{$x=y$}\\
  \fff \ow,
\endcases
$$
\ie, total equality, represents the situation
that equality is ``decidable" or ``computable"
at sort $s$,
for example, when $s=\nats$.
Second, the case
$$
\eqs^A_s(x,y) \ = \ 
\cases
  \ttt \ift{$x=y$}\\
  \ua \ow
\endcases
$$
represents typically the situation that
that equality is ``semidecidable".
An example is given by the initial {\it term algebra\/} 
of an r.e. equational theory.
Third, the case
$$
\eqs^A_s(x,y) \ = \ 
\cases
  \ua \ift{$x=y$}\\
  \fff \ow,
\endcases
$$
represents typically the situation that
that equality is ``co-semidecidable".
Examples are given by the data types
of {\it streams\/} and {\it real numbers\/},
as mentioned in 1.1.3; see Example 2.2.4($c$) below.
\endpr

\smskipn
Note that any many-sorted signature \Sig\
can be {\it standardised} to a signature \,\SigBBB\
\,by adjoining the sort \,\bools\ \,together
with the standard boolean operations;
and, correspondingly, any algebra $A$ can be standardised
to an algebra \ABBB\
\,by adjoining the algebra \BBB\
as well as the conditional and equality operators.

\Examplesn{2.2.4 \,(Standard algebras)}
\sn
($a$) The simplest standard algebra is the algebra \BBB\ of the booleans
(Example 2.1.1($a$)).
\smskipn
($b$) A {\it standard algebra of naturals\/} \,\NNN\
\,is formed by standardising the algebra \,\NNNo\ 
\linebreak
(Example 2.1.1($b$)),
with (total) equality and order operations on \NN:
$$
\boxed{
\matrix \format\l&\quad\l\\
\algebras &\NNN \\
\imports &\NNNo, \,\BBB\\
\functionss
&\ifnatN:\BB\times\NN^2\to\NN,\\
&\eqnatN,\,\lsnatN:\NN^2\to\BB\\
\ends&
\endmatrix
}
$$
($c$) {\it A standard partial algebra \RRR\ on the reals\/} 
is formed similarly by standardising the field \RRRi\ 
(Example 2.1.1($d$)),
with partial equality and order operations on \RR:
$$
\boxed{
\matrix \format\l&\quad\l\\
\algebras &\RRR \\
\imports &\RRRi, \,\BBB\\
\functionss
&\ifrealR:\BB\times\RR^2\pto\RR,\\
&\eqrealR,\,\lsrealR:\RR^2\pto\BB\\
\ends&
\endmatrix
}
$$
where 
\TOL
$$
\eqrealR(x,y) \ = \
\cases
\ua \tif{$x=y$}\\
\fff \tif{$x\ne y$}.
\endcases
$$
$$
\lsrealR(x,y) \ = \
\cases
\ttt \tif{$x<y$}\\
\fff \tif{$x>y$}\\
\ua \tif{$x=y$},
\endcases
\tag"and"
$$
\TOR

\Discussionn{2.2.5 \,(Semicomputability and co-semicomputability)}
The significance of the partial equality 
and order operations in Example ($c$) above, 
in connection with computability
and continuity, has been touched on in 1.1.3.
The {\it continuity\/} of partial functions will be discussed in \S2.5
(and see in particular Example 2.5.3($b$)).
Regarding {\it computability\/},
these definitions are intended to 
reflect, or capture the intuition of, 
the ``{\it semicomputability\/}" of order and the
``{\it co-semicomputability\/}" of equality
on (a concrete model of) the reals.
For given two reals $x$ and $y$, 
represented (say) by their infinite decimal expansions,
suppose their decimal digits are being read systematically,
the $n$-th digit of both at step $n$.
Then if $x\ne y$ or $x<y$,
this will become apparent after finitely many steps,
but no finite number of steps can confirm that $x=y$.
\endpr

Throughout this paper, we will assume the following,
unless specifically noted to the contrary.

\Assumptionn{2.2.6 \,(Standardness Assumption)}
The signature \Sig\ and \Sig-algebra $A$ are standard.
\endpr

\shead{2.3}{Adding counters: \ N-standard signatures and algebras}
The standard algebra \NNN\ of naturals (Example 2.2.4($b$))
plays, like \BBB, an essential role in computation.
This motivates the following definition.

\Defn{2.3.1 \ (N-standard signature)}
A signature \Sig\ is {\it N-standard\/} if
\itemm{($i$)} 
it is standard, \,and
\itemm{($ii$)} 
it contains the standard signature of naturals (Example 2.2.4($b$)), \ie,
\,$\Sig(\NNN) \ \sseq \Sig$.
\endpr

\Defn{2.3.2 \ (N-standard algebra)}
Given an N-standard signature \Sig,
a corresponding \Sig-algebra $A$ is {\it N-standard} if 
it is an expansion of \NNN.
\endpr

Note that any standard signature \Sig\ can be
N-{\it standardised} to a signature \SigN\ 
by adjoining the sort \nats\ and the operations 0, \Ss,
\eqnat, \lsnat\ and \ifnat.
Correspondingly,
any standard \Sig-algebra $A$ can be N-{\it standardised} 
to an algebra \AN\ 
by adjoining the carrier \NN\
together with the corresponding standard functions.

\Examplesn{2.3.3 \,(N-standard algebras)}
\itemm{($a$)\,}
The simplest N-standard algebra is the algebra \NNN\ 
(Example 2.2.4($b$)).
\itemm{($b$)\,}
We can N-standardise the standard real algebra
\,\RRR\ (Example 2.2.4($c$))
\,to form the algebra \,\RRRN.
\endpr

\shead{2.4}{Adding arrays: \ Algebras \,\Ax\ of signature \Sigx}
The significance of arrays for computation
is that they provide
{\it finite but unbounded memory}.

Given a standard signature \Sig,
and standard \Sig-algebra $A$, we expand \Sig\ and $A$
in two stages:

\itemm{($1^\circ$)}
N-standardise these to form \SigN\ and \AN, as in \S2.3.

\itemm{($2^\circ$)}
Define, for each sort $s$ of \Sig, the carrier \Axs\
to be the set of {\it finite sequences\/} or {\it arrays\/} \ax\
over \As, of ``starred sort" \sx.

The resulting algebras \Ax\ have signature \Sigx,
which extends \SigN\
by including,
for each sort $s$ of \Sig, the new starred sorts \sx,
and certain new function symbols.
Details are given in \cite[\S2.7]{tz:hb}
and (an equivalent but simpler version) in \cite[\S2.4]{tz:top}.

The reason for introducing starred sorts
is the lack of effective coding
of finite sequences within abstract algebras in general.

\shead{2.5}{Topological partial algebras}
We now add topologies to our partial algebras,
with the requirement of continuity 
for the basic partial functions.
Background information on topology can be obtained
from any standard text, \eg,
\cite{kelley:book,dugundji:book,engelking:book}.
\Defn{2.5.1}
Given two topological spaces $X$ and $Y$,
a partial function 
\nl
$f:X\pto Y$ 
\,is {\it continuous} if for every open \,$V\sseq Y$,
$$
f^{-1}[V] \ \eqdf \ \curly{x\in X \mid x\in\dom{f} \ \tx{and} \ f(x)\in Y}
$$
is open in $X$.
\endpr
\Defn{2.5.2}
($a$)
\,A {\it topological partial \Sig-algebra} is a partial \Sig-algebra
with topologies on the carriers
such that each of the basic \Sig-functions 
is continuous.
\smskipn($b$)
\,An (N-){\it standard topological partial algebra} 
is a topological partial algebra which is also an (N-)standard partial algebra,
such that the carriers \BB\ (and \NN) have the discrete topology.  
\endpr

\Examplesn{2.5.3}
($a$) ({\it Discrete algebras.\/})
The standard algebras \,\BBB\ \,and \,\NNN\ \,of booleans and naturals
respectively \,(\S\S2.1, 2.2) 
\,are topological (total) algebras under the discrete topology.
All functions on them are trivially continuous, since the 
carriers are discrete.
\sn
($b$) ({\it Partial real algebra.\/})
An important standard
topological partial algebra for our purpose is the real algebra \RRR\
(Example 2.2.4($c$)), or its N-standardised version \RRRN\ 
(Example 2.3.3($b$)),
in which 
\RR\ has its usual topology, and \BB\ and \NN\ the discrete topology.
Recall our earlier discussion (1.1.3) of partiality of tests in connection
with continuity, and
note that the partial operations 
\,\eqrealR\ \,and \,\lsrealR\ 
\,are continuous, in the sense of Definition 2.5.1.
\sn
($c$) ({\it Partial interval algebras.\/})
\ Another useful class of standard topological partial algebras are 
of the form
$$
\boxed{
\matrix \format\l&\quad\l\\
\algebras &\III \\
\imports &\RRR\\
\carrierss & I\\
\functionss&\iI:I\to\RR,\\
&F_1:I^{m_1}\to I,\\
&\quad\dots\\
&F_k:I^{m_k}\to I\\
\ends&
\endmatrix
}
$$
where $I$ is the closed interval \,$[0,1]$
\,(with its usual topology),
\ \iI\ is the embedding of $I$ into \RR,
\,and
\,$F_i: I^{m_i}\to I$
\,are continuous partial functions.
These are called {\it (partial) interval algebras on} $I$.
There are also N-standard versions:
$$
\boxed{
\matrix \format\l&\quad\l\\
\algebras &\IIIN \\
\imports &\RRRN\\
\carrierss & I\\
\functionss&\iI:I\to\RR,\\
&\dots\\
\ends&
\endmatrix
}
$$
\endpr

\mn($d$) ({\it N-standard total real algebra\/}.)
\,The algebra \RRRtN\ is
(``$t$" for ``total topological"), defined by
$$
\boxed{
\matrix \format\l&\quad\l\\
\algebras &\RRRtN \\
\imports &\RRRo, \,\NNN,\,\BBB\\
\functionss&\ifrealR:\BB\times\RR^2\to\RR,\\
&\divNR:\RR\times\NN\to\RR,\\
\ends&
\endmatrix
}
$$
Here \RRRo\ is the ring of reals (\S2.1.1($c$)),
\NNN\ is the standard algebra of naturals (2.2.4($b$)),
and \divN\ is division of reals by naturals.

Note that \RRRtN\ does not contain (total) boolean-valued 
functions $<$ or $=$ on the reals,
since they are not continuous
(\cf\ the partial functions \eqreal\ and \lsreal\ of \RRR).
It is therefore not an expansion of \RRR.


\Defn{2.5.4 \ (Extensions of topology to \AN\ and \Ax)}
Corresponding to the various algebraic expansions of $A$
detailed in \S\S2.3 and 2.4, there are induced topological expansions.
\sn($a$)
The topological partial N-standard algebra \AN, of signature \SigN,
is constructed from $A$ by giving the new carrier \NN\ the discrete topology.
\smskipn($b$)
The topological partial array algebra \Ax, of signature \Sigx,
is constructed from \AN\ as follows.
Viewing the elements of \Axs\ as (essentially)
arrays of elements of \As\ of finite length,
we can give
\Axs\ the {\it disjoint union} topology
of the sets \,$(\As)^n$ \,of arrays of length $n$,
for all $n\ge0$,
where each set $(\As)^n$ is given the {\it product topology\/}
of the sets \As.

The topology on \Ax\ can also be described as follows.
The {\it basic open sets\/} in \Axs\ are of the form
$$
\curly{\ax \in \Axs \mid \Lgths(\ax)> i_n \ \ \tx{and} 
\ \ \ax[i_1]\in U_1,\ \dots,\ \ax[i_n]\in U_n}
$$
for some $n>0$, \,$i_1<\dots<i_n$ \,and
open sets \,$U_1,\dots,U_n\sseq\As$.

It is easy to check that \Ax\ is indeed a topological algebra,
\ie, all the new functions of \Ax\ are continuous.

\newpage

\shead{2.6}{Metric algebra}
A particular type of topological algebra is a {\it metric 
partial algebra}.
This is
a many-sorted standard partial algebra
with an associated metric:
$$
\boxed{
\matrix \format\l&\quad\l\\
\algebras &A\\
\imports &\BBB, \,\RRR\\
\carrierss & \tup{A}1r,\\
\functionss&F^A_1:A^{u_1}\to A_{s_1},\\
&\quad\dots\\
&F^A_k:A^{u_k}\to A_{s_k},\\
&\ds^A_1:A_1^2\to\RR,\\
&\quad\dots\\
&\ds^A_r:A_r^2\to\RR\\
\ends&
\endmatrix
}
$$
where \,\BBB\ \ \,and \,\RRR\
\,are respectively the algebras of booleans and reals 
(Examples 2.1.1($a$), 2.2.4($c$)),
the carriers \,\tup{A}1r\ \,are metric spaces with metrics
\,$\ds^A_1,\dots,\ds^A_r$ \,respectively,
\,\tup{F}1k \,are the \Sig-function symbols
other than \,\tup{\ds}1k,
\,and the (partial) functions
$F^A_i$ are all continuous with respect to these metrics,
where continuity of a partial function is understood as in
Definition 2.5.1.

Clearly,
metric algebras can be viewed as special cases of 
{\it topological partial algebras\/}.

Note that 
the carrier \BB\ (as well as \NN, if present)
has the {\it discrete metric},
defined by
$$
\ds(x,y) \ = \ 
\cases
0 \tif{$x=y$}\\
1 \tif{$x\ne y$},
\endcases
$$
which induces the discrete topology.

We will often speak of a ``metric algebra $A$",
without stating the metric explicitly.

\Examplen{2.6.1}
The partial and total
real algebras \RRR, \RRRN\ and \RRRtN\ (Examples 2.5.3)
can be recast as metric algebras in an obvious way.
\endpr

\Remarkn{2.6.2 \ (Extension of metric to \Ax)}
A metric algebra $A$ can be expanded to a metric
algebra \Ax\ of arrays over $A$.
Namely, given a metric \,\dss\ \,on \As,
we define a (bounded) metric \,\dsxs\ \,on \Axs\ as follows:
\,for \,$\ax = (\tup{a}1k), \,\bx=(\tup{b}1l)\in \Axs$:
$$
\dsxs(\ax,\,\bx)\ =
\cases
1 \ift{$k\ne l$}\\
\min\big(1,\ \max_{i=0}^{k-1}\dss(\ax[i],\bx[i])\big) \ow
\endcases 
$$
This gives the same topology on \Ax\
as that induced by the topology on $A$ (Definition 2.5.4)
\cite{engelking:book}.

\Remarkn{2.6.3 \ (Product metric on $A$)}
If $A$ is a \Sig-metric algebra, then
for each \Sig-product sort \,$u=\tuptimes{s}{1}{m}$,
\,we can define a metric  
\,\dsu\ \,on \Au\ by
$$
d_u((x_1,\dots,x_m),(y_1,\dots,y_m)) \ = 
\ \max_{i=1}^m\bigl(d_{s_i}(x_i,y_i)\bigr)
$$
or more generally, by the $\ell_p$ metric
$$
d_u((x_1,\dots,x_m),(y_1,\dots,y_m)) \ = 
\ \bigl(\sum_{i=1}^m(d_{s_i}(x_i,y_i))^p\bigr)^{1/p}
\tag{($1\le p \le\infty$)}
$$
where $p=\infty$ corresponds to the ``max" metric.
This induces the product topology on \Au.
\endpr

\shead{2.7}{W-continuity: \ Another notion of continuity of partial functions}
Recall our definition (2.5.1) of continuity of partial functions:
\,$f:X\pto Y$ 
\,is continuous if for every open \,$V\sseq Y$,
\ $f^{-1}[V]$ \ is open in $X$.

This is not the only reasonable definition.
Another definition, used in \cite{weih:book} and 
\cite{brattka96,brattka:thesis}
(henceforth ``W-continuity"),
amounts to saying that $f$
is continuous iff its restriction to its domain
$$
f\rest \dom{f}:\dom{f}\to Y
$$
is continuous (as a total function),
where \,\dom{f} \,has the topology as a subspace of $A$;
or, equivalently, iff for every open \,$V\sseq Y$,
\ $f^{-1}[V]$ \ is open in \,\dom{f}.

The following is easily checked:

\Propn{2.7.1}
$f$ is continuous \ \ifff\ \ $f$ is W-continuous \,and \,\dom{f} is open.
\endpr

\Remarkn{2.7.2}
It is instructive to express these two notions of continuity
in terms of metric spaces. Suppose 
\,$f:X\pto Y$ \,where $X$ and $Y$ are metric spaces.  Then
\itemm{($a$)}
$f$ is continuous iff
$$
\all a\in \dom{f} \,\all\eps>0 \,\ex\del>0 \,\all x\in \Bb(a,\del)
\,\bigl(x\in \dom{f}\con f(x)\in \Bb(f(a),\eps)\bigr).
$$
\itemm{($b$)}
$f$ is W-continuous iff
$$
\all a\in \dom{f} \,\all\eps>0 \,\ex\del>0 \,\all x\in \Bb(a,\del)
\,\bigl(x\in \dom{f}\imp f(x)\in \Bb(f(a),\eps)\bigr).
$$

\n
Here \,$\Bb(a,\del)$
\,is the open ball with centre $a$ and radius \del.
\endpr

\Examplen{2.7.3}
Consider the partial function \, $f\: \RR\ \pto \ \RR$
\,defined by
$$
f(x) \ = \ 
\cases
  0 \ift{$x$ is an integer}\\
  \ua \ow.
\endcases
$$
Then $f$ is W-continuous, but not continuous.
\endpr

\newpage

\Shead3{`\Whilebig' programming with countable choice}
The programming language $\WhileCC = \WhileCCSig$
\,is an extension of \,\WhileSig\ \cite[\S2.1, 2.13]{tz:hb}
\,with an extra \,\qchooses\ \,rule of term formation.
We give the complete definition 
of its syntax and semantics, using the 
{\it algebraic operational semantics\/} of \cite{tz:hb}.

Assume \Sig\ is an N-standard signature,
and $A$ is an N-standard \Sig-algebra.

\shead{3.1}{Syntax of \WhileCCSig}
We define four syntactic classes:
{\it variables}, {\it terms}, {\it statements} and
{\it procedures}.

\smskipn($a$)
$\Var = \VarSig$
\,is the class of \Sig-{\it program variables\/}, and
for each \Sig-sort $s$, 
\,\Vars\
\,is the class of program variables of sort $s$:
\,$\att^s,\btt^s,\dots,\xtt^s,\ytt^s\dots$.

\smskipn($b$)
$\PTerm \,= \PTermSig$
\ is the class of \Sig-{\it program terms\/} \ $t,\dots$,
\ and for each \Sig-sort $s$,
\,\PTerms\
\,is the class of program terms of sort $s$.
These are generated by the rules
$$
t \ ::= \ \xtt^s \mid F(\tup{t}{1}{n}) \mid \chooses \ \ztt^\natss:b
$$
where \,$s,\tup{s}{1}{n}$ \,are \Sig-sorts,
\ $F: \tuptimes{s}{1}{n}\to s$ \ is a \Sig-function symbol,
\,$t_i\in \PTerm_{s_i}$ \,for \,$i=1,\dots,n$ \,($n\ge0$),
and $b$ is  a boolean term, \ie, a term of sort \,\bools.

Think of \,\qchooses\ \,as a generalisation of
the {\it constructive least number operator\/}
\,$\leasts\ \ztt:b$
\,which has the value $k$ in case 
\,$b[\ztt/k]$ \,is true
and \,$b[\ztt/i]$ \,is defined and false
for all $i<k$, \,and is undefined in case no such $k$ exists.

Here \,`\choosezb' \,selects {\it some\/} value $k$ such that
\,$b[\ztt/k]$
\,is true, if any such $k$ exists (and is undefined otherwise).
Which value is selected depends, in general, on the {\it implementation\/}
of the algebra	$A$.
In our abstract semantics,
we will give the meaning as the set of
{\it all possible $k$'s\/}
(hence ``countable choice").
Any concrete model will select a particular $k$, according to the 
implementation.

Note that
the program terms extend the algebraic terms 
(\ie, the terms over the signature \Sig)
by
including in their construction the \,\qchooses\ \,operator,
which is not an operation of \Sig.
An alternative formulation would be to have \,\qchooses\,
\,{\it not\/} as part of the term construction,
but rather as a new atomic program statement: 
\,`$\chooses\ \ztt:b$'.
We prefer the present treatment, as it leads to the construction
of {\it many-valued term semantics\/} (as we will see), 
which is interesting in itself,
and which we would get anyway if we were to extend our
syntax to include many-valued function procedure calls
in our term construction.

We write \,$t:s$ \,to indicate that \,$t \in \PTerms$,
\,and for \,$u = \tuptimes{s}{1}{m}$,
\,we write \,$t:u$
\,to indicate
that $t$ is a $u$-{\it tuple} of program terms,
\ie, a tuple of program terms of sorts \ \tup{s}{1}{m}.

We also use the notation \,$b,\dots$
\,for boolean terms.

\sn($c$)
$\AtSt = \AtStSig$
\ is the class of {\it atomic statements\/} $\Sat,\dots$ \, defined by
$$
\Sat \ ::= \ \skips \br \divs \br \xtt:= t 
$$
where 
\,`\divs' \,stands for ``divergence" (non-teremination),
and 
\,$\xtt:= t$
\,is a
{\it concurrent assignment},
\ where for some product type $u$,
\ $t:u$
\,and \xtt\ is a $u$-tuple of {\it distinct\/} variables.

\smskipn($d$)
$\Stmt = \StmtSig$
\,is the class of statements \,$S,\dots$,
\,generated by the rules
$$
S \ ::= \ \Sat \br \ S_1;S_2 \br
\ifs\ b \ \thens\ S_1 \ \elses\ S_2 \ \fis
\br \whiles\ b \ \dos\ S \ \ods
$$

\smskipn($e$)
$\Proc = \ProcSig$
\ is the class of function procedures
\ $P,Q,\dots$.
\ These have the form
$$
P \ \ident \ \funcs \ \ins \ \att\ \outs \ \btt\ \auxs \ \ctt\
\begins \ S\ \ends
$$
where \att, \btt\ and \ctt\ are lists of {\it input variables},
{\it output variables} and {\it auxiliary} (or
{\it local) variables} respectively,
and $S$ is the {\it body}.
Further, we stipulate:
\bull
\att, \btt\ and \ctt\ \,each consist of distinct variables,
and they are pairwise disjoint,
\bull
all variables occurring in $S$ must be
among \,\att, \btt\ or \ctt,
\bull
the {\it input variables\/} \att\ must not occur on
the lhs of assignments in $S$,
\bull
{\it initialisation condition}: \,$S$ has the form \,$\Sinit;S'$,
\,where \Sinit\ is a concurrent assignment
which initialises all the {\it output\/} and {\it auxiliary variables\/},
\ie, assigns to each variable in \,\btt\ \,and \,\ctt\  
\,the default term (2.1.2)
of the same sort.

If \,$\att:u$ \,and \,$\btt:v$, then
$P$ is said to have {\it type} \utov,
\ written \ $P: \utov$.
Its {\it input type} is $u$.

\shead{3.2}{Algebraic operational semantics of \WhileCC}
We will interpret programs as countably-many-valued
state transformations,
and function procedures as
countably-many-valued functions on $A$.
Our approach
follows the {\it algebraic operational semantics\/}
of \cite[\S\S3.4]{tz:hb}.
First we need some definitions and notation
for many-valued functions.

\Notationn{3.2.1}
\itemm{($a$)}
$\PPPom(X)$ \,is the set of all countable subsets of a set $X$,
including the empty set.
\itemm{($b$)}
$\PPPomp(X)$ \,is the set of all countable {\it non-empty\/}
subsets of $X$.
\itemm{($c$)}
We write \,\Yua\ \,for \,$Y\cup\curly{\ua}$,
\,where `\ua' denotes divergence.
\itemm{($d$)}
We write \,$f:X\toto Y$ \,for
\,$f:X\to\PPPom(Y)$.
\itemm{($e$)}
\,We write \,$f:X\totop Y$ \,for
\,$f:X\to\PPPomp(Y)$.
\endpr

We will interpret a \WhileCC\ \,procedure
$$
P:\utos
$$
as a countably-many-valued function \PA\ from
\Au\ to \Asua, \,\ie, as a function
$$
\PA:\ \Au\ \to\ \PPPom(\Asua)
$$
or, in the above notation:
$$
\PA: \ \Au\ \totop \ \Asua.
$$

\Remarkn{3.2.2 \ (Significance of `\ua')}
Notice that an output of, say, \,$\{2,5,\ua\}$
\,is different from \,$\{2,5\}$,
since the former indicates the possibility of divergence.
So a semantic function will have, for inputs not in its domain,
\,`\ua' \,as a possible output value.
\endpr

\Defn{3.2.3 \ (States)}
($a$) \,For each \Sig-algebra $A$, a {\it state} on $A$ is a family
\nl
\ang{\sigs\mid \sinSortSig}
\,of functions
$$
\sigs : \Vars \to \As.
$$
Let \StateA\ be the set of states on $A$,
with elements \ $\sig,\dots$.
\sn
($b$) \,Let \sig\ be a state over $A$,
\ $\xtt \ident (\tup{\xtt}{1}{n}):u$
\ and
\ $a = (\tup{a}{1}{n}) \in \Au$
\ (for $n\ge1$).
The {\it variant} \,$\sig\{\xtt / a\}$ \,of \sig\
is the state over $A$ formed from \sig\ by replacing its value
at $\xtt_i$ by $a_i$ for $i=1,\dots,n$.

We give a brief overview of 
{\it algebraic operational semantics\/}.
This was used in \cite{tz:book} for 
deterministic imperative languages with \qwhiles\
and recursion (see \cite{tz:hb} for the case of \WhileSig),
but it can be applied to a wide variety of imperative languages.
It has also been used to analyse compiler correctness \cite{stephenson:thesis}.
It can also be adapted, as we will see, to a nondeterministic language
such as \WhileCCx.

Assume ($i$) we have a meaning function for atomic statements
$$
\angg{\Sat} :\ \StateA \ \totop \ \StateAua,
$$
and ($ii$) we have defined a 
pair of functions
$$
\align
  \First&:\ \Stmt \to \AtSt\\
  \RestA&:\ \Stmt \times \StateA \to \Stmt ,
\endalign
$$
where, for a statement $S$ and state \sig,

{\displaytext
\ $\First (S)$ \ is an atomic statement
which gives the {\it first} step in the execution of $S$ 
(in any state),
and \ $\RestA (S,\sig)$ \ is a statement
(or, in the present context, a finite set of statements)
which gives the {\it rest} of the execution in state \sig.

}
\n
From these we define the
{\it computation step\/} function
$$
\CompStepA : \ \Stmt\times\StateA \totop \ \StateAua
$$
\TOL
$$
\CompStepA (S,\sig) \ = \ \angg{\First(S)}^A\sig.
\tag"by"
$$
\TOR
from which, in turn, we can define 
(for the deterministic language of \cite{tz:hb})
a {\it computation sequence\/} 
or (for the present language)
a {\it computation tree\/}.
The aim is to define a {\it computation tree stage\/} function
$$
\CompTreeStageA: \ \Stmt\times\StateA\times\NN \ \totop \ (\StateAua)^{<\om}
$$
\,where \,$\CompTreeStageA(S,\sig,n)$ \,represents the 
first $n$ stages of \,\CompTreeASsig.
Here 
\,$(\StateAua)^{<\om}$
\,denotes the set of finite sequences 
from 
\,\StateAua,
\,interpreted as finite initial segments of the paths through 
the computation tree.
From this, in turn, are defined the semantics 
of statements and procedures.

The intuition behind these semantics
is that 

{\displaytext
for any input \xinAu,
\,$\PA(x)$ is the set of all possible outcomes 
(including divergence),
for all possible implementations of the 
\,\qchooses\ \,construct,
including non-constructive implementations!

}
\n
For if (for a given input $x$) the only infinite paths through 
the semantic computation tree are non-constructive,
then $\PA(a)$ will still include `\ua'.

We now turn to the details of these definitions.

\mn{\bf ({\bi a}) \,Semantics of program terms.}
\ The meaning of
\,$t\in \PTerm_s$ \,is a function
$$
\tA :\ \StateA \totop \ \Asua.
$$
The definition is by structural induction on $t$:
$$
\align
\bb{\xtt}^A\sig \ =&\ \{\,\sig(\xtt)\,\} \\
\bb{c}^A\sig \ =&\ \{\,c^A\,\} \\
\bb{F(\tup{t}1m)}^A\sig
\ =&\quad\ \,\curly{y\mid 
\ex x_1\in A\cap\bb{t_1}\sig\,\dots\,\ex x_m\in A\cap\bb{t_m}\sig\,:
\,\FA(\tup{x}1m)\da y}\\
&\cup \ \{\,\ua \,\mid \,
\ex x_1\in A\cap\bb{t_1}\sig\,\dots \,\ex x_m\in A\cap\bb{t_m}\sig\,:
\,\FA(\tup{x}1m)\,\ua \,\}\\
&\cup \ \{\,\ua 
\mid \ua \in \bb{t_i}^A\sig\ \,\tx{for some $i$, $1\le i \le m$}\,\}\\
\bb{\ifs(b,t_1, t_2)}^A\sig
\ =&\quad\ \,\curly{y \mid 
\big(\ttt\in\bb{b}^A\sig \,\con\,y\in\bb{t_1}^A\sig\big)
\ \dis
\ \big(\fff\in\bb{b}^A\sig \,\con\,y\in\bb{t_2}^A\sig\big)}\\
&\cup \ \{\,\ua \mid \ua \in \bb{b}^A\sig \,\}\\
\bb{\choosezb}^A\sig
\ =&\quad\ \,\curly{\ninNN  \mid \ttt\in \bA\sigzn}\\
&\cup \ \curly{\ua \mid \all\ninNN
\big(\fff\in\bA\sigzn\ \dis\ \ua\in\bA\sigzn\big)}.
\endalign
$$
Notice that \,$\bb{\choosezb}^A\sig$
\,could include both natural numbers and `\ua',
since for any $n$, \,$\bA\sigzn$ \,could include both \,\ttt\ \,and \,\fff.

\mn
{\bf ({\bi b}) \ Semantics of atomic statements}.
\ The meaning of an atomic statement \,$\Sat\in\AtSt$ \,is a function
$$
\angg{\Sat} :\ \StateA \ \totop \ \StateAua
$$
defined by:
$$
\align
\angg{\skips}^A\sig \ =& \ \curl{\sig}\\
\angg{\divs}^A\sig \ =& \ \curly{\ua}\\
\angg{\xtt:=t}^A\sig \ =& 
\quad\ \,\curly{\sig\{\xtt/a\}\mid a\in A\cap\tA\sig}\\
&\cup\ \curly{\ua\mid \ua\in\tA\sig}
\endalign
$$
\mn
{\bf ({\bi c}) \,The \,\First\ \,and\ \,\Rest\ \,operations.}
\ The operation
$$
\First: \ \Stmt \ \to \ \AtSt
$$
is defined exactly as in \cite[\S3.5]{tz:hb},
namely:
$$
\First(S) \ = \
\cases
S \ift{$S$ is atomic}\\
\First(S_1) \ift{$S \ident S_1;S_2$}\\
\skips \ow.
\endcases
$$
The operation
$$
\RestA:\ \Stmt \times \StateA \ \totop \ \Stmt,
$$
is defined as follows
(\cf\ \cite[\S3.5]{tz:hb}):

\sn
{\it Case 1.}   \ $S$ is atomic. \ Then
$$
\RestA (S,\sig) \ = \ \curly{\skips}.
$$
{\it Case 2.}  \ $S \ \ident \ S_1;S_2$.
\sn\indent
{\it Case 2a.} \ $S_1$ is atomic. Then
$$
\RestA(S,\sig) \ = \ \curly{S_2}.
$$
\indent
{\it Case 2b.} \ $S_1$ is not atomic. Then
$$
\RestA(S,\sig) \ = \ 
\curly{S';S_2 \mid S' \in \RestA(S_1,\sig)}
\ \cup\ \curly{\divs\mid\divs\in\RestA(S_1,\sig)}.
$$
{\it Case 3.} \ $S \ \ident \ \ifs \ b \ \thens \ S_1 \ \elses \ S_2 \ \fis$.
\ Then \ $\RestA (S,\sig)$ \ contains all of:
$$
\cases
S_1 \tif{$\ttt\in\bA\sig$},\\
S_2 \tif{$\fff\in\bA\sig$},\\
\divs \tif{$\ua\in\bA\sig$}.
\endcases
$$
Note that more than one condition may hold.

\sn
{\it Case 4.} \ $S \ \ident \ \whiles\ b \ \dos \ S_0 \ \ods$.
\ Then \ $\RestA(S,\sig)$ \ contains all of:
$$
\cases
S_0;S \tif{$\ttt\in\bA\sig$},\\
\skips \tif{$\fff\in\bA\sig$},\\
\divs \tif{$\ua\in\bA\sig$}.
\endcases
$$
Note again that more than one condition may hold.

\mn
{\bf ({\bi d}) \,Computation step.}
\ From the \First\ function we can define the 
computation step function
$$
\CompStepA : \ \Stmt\times\StateA \totop \ \StateAua
$$
which is like the one-step computation function \,\CompIA\
\,of \cite[\S3.4]{tz:hb}, except for being multi-valued:
$$
\CompStepA (S,\sig) \ = \ \angg{\First(S)}^A\sig.
$$
\mn
{\bf ({\bi e}) \,The computation tree.}
\ The {\it computation sequence\/},
which is basic to the semantics
of \,\While\ \,computations in \cite{tz:hb},
is replaced here by a {\it computation tree\/}
$$
\CompTreeA (S,\sig)
$$
of a statement $S$ at a state \sig.
This is an \om-branching tree,
branching 
according to all possible outcomes (\ie, ``output states")
of the one-step computation function \,\CompStepA.
Each node of this tree is labelled by
either a state
or `\ua'.

Any actual (``concrete") computation of statement $S$ at state \sig\
corresponds to 
one of the paths through this tree.
The possibilities for any such path are:

\itemm{($i$)} it is finite, ending in a leaf containing a state:
the final state of the computation;

\itemm{($ii$)} it is finite, ending in a leaf containing `\ua'
(local divergence);

\itemm{($iii$)} it is infinite (global divergence).

Correspondingly, the function \,\CompA\
\,of \cite[\S3.4]{tz:hb} is replaced by a {\it computation tree stage\/}
function
$$
\CompTreeStageA: \ \Stmt\times\StateA\times\NN \ \totop \ (\StateAua)^{<\om}
$$
\,where \,$\CompTreeStageA(S,\sig,n)$ \,represents the 
first $n$ stages of \,\CompTreeASsig.
This is defined (like \CompA) by a simple recursion (``tail recursion")
on $n$:
\sn
{\it Basis:\/}
\,$\CompTreeStageA(S,\sig,0) \ = \ \curl{\sig}$,
\ \ie, \,just the root labelled by \,\sig.
\sn
{\it Induction step:\/}
\,$\CompTreeStageA(S,\sig,n)$
\,is formed by attaching to the root \,\curl{\sig}
\,the following:
\itemm{($i$)\,}
for $S$ atomic:
\,the leaf \,\curl{\sigp},
\,for each \,$\sig' \in \angg{S}^A\sig$
\,(where \,\sigp\ \,may be a state or \,\ua);
\itemm{($ii$)\,}
for $S$ not atomic:
\nl
\,the subtree \,$\CompTreeStageA(S',\sig',n)$,
\,for each \,$\sig' \in \CompStepA(S,\sig)$ \,$(\sig'\ne\ua)$
\,and \,$S'\in \RestA(S,\sig)$, \,as well as
\,the leaf \,\curly{\ua} \,if \,$\tx{`\ua'}\in \CompStepA(S,\sig)$.

Then \,\CompTreeASsig\ \,is defined as the ``limit" over $n$ of
\,$\CompTreeStageA(S,\sig,n)$.

Note that only the leaves of \,\CompTreeASsig\ 
\,may contain 
`\ua', indicating ``local divergence".

\mn
{\bf ({\bi f}) \,Semantics of statements.}
\ From the semantic computation tree we can easily define the 
i/o semantics of statements
$$
\SA :\ \StateA \ \totop \ \StateAua.
$$
Namely, 

{\displaytext
\SA\sig\ \,is the set of states and/or `\ua'
\,at all leaves in 
\CompTreeASsig, \,together
with \,`\ua' \,if \,\CompTreeASsig\ \,has an infinite path.

}

\mn
{\bf ({\bi g}) \,Semantics of procedures.}
\ Finally, if
$$
P \ \ident \ \funcs\ \ins \ \att\ \outs \ \btt\ \auxs \ \ctt
\ \begins \ S\ \ends
\tag1
$$
is a procedure of type \utov,
then its meaning in $A$ is a function
$$
\PA : \ \Au \ \totop \ \Av^\ua
$$
defined as follows (\cf\ \cite[\S3.6]{tz:hb}).
For \xinAu,
$$
\PA(x) \ = \ \curly{\sig'(\btt)\mid \sig'\in \SA\sig}
\ \cup \ \curly{\ua \mid \ua\in\SA\sig}
$$
where \sig\ is any state on $A$ such that
\,$\sig[\att] = x$.

\Remarkn{3.2.4}
From the initialisation condition (\S3.1($e$))
it follows by a ``functionality lemma" (\cf\ \cite[3.6.1]{tz:hb})
that \,\PA\ \,is well defined.
\endpr

\Defn{3.2.5}
A \WhileCC\ \,procedure 
\,$P:\utov$
\,is {\it deterministic\/} on $A$
if for all \xinAu,
\,$\PA(x)$ \,is a singleton.
\endpr

\Remarkn{3.2.6 \,(Two concepts of deterministic computation)}
One can distinguish between two notions of deterministic computation:
($i$) {\it strong deterministic computation\/}, the common concept, 
in which each step of the computation is determinate; \,and
($ii$) {\it weak deterministic computation\/}, 
in which
the output (or divergence) is 
uniquely determined by (\ie, a unique function of) the input,
but the steps in the computation are {\it not\/} determinate.
A good example of ($ii$) is the Gaussian elimination algorithm (\S1.2.1, \S4.1)
which, although defining a unique function (the inverse of a matrix),
incorporates the (nondeterministic!) pivot function
as a subroutine.
In Definition 3.2.5 and elsewhere in this paper,
we are concerned with the weak sense of deterministic computation.
\endpr

\Defn{3.2.7}
($a$) \,A many-valued function 
\,$F : \ \Au \ \totop \ \Asua$
\,is \WhileCC\ \,computable on $A$
if there is a \WhileCC\ \,procedure $P$
such that \,$F = \PA$.
\sn
($b$) \,A partial function \,$F:\Au\pto \As$
\,is \,\WhileCC\ \,{\it computable\/} on $A$
if there is a deterministic \,\WhileCC\ \,procedure 
\,$P:\utos$
\,such that for all \xinAu,
\itemm{($i$)}
\ $F(x) \da y \ \impp \ \PA(x) = \{y\}$,
\itemm{($ii$)}
\ $F(x) \ua \ \impp \ \PA(x) = \{\ua\}$,
\endpr

\Remarkn{3.2.8 \,(Many-valued algebras)}
As we have seen, the semantics for \WhileCC\ procedures
is given by countably many-valued functions.
If we were to start with algebras with many-valued basic operations,
as in \cite{brattka96,brattka:thesis}, the algebraic operational semantics
could handle this just as easily, by adapting the 
clause for the basic \Sig-function $F$ in part ($a$) 
(``Semantics of program terms") 
of the semantic definition above.
\endpr

\shead{3.3}{The language \WhileCCxSig}
In \cite{tz:top,tz:hb} we worked with
the language \Whilex\ \,rather than \While,
which can be viewed as \While\ augmented by 
auxiliary array and \,\nats\ \,variables \cite[\S3.13]{tz:hb}.
The importance of \Whilex\ 
computability lies in the fact
that it forms the basis
for a generalised Church-Turing Thesis for computability
on abstract many-sorted algebras \cite[\S8]{tz:hb}.

Here, similarly,
we will work with the language 
$\WhileCCx = \WhileCCxSig$,
which may be thought of as
\WhileCCSig\ \,augmented by auxiliary array and \,\nats\ \,variables
(or as \WhilexSig\ \,augmented by the \,\qchooses\ \,construct).
More precisely:

\Defn{3.3.1 \,(The \WhileCCxSig\ language)}
A \WhileCCxSig\ procedure is a \WhileCC(\Sigx) procedure
in which the input and output variables have sorts in \Sig\ only.
(However the auxiliary variables may have starred sorts or sort \,\nats.)
\endpr

\n
Thus a \WhileCCxSig\ procedure defines a countably-many-valued
function on any standard \Sig-algebra.

\shead{3.4}{Some computability issues in the semantics of 
\WhileCCx\ procedures}
Some interesting issues in the semantics of \WhileCCx\ arise 
already in the case of computation over the algebra \NNN\ 
of naturals (Example 2.2.4($b$)).

\itemm{{\bf (a)}}
{\bf Elimination of `\chooses' from deterministic \WhileCCx\ programs
over total algebras}

\n
The `\chooses' operator can be eliminated from deterministic
\WhileCCx\ \,pro\-ce\-dures 
(\cf\ Definition 3.2.5 and Remark 3.2.6) over total algebras.

\Propn{3.4.1}
For any total \Sig-algebra $A$ and \,$f\: \Au\pto\As$,
\sn
\ce{$f$ is \WhileCCx\ computable over $A$
\ $\ifff$
\ $f$ is \Whilex\ computable over $A$.}
\endpr

\Pf
($\tto$)
Let $P$ be a deterministic \WhileCCx\ \,procedure over $A$
which computes $f$.
Since $A$ is total,
evaluation of any boolean term $b$ over $A$
(relative to a state) converges to \ttt\ or \fff\ in $A$.
Further, since $P$ is deterministic,
its output for a given input is independent of the implementation.
Hence every `\chooses' term in $P$ of the form
\,$\chooses \ \ztt: b[\ztt]$
\, can be replaced by a 
\,\qwhiles\ \,loop which 
tests \,$b[0],\,b[1], \,b[2],\ \dots$
\,in turn, \ie,
finds
the {\it least\/} $k$ for which \,$b[k]$
\,is true, if it exists, and diverges otherwise.
\endpf

Applying this to the total algebra \NNN, and
recalling that
\Whilex\ computability over \NNN\ 
is equivalent to {\it partial recursiveness\/} 
(\ie, classical computbility) over \NN\ \cite{tz:hb},
we have:

\Corn{3.4.2}
For any \,$f\: \NN^m\pto\NN$,
\sn
\ce{$f$ is \WhileCCx\ computable over \NNN\ 
\ $\ifff$
\ $f$ is partial recursive over \NN.}
\endpr

\mn
{\bf(b) \,Recursive and non-recursive implementations}

\n
The semantics \PA\ of a procedure $P$
is given, for an input $x$,
by {\it all paths\/} of the computation tree 
\,$T=\CompTreeASsig$ \,(where $S$ is the body of $P$)
\,representing {\it all possible computation sequences\/}
for $S$ starting at state \sig, where
\,$\sig[\att] = x$,
\ie, all possible implementations of 
instances of the `\chooses' construct occurring in the execution of $S$
starting at \sig.
This gives rise to interesting computation-theoretic issues
even in the simple case that $A = \NNN$.
In this case we can assume that $T$
is coded as a subset of \NN\
in a standard way.
Now any path of $T$ ending in a leaf is finite,
and therefore (trivially) recursive.
An infinite path or computation sequence
(leading to divergence), however,
may or may not be recursive.

\Propn{3.4.3}
There is  a \WhileCCx(\NNN)\ \,procedure $P$
such that its computation tree 
has infinite paths, 
but no recursive infinite paths.
\endpr

\Pf
Our construction of $P$ is
based on the construction of a recursive tree
with infinite paths, but no recursive infinite paths 
\cite[V.5.25]{odifreddi:book}.
Let $A$ and $B$ be two disjoint r.e., recursively inseparable sets,
and suppose \,$A = \ran{f}$ \,and \,$B = \ran{g}$ 
\,where $f$ and $g$ are total recursive functions.
The procedure $P$ can be written in pseudo-code as:
$$
\align
&\funcs \ \,\auxs \ \,\ntt,\,\ktt:\nats,\\
&\qqquad \,\choicesttx:\nats\str,\quad
  \curly{\tx{array recording all choices up to present stage \,\ntt}}\\
&\qqquad \,\haltt:\bools\\
&\begins\\
&\qquad\ntt\,:=\,0;\\
&\qquad\choicesttx\,:=\Nulls;\\
&\qquad\haltt\,:=\falses;\\
&\qquad\whiles\ \,\nots\ \,\haltt\ \,\dos\\
&\qqquad \ntt\,:=\,\ntt+1;\\
&\qqquad\choicesttx\,:=\, \Newlengths(\choicesttx, \,\ntt+1);\\
&\qqquad\choicesttx[\ntt]\,:=\, 
  \chooses \ \ztt: (\ztt = 0 \ \,\ors\ \,\ztt = 1);\\
&\qqquad\fors \ \,\ktt \,:= \,0 \ \,\tos\ \,\ntt-1\ \,\dos\\
&\qqquad\qquad
\ifs\ (\choicesttx[\ktt]=0 \ \,\ands\ \,\ktt\in\curly{f(0),\dots,f(\ntt-1)}) 
\ \,\ors\\
&\qqquad\qquad\quad
(\choicesttx[\ktt]=1 \ \,\ands\ \,\ktt\in\curly{g(0),\dots,g(\ntt-1)})\\
&\qqquad\qquad\thens \ \,\haltt\,:=\,\trues\\
&\qqquad\ods\\
&\qquad \ods\\
&\ends.
\endalign
$$
Let
\,$\al_0,\,\al_1,\,\al_2,\,\dots$
\,be the successive values (0 or 1) given by the `\chooses' \,operator
in some given implementation of $P$.
Note that at stage $n$,
$$
\choicesttx[k] \ = \ \al_k \qquad \tx{for} \ \,k=0,\dots,n-1.
$$
Further, the execution diverges if, and only if, 
the set \,$C \eqdf \curly{k\mid \al_k = 1}$
{\it separates\/} $A$ and $B$ (\ie, \,$A\sseq C$ \,and \,$C\cap B = \nil$),
in which case $C$, and hence its characteristic function 
\,$\al = (\al_0, \al_1,\al_2,\dots)$,
are {\it non-recursive\/}.

Note finally that for any given 
sequence \al\ of choices,
\al\ is effectively obtainable 
from the corresponding computation sequence or path,
\ie, \al\ is recursive in that path
(with a standard coding of the computation tree).
Hence, since any infinite sequence \al\ is non-recursive,
so is the corresponding infinite path.
\endpf

\Remarksn{3.4.4}
(1) \,Clearly, $P$ as defined above
is not semantically equivalent to a \Whilex(\NNN) \,procedure.
This does not contradict Proposition 3.4.1, 
since $P$ is not deterministic.
\sn
(2) \,According to
our semantics above (\S3.2), for $P$ as defined above,
\,$\ua\in\PA()$,
\,\ie, divergence is possible.
However, if we were to restrict all computation sequences
to be recursive, then divergence would not be a possible outcome
for $\PA()$.
The semantics, as we give it (\ie, all possible computation sequences
are included, whether recursive or not) is simpler than this alternative.
In any case, as we will see,
this choice will not affect continuity considerations
(\cf\ Lemmas 5.1.7 and 5.2.1).
\endpr

\shead{3.5}{Approximable \,\WhileCCx\ \,computability}
The basic notion of computability that we will be using
in working with metric algebras
is not so much computability,
as rather {\it computable approximability on metric algebras\/}, 
as discussed in \cite[\S9]{tz:top}. We have to adapt the definition
given there to the nondeterministic case
with countable choice.

Let $A$ be a metric \Sig-algebra,
$u$ a \Sig-product type and $s$ a \Sig-type.
Let \,$P:\nats\times u \ \to \ s$
\,be a \WhileCCxSigN\ \,procedure.
Put
$$
\PAn \ \eqdf \ \PA(n,\,\cdot\,):\ \Au \ \totop \ \Asua.
$$
Note that that 
for all \xinAu, \,$\PAn(x)\ne \nil$.

\Defn{3.5.1 \,(\WhileCCx\ \,approximability to a single-valued function)}
\nl
Let \,$F:\Au\pto \As$
\,be a single-valued partial function on $A$.
\sn
($a$)
\,$F$ is {\it \WhileCCx\ approximable\/} by $P$ on $A$
if for all \,\ninNN\ \,and \,all \xinAu:
$$
x \in \dom{F} \ \impp\ \ua \,\notin \,\PAn(x)\ \sseq\ \Bb(F(x),\,2^{-n}).
\tag1
$$
($b$) $F$ is {\it strictly \WhileCCx\ approximable\/} by $P$ on $A$
if in addition to (1), 
$$
x \notin \dom{F} \ \impp\ \PAn(x)=\curly{\ua}.
\tag2
$$

\Remarkn{3.5.2}
If $F$ is strictly approximable by $P$, then 
(from (1) and (2))
for all \xinAu\ and all $n$:
$$
F(x)\ua \ \ \ifff\ \ \ua \in \PAn(x) \ \ \ifff\ \ \PAn(x) = \curly{\ua}.
$$
\endpr

Clearly, \WhileCCx\ \,{\it computability\/} is a special case of
\WhileCCx\ \,{\it approximability\/}.
\endpr

\Defn{3.5.3 \ (\WhileCCx\ \,approximability to a many-valued function)}
\nl
Let \,$F:\Au\toto \As$
\,be a countably-many-valued function on $A$.
\sn
\itemm{($a$)}
\,$F$ is {\it \WhileCCx\ approximable\/} by $P$ on $A$
if for all \,\ninNN\ \,and \,all \xinAu:
$$
\aligned
F(x)\ne\nil \ \ \impp\ \ \ua \,\notin \,\PAn(x)\ &\sseq\ 
  \bigcup_{y\in F(x)}\Bb(y,\,2^{-n}) \\
\tx{and} \quad F(x) \ &\sseq\ \bigcup_{y\in\PAn(x)}\Bb(y,\,2^{-n}).
\endaligned
\tag3
$$
Note that 
(assuming $\ua\notin\PAn(x)$)
the r.h.s. of (3) implies 
$$
\dsH(\ol{F(x)},\,\ol{\PAn(x)}) \,\le \,2^{-n},
\tag4
$$
and is implied by 
$$
\dsH(\ol{F(x)},\,\ol{\PAn(x)}) \,< \,2^{-n},
\tag5
$$
where \,$\ol{X}$ \,denotes the closure of $X$,
and \dsH\ is the {\it Hausdorff metric\/} on the set of
closed, bounded non-empty subsets of \As\ \cite[4.5.23]{engelking:book}.
(Actually, the Hausdorff metric applies only to the space of closed 
{\it bounded\/} subsets of a given metric space,
so (4) and (5) should be taken as heuristic statements.)

\n
In other words 
(assuming $F(x)\ne0$), for all \xinAu\ and all $n$,
\,{\it each\/} output of $F(x)$ lies within $2^{-n}$ of
some output of $\PAn(x)$,
and vice versa.

\itemm{($b$)}
$F$ is {\it strictly \WhileCCx\ approximable\/} by $P$ on $A$
if in addition,
$$
\F(x) = \nil \ \ \impp\ \ \PAn(x) = \curly{\ua}.
$$
\endpr

\Remarkn{3.5.4}
(\Cf\ Remark 3.5.2.)
\,If $F$ is strictly approximable by $P$, then 
for all \xinAu\ and all $n$:
$$
F(x)=\nil \ \ \ifff\ \ \ua \in \PAn(x) \ \ \ifff\ \ \PAn(x) = \curly{\ua}.
$$
\endpr
 
\bn

\itemm{\bbf4\ \ }{\bbf Examples of \WhileCCxbig\ computations and approximating
\nl
computations}


\shead{4.1}{Discussion: \,Use of `\chooses' for searching and dovetailing}
Following the examples in Section 1,
the `\chooses' construct was introduced to compute many-valued functions.
Technically, the `\chooses' construct strengthens the power
of the \While\ \,language in performing searches.
In a {\it partial algebra\/}, simple searches 
(\eg, ``find some $x_k$ in an effectively enumerated set 
\,$X = \curly{x_0,x_1,x_2,\dots}$ 
\,satisfying $b(x_k)$")
will obviously fail in general if the search simply
follows the given enumeration of $X$ 
(\ie, testing in turn whether \,$b(x_0)$, $b(x_1)$, $b(x_2)$, \dots 
holds),
since the computation of the boolean predicate $b(x)$ may not terminate
for some $x$.

This problem is overcome, at the {\it concrete model\/} level,
by the use of scheduling techniques such as 
{\it interleaving\/} or ``{\it dovetailing\/}":
\,at stage $n$, do $n$ steps in testing whether $b(x_i)$
holds, for $i=0,\dots,n$.

An important function of the `\chooses' construct, which 
will recur in our examples, is to simulate such scheduling techniques
at the {\it abstract model\/} level.
This allows searches over any countable subset $X$
of an algebra $A$ that has a computable enumeration
$$
\enumX:\ \NN\ \to\ X,
$$
since we can search $X$ in $A$ by assignments such as
$$
\xtt\ :=\ \enumX(\chooses\ \ztt:b(\enumX(\ztt))).
$$
\endpr

\shead{4.2}{Examples}
We now illustrate the use of the \WhileCCx\ \,language
in topological partial algebras
with examples,
which involve computations which are either many-valued,
or approximating, or both.
The examples given in \S1.2 to motivate many-valued abstract computation
are a good place to start.
They can be displayed in the table:

\bn
\ruledtable
\| Exact computation | Approximating computation \crthick
Single-valued   \| Gaussian elimination | $e^x$, sin($x$), etc.    \cr
Many-valued \| Approx. points in metric algebra | All simple roots of polynomial
\endruledtable
\sn

Examples 4.2.1, 4.2.2 and 4.2.4 below are all based on the 
metric algebra derived from \,\RRRN\ \,(Example 2.3.3($b$)).

\Examplen{4.2.1 \,(Gaussian elimination)}
This is a single-valued exact computation.
The algorithm can be found in any standard 
text of numerical computation,
\eg, \cite{heath:book}.
It is deterministic, but only in the weak sense
(\cf\ Remark 3.2.6), since it contains, as an
essential component, the computation of the {\it pivot\/} function (\S1.2),
which is many-valued,  and can be formalised simply with the
`\chooses' construct:
$$
\align
&\funcs \ \ \ins\ \ \xtt_1,\dots,\xtt_n\:\reals\\
&\qquad\outs\ \ \itt\:\nats\\
&\qquad\auxs\ \ \ktt\:\nats\\
&\begins\\
&\qquad\itt\,:= 
  \,\chooses \ \ktt\: (\ktt = 1 \ \,\ands\ \,\xtt_1 \ne 0) \ \,\ors\\
  &\qqqquad\quad          (\ktt = 2 \ \,\ands\ \,\xtt_2 \ne 0) \ \ors\\
  &\qqqquad\qqquad \dots\\
  &\qqqquad\quad         (\ktt = n \ \,\ands\ \,\xtt_n \ne 0)\\
&\ends.
\endalign
$$

\Examplen{4.2.2 \,(Approximations to \,${\bk e}^{\bk x}$)}
On the N-standard interval algebra \,\IIIN\ (Example 2.5.3($c$))
we give a \While\ procedure to approximate the function $e^x$ on $I$.
$$
\align
&\funcs \ \ \ins\ \ \ntt\:\nats,\qquad\curly{\tx{degree of approximation}}\\
&\qqquad \xtt\:\intvls\\
&\qquad\outs\ \ \stt\:\reals \qquad\,\curly{\tx{partial sum of power series}}\\
&\qquad\auxs \ \ \ytt\:\reals,\quad\ \,\curly{\tx{current term of series}}\\
&\qqquad \,\ktt\:\nats\qquad\,\curly{\tx{counter}}\\
&\begins\\
&\qquad\ktt\,:=\,0;\\
&\qquad\ytt\,:=\,1;\\
&\qquad\stt\,:=\, 1;\\
&\qquad\whiles\ \,\ktt<2^{\ntt+1}\ \,\dos\\
&\qqquad \ktt\,:=\,\ktt+1;\\
&\qqquad\ytt\,:=\, \ytt\times\iI(\xtt)/\,\iN(\ktt);
    \qquad\curly{\ytt=\xtt^\ktt/\ktt\,!}\\
&\qqquad\stt\,:=\,\stt+\ytt \qqquad\qquad\quad
\curly{\stt = \sum_{i=0}^\ktt \xtt^i/i\,!}\\
&\qquad \ods\\
&\ends
\endalign
$$                                                                              
where \,$\iI:I\to\RR$
\,is the embedding of $I$ in \RR,
which is primitive in \Sig(\IIIN),
and \,$\iN:\NN\to\RR$ \,is the embedding of \NN\ in \RR,
which is easily definable in \While(\RRRN).

Denoting the above function procedure by $P$, 
and \IIIN\ by $A$,
we have the semantics
$$
P_n^A\: I \ \to\ \RR
$$
with 
$$
P_n^A(x) \ = \ \sum_{i=0}^{2^{n+1}} {{x^i}\over{i\,!}}
$$
and so for all $x\in I$,
$$
\ds(P_n^A(x),\,e^x)\ < \ 2^{-n},
$$
\ie, $e^x$ is \While\ approximable on \,\IIIN\ by $P$.

This computation of $e^x$ is single-valued, but approximating.

\Examplen{4.2.3 \,(``Choosing" a member of an enumerated subspace
close to an arbitrary element of a metric algebra)}
Given a metric algebra $A$ with a countable dense
subspace $C$, \ and an enumeration of $C$
$$
\enumC\: \NN\ \onto\ C
$$
in the signature,
we want to compute a function
$$
F\:A\times\NN\ \to\ C
$$
such that
$$
F(a,n) \ =\ \tx{``some" $x\in C$ \,such that \,$\ds(a,x)\,<\,2^{-n}$}.
$$
This is a generalised version of the problem
of approximating reals by rationals (Example 1.2.3).

Here is a \WhileCCx\ procedure (in pseudo-code) 
for an exact computation of this function.
(Note that the real-valued function \,$2^{-n}$ \,is \While\ computable on
\,\RRRN, \,and hence on $A$.)
$$
\align
&\funcs \ \ins\ \ \ \att:\spaces,\\
&\qqquad \ntt:\nats\\
&\qquad\outs\ \ \xtt:\spaces\\
&\qquad\auxs\ \ \ktt:\nats\\
&\begins\\
&\qquad\xtt\,:= 
  \,\enumC \big(\chooses \ \ktt: \ds(\att, \enumC(\ktt))<2^{-\ntt}\big)\\
&\ends
\endalign
$$                                                                              
This computation is many-valued, but exact.

\Examplen{4.2.4 \,(Finding simple roots of a polynomial)}
We construct a 
\WhileCC\ procedure to approximate ``some" simple root 
of a polynomial $p(X)$ with real coefficients,
using the method of bisection.
By a {\it simple  root of\/} $p(X)$ we mean a real root 
at which $p(X)$ changes sign.
(See \cite{heath:book}. In practice, a hybrid method is generally used,
involving bisection, Newton's method, etc.)

Fundamental to the bisection method is the concept 
of 
a {\it bracket\/} for $p(X)$,
which means an
interval \,$[a,b]$
\,such that $p(a)$ and $p(b)$ have opposite signs.
By {\it rational bracket\/},
we mean a bracket with rational endpoints.

We note the following:

\itemm{(1)}
Any bracket for $p$ contains a 
root of $p$
(by the Intermediate Value Theorem),
in fact a simple root of $p$.

\itemm{(2)}
Conversely, any simple root of $p$
is contained in
a rational bracket for $p$
of arbitrarily small width.

\itemm{(3)}
If $x$ is a simple root of $p$,
then any bracket for $p$ of sufficiently small width
which contains $x$,
contains no other simple root of $p$.

\itemm{(4)}
If $[a,b]$ is a bracket for $p$,
then, putting \,$m = (a+b)/2$,
exactly one of the following holds:
\itemmm{$(i)$}
$p(m)= 0$; then $m$ is a root of $p$ (not necessarily simple);
\itemmm{$(ii)$}
$p(m)$ has the same sign as $p(a)$;
then $[m,b]$ is a bracket for $p$;
\itemmm{$(iii)$}
$p(m)$ has the same sign as $p(b)$;
then $[a,m]$ is a bracket for $p$.

It follows from the above that
starting with any
rational bracket $J$ for $p$,
we can, by repeated bisection, 
find a nested sequence of rational brackets
$$
J = J_0, \ J_1,\ J_2,\dots
\qquad
\tx{where}
\qquad
\bigcap_{n=0}^\infty J_n = \curl{x}
$$
for some simple root $x$ of $p$.
Then, letting $r_n$ be the left-hand endpoint
of $J_n$, we have a fast Cauchy sequence
\,$\ang{r_n}_n$
\,with limit $x$.

One complication with our algorithm
is the occurrence of case ($i$) in (4) above, 
\ie, the case that the midpoint $m$ of the bracket
is itself a root of $p$,
since by the co-semicomputabil\-ity of
equality (Discussion 2.2.5) on \RR\ we can only verify when
$f(m)\ne 0$, not when $f(m)=0$.
We therefore proceed as follows.
By means of the `\chooses' construct,
we search in the middle third (say) of the bracket $[a,b]$
for a ``division point", \ie, a rational point $d$ such that $f(d)\ne0$,
producing 
either $[a,d]$ or $[d,b]$ as a sub-bracket.

This new bracket may not halve the width
of $[a,b]$; in the worst case its width
is $2/3$ the width of $[a,b]$.
However a second iteration of this procedure
leads to a bracket of width at most
$(2/3)^2 < 1/2$ the width of $[a,b]$,
and so $2n$ iterations lead to a bracket of width
less than $2^{-n}\,\times$ the width of $[a,b]$.

This new bracket may not halve the width
of $[a,b]$; in the worst case its width
is $2/3(b-a)$.
However a second iteration of this procedure
leads to a bracket of width at most
$(2/3)^2 < 1/2$ the width of $[a,b]$,
and so $2n$ iterations lead to a bracket of width
less than $2^{-n}(b-a)$.

For convenience,
we will use the following two conservative extensions
to our ``official" programming notation:

\itemm{($a$)}
Simultaneously choosing two naturals with a single condition:
$$
\ktt_1,\ktt_2\ := \ \chooses\ \ztt_1, \ztt_2: b[\ztt_1,\ztt_2]
$$
which is easily expressible in \WhileCC\
\,by the use of a primitive recursive pairing function
\pairs\ on \NN\ and its inverses \,$\projs_1, \projs_2$:
$$
\align
\ktt\ &:=\ \chooses\ \ztt: b[\projs_1(\ztt),\projs_2(\ztt)];\\
\ktt_1,\ktt_2\ &:=\ \projs_1(\ktt), \projs_2(\ktt)
\endalign
$$
\itemm{($b$)}
Choosing a rational (of type \reals) satisfying a boolean condition:
$$
\qtt\ := \ \chooses\ \rtt^\realss: 
\big(\tx{``\rtt\ is rational"} \ \ands \ b[\rtt]\big)
$$
Let \,$\rats: \NN\to \RR$
\,be a \While-computable enumeration
of the rationals in \RR. 
Then this can be interpreted as:
$$
\qtt := \rats\big(\chooses\ \ktt: b[\rats(\ktt)]\big)
$$

\n
Finally, a polynomial $p(X)$  over \RR\
will be represented by an element \px\ of \RRx:
$$
\px\ = \ (\tup{a}0{n-1})\ = \ \sum_{i=0}^{n-1}a_i X^{n-i}
$$
Its evaluation at a point $c$,
denoted by $\px(c)$, 
is easily seen to be
\While(\RRR) computable in \px\ and $c$.

Now we give a \WhileCCx\  procedure for approximably computing
some simple root of an
input polynomial, in the signature of \RRR.
$$
\align
&\funcs \ \ins\quad \ \ \ntt:\nats,\qquad
  \ \curly{\tx{degree of approximation}}\\
&\qqquad \,\pttx:\reals\str\
  \qquad\curly{\tx{input polynomial, given by list of coefficients}}\\
&\qquad\outs\ \ \ \ \xtt:\reals \qquad\ \curly{\tx{approximation to root}}\\
&\qquad\auxs\ \att,\btt:\reals, \qquad\curly{\tx{endpoints of bracket}}\\
&\qqquad\ \ \,\dtt:\reals,\qquad\curly{\tx{division point of bracket}}\\
&\qqquad \ \ \ \ktt:\nats\qquad\ \curly{\tx{counter}}\\
&\begins\\
&\qquad\ktt\,:=\,0;\\
&\qquad
  \att, \btt\,:=\, \chooses\ \att, \btt: (\tx{``\att\ and \btt\ are rational"} 
  \ \ \ands\ \ \att < \btt < \att+1 \ \ \ands\\
& \qqqquad\qqquad\ \ (\pttx(\att)>0 \ \,\ands\ \,\pttx(\btt)<0)\\
& \qqqquad\qquad\quad
   \ \ors\ \ (\pttx(\att)<0 \ \,\ands\ \,\pttx(\btt)>0));\\
&\qquad\whiles\ \,\ktt<2\ntt\ \,\dos\\
&\qqquad \ktt\,:=\,\ktt+1;\\
&\qqquad \dtt\,:=\, \chooses\ \dtt: (\tx{``\dtt\ is rational"}\ \ \ands
  \ \ (2\att+\btt)/3 < \dtt < (\att+2\btt)/3 \\
&\qqqquad\qqquad \ands\ \ \pttx(\dtt)\ne 0);\\
&\qqquad \ifs\ \ (f(\dtt)>0\ \,\ands\ \,f(\att)>0)
  \ \ \ors \ \ (f(\dtt)<0\ \,\ands\ \,f(\att)<0)\\
&\qqquad\qquad\thens \ \ \att,\btt\,:=\, \dtt,\btt\qquad
    \curly{\tx{new bracket on right part of old}}\\
&\qqquad\qquad\elses \ \ \,\,\att,\btt\,:=\, \att,\dtt\qquad
    \curly{\tx{new bracket on left part of old}}\\
&\qqquad \fis\\
&\qquad \ods;\\
&\qquad \xtt\,:= \,\att \qqqquad\qquad
   \ \ \curly{\tx{$\xtt\,:=\,\btt$ \,would also work here}}\\
&\ends.
\endalign
$$                                                                              
For input natural $n$ and polynomial $p$,
the output is within $2^{-n}$ of some simple root of $p$.
Further, for {\it any\/} simple root $e$ of $p$, there is
{\it some\/} implementation of the `\chooses' operator
which will give an output within $2^{-n}$ of $e$.
Finally, the computation will diverge
if, and only if, $p$ has no simple roots. 

This computation is both many-valued and approximating.

\newpage

\itemm{\bbf5\ \ }{\bbf Countably-many-valued functions; \ 
Continuity of \WhileCCxbig
\nl
computable functions}

\mn
In this section we discuss the continuity of
countably-many-valued functions, and then prove that the
countably-many-valued functions
computed by \WhileCCx\ programs
are continuous.

\shead{5.1}
{Topology and continuity with countably many values and 
`$\bs \uparrow$'}
Recall Notation 3.2.1.

\Defn{5.1.1 \ (Totality)}
The function \,$f:X\toto Y$
\,is said to be {\it total\/} if for all \,$x\in X$, \,$f(x)$ 
is a {\it non-empty\/} subset of $Y$,
\ie, if \,$f:X\totop Y$.
\endpr

Our semantic functions (in Section 6) will typically be of the form
$$
\Phi: \ \Au \ \totop \ \Avua.
\tag1
$$

\Remarkn{5.1.2}
We think of the ``deterministic version" of (1)
as being a total function \,\Ph, \,where
for each $x\in X$, $\Ph(x)$ is a {\it singleton\/},
containing either an element of \Av\ (to indicate convergence)
or `\ua' (to indicate divergence).
(Recall Remark 3.2.2.)
\endpr

We must now consider what it means for such a function (1)
to be {\it continuous\/}.

\Defn{5.1.3 \ (Continuity)}
Let \,$f\:X\toto Y$, where $X$ and $Y$ are topological spaces.
\sn($a$)
For any \,$V\sseq Y$,
$$
f^{-1}[V] \ \eqdf\ \ \curly{x\in X \br f(x)\cap V \ne \nil},
$$
\ie, \,$x\in f^{-1}[V]$ \,iff 
at least one of the elements of \,$f(x)$ \,lies in $V$.
\sn($b$)
$f$ is {\it continuous\/}
(w.r.t. $X$ and $Y$) iff for all open  \,$V\sseq Y$,
\,$f^{-1}[V]$ \,is open in $X$.
\endpr

\Remarksn{5.1.4}
($a$) For metric spaces $X$ and $Y$, Definition 5.1.3($b$)
becomes: 
\nl
$f\:X\toto Y$ \,is continuous iff
$$
\all a\in X \,\all b\in f(a)\,\all\eps>0 \,\ex\del>0 \,\all x\in \Bb(a,\del)
\,\bigl(f(x)\cap \Bb(b,\eps)\ne\nil\bigr).
$$
($b$) Definition 5.1.3($b$) reduces to 
the standard definition of continuity for 
total single-valued functions from $X$ to $Y$.
\sn
($c$) It also reduces to the definition of continuity for partial
single-valued functions (Definition 2.5.1 and Remark 2.7.2($a$)),
as we will see below (Remark 5.1.9).
We must first see how to extend the topology on $Y$ to that on \Yua\
(Definition 5.1.6 below).
\endpr

\Defn{5.1.5}
For two functions
\ $f\:  X \toto Y$,
\ $g\:  X \toto Y$,
\ we define 
$$
f\,\sqsseq \,g \ \ \ifffdf\ 
\ \tx{for all} \ x\in X, \ f(x) \,\sseq \,g(x).
$$
\endpr

\Defn{5.1.6 \,(Topology on \,\Yua)}
We extend the topology on $Y$ to 
\,\Yua\ ($= Y\cup\curly{\ua}$)
by specifying that the only open set containing \,\curly{\ua}
\,is \,\Yua. (So \,\Yua\ \,is a ``one-point compactification" of $Y$.)
\endpr

Now, given a function
\,$f : \ X \toto \Yua$,
\,we define functions
\TOL
$$
\align
\fua:& \ X \ \toto \ \Yua\\
\fmin:& \ X \ \toto \ Y
\tag"and"
\endalign
$$
by 
$$
\align
\fua(x) \ &= \ f(x)\cup\curly{\ua}\\
\fmin(x) \ &= \ f(x) \backslash \curly{\ua}.
\endalign
$$
\TOR
In other words, \,\fua\ \,{\it adds\/} \,`\ua' \,to the set \,$f(x)$
\,for each \,$x\in X$
\,and \,\fmin\ \,{\it removes\/} \,`\ua' \,from every such set.
This changes the semantics of $f$ 
(see Remark 3.2.2), but not its {\it continuity properties\/},
as will be seen from the following technical lemma,
which will be used
in the proof of continuity 
of computable functions below (\S5.2).

\Lemman{5.1.7}
Let 
\ $f:\ X\ \toto \ Y$ 
 and
\ $g:\ X\ \totop \ \Yua$
\ be any two functions such that 
$$
f\ \sqsseq\ g \ \sqsseq\ \fua,
$$
\ie, for all $x\in X$,
\,$g(x) \ne\nil$,
\,and either \,$g(x) = f(x)$ or $g(x) = f(x)\cup \curly{\ua}$.
Then
$$
f\ \ \tx{is continuous} \ \ \ifff\ \ g\ \ \tx{is continuous}.
$$
\endpr
\Pf
($\tto$)
\ Suppose $f$ is continuous.  
We must show $g$ is continuous.
Let $V$ be an open subset of \Yua.
We must show \,$g^{-1}[V]$ \,is open in $X$.
There are two cases, according as \ua\ is in $V$ or not.
\sn
{\it Case 1}: \,$\ua\notin V$, \ie, $V\sseq Y$.
\ Then $V$ is also open in $Y$ (by definition of the topology on \Yua).
Hence \,$f^{-1}[V]$ \,is open in $X$, and hence
$$
\align
g^{-1}[V] \ &= \ \curly{x\in X \mid g(x)\cap V\ne\nil}\\
&= \ \curly{x\in X\mid f(x)\cap V\ne\nil}\qquad\tx{since $\ua\notin V$}\\
&= \ f^{-1}[V]
\endalign
$$
is open in $X$.
\sn
{\it Case 2}: \,$\ua\in V$.
\ Then $V = \Yua$
(by definition of the topology on \Yua).
Hence 
$$
g^{-1}[V] \ = \ g^{-1}[\Yua] \ = \ X \qquad \tx{(since $g$ is total)},
$$
which is open in $X$.
\sn
($\ffrom$)
\ Suppose $g$ is continuous.
We must show $f$ is continuous.
Let $V$ be an open subset of $Y$.
We must show \,$f^{-1}[V]$ \,is open in $X$.
Since $V$ is also open in \Yua\ 
(by definition of the topology on \Yua),
$g^{-1}[V]$ \,is open in $X$.  Hence
$$
\align
f^{-1}[V] \ &= \ \curly{x\in X \mid f(x)\cap V\ne\nil}\\
&= \ \curly{x\in X\mid g(x)\cap V\ne\nil}\qquad\tx{since $\ua\notin V$}\\
&= \ g^{-1}[V]
\endalign
$$
is open in $X$.
\endpf

\Corn{5.1.8}
Suppose 
\ $f: X\ \totop \Yua$ \ (\ie, $f$ is total). Then
$$
f\ \ \tx{is continuous} \ \ \ifff\ \ \fmin\ \ \tx{is continuous}
\ \ \ifff\ \ \fua\ \ \tx{is continuous}.
$$
\endpr

\Pf
Apply Lemma 5.1.7 twice:
once with \,\fmin\ \,and $f$, 
and once with \,\fmin\ \,and \,\fua.
\endpf

\Remarkn{5.1.9 \,(Justification of Remark 5.1.4($c$))}
Let \,$f:X\pto Y$ 
\,be a single-valued partial function.
Define 
\sn
($a$) \,$\fcheck\:X\toto Y$ \,by
$$
\fcheck(x) \ = \
\cases
\curly{f(x)} \ift{$x\in\dom{f}$}\\
\nil \ow
\endcases.
$$
($b$) \,$\fhat\:X\totop \Yua$ \,by
$$
\fhat(x) \ = \
\cases
\curly{f(x)} \ift{$x\in\dom{f}$}\\
\curly{\ua} \ow.
\endcases
$$
(We can view either \,\fcheck\ \,or \,\fhat\ \,as 
``representing" $f$ in the present context, \cf\ Remark 5.1.2.)
\,Then
$$
\align
&f\ \ \tx{is continuous \ (according to Def\. 2.5.1)} \\
\ifff\ \ &\fcheck\ \ \tx{is continuous \ (according to Def\. 5.1.3)}\\
\ifff\ \ &\fhat\ \ \tx{is continuous \ (according to Def\. 5.1.3)}
\endalign
$$
The equivalence of the continuity of $f$ and \,\fcheck\ 
\,follows immediately from the definitions.
The equivalence of the continuity of \,\fcheck\ \,and \,\fhat\
\,follows from Lemma 5.1.7.
\endpr

\Remarkn{5.1.10 \,(Comparison with W-continuity)}
As in \S2.7, we can consider another notion of continuity 
for functions \,$f:X\toto Y$ \,by modifying Definition 5.1.3($b$);
\,namely,
$f$ is W-continuous
iff for all open  \,$V\sseq Y$,
\,$f^{-1}[V]$ \,is open in \dom{f}.
Note that Lemma 5.1.7, and
the equivalences given in Remark 5.1.9, 
also hold for W-continuity.
\endpr

\Lemman{5.1.11}
Given \,$f:X\toto \Yua$, \,extend it to
\,$\ftil:\Xua\toto\Yua$ \,by stipulating that \,$\ftil(\ua) = \ua$.
If $f$ is continuous and total, then \ftil\ is continuous.
\endpr

\Pf
Let $V$ be an open subset of \Yua.
We must show \,$\ftil^{-1}[V]$ \,is open in \Xua.
There are two cases:
\sn
{\it Case 1}: \,$\ua\notin V$, \ie, $V\sseq Y$.
Then \,$\ftil^{-1}[V] = f^{-1}[V]$,
\,which is open in $X$, and hence in \Xua.
\sn
{\it Case 2}: \,$\ua\in V$.
\ Then $V = \Yua$
(by definition of the topology on \Yua).
Hence 
$$
\align
\ftil^{-1}[V] \ &= \ \ftil^{-1}[\Yua] \\
&= \ \dom{f}\cup \curly{\ua}\\
&= \ X\cup\curly{\ua} \qquad\tx{(since $f$ is total)}
\endalign
$$
which is open in \Xua.
\endpf

\Defn{5.1.12 \,(Composition)}
\nl
($a$) Suppose \ $f:X\toto Y$ \ and \ $g:Y\toto Z$.
\ We define \ $g\circ f:X\toto Z$ \ by
$$
(g\circ f) (x)\ = \ \bigcup\curly{g(y)\mid y\in f(x)}.
$$
($b$)
Suppose \ $f:X\toto \Yua$ \ and \ $g:Y\toto\Zua$.
\ We define \ $g\circ f:X\totop \Zua$ \ by
$$
\align
(g\circ f) (x)\ = \ &
\quad\ \,\bigcup\curly{g(y)\mid y\in f(x)\cap Y}\\
&\cup\ \curly{\ua \br \ua \in f(x)}
\endalign
$$
\endpr

\Propn{5.1.13 \,(Continuity of composition)}
\sn
($a$) 
If \ $f:X\toto Y$ \ and \ $g:Y\toto Z$ \ are continuous,
then so is \ $g\circ f:X\toto Z$.
\sn
($b$)
If \ $f:X\totop \Yua$ \ and \ $g:Y\totop \Zua$ \ are continuous,
then so is \ $g\circ f:X\totop \Zua$.
\endpr

\Pf
($a$) Just note that 
for \,$W\sseq Z$,
$$
(g\circ f)^{-1}[W] \ = \ f^{-1}[g^{-1}[W]].
$$
($b$) 
We give two proofs:
\ ($i$) Note that
$$
(g\circ f)^- \ = \ g^- \circ f^- :\ X \ \toto \ Z
$$ 
and use part ($a$) and Corollary 5.1.8.
\sn
($ii$)
Note that
for \,$W\sseq \Zua$,
$$
(g\circ f)^{-1}[W] \ = \ f^{-1}[\gtil^{-1}[W]]
$$
(in the notation of Lemma 5.1.11), and apply Lemma 5.1.11.
\endpf

\Defn{5.1.14 \,(Union of functions)}
Let \,$f_i:X\toto \Yua$ \,be a family of functions for \iinI.
Suppose for all \,$x\in X$, \,$\bigcup_\iinI f_i(x)$ 
\,is countable.
Then we define 
$$
\bigsqcup_\iinI f_i:\ X \ \toto \ \Yua
$$
by 
$$
\big(\bigsqcup_\iinI f_i\big)(x) \ = \ \bigcup_\iinI f_i(x).
$$
\endpr

\Lemman{5.1.15}
If \,$f_i:X\toto \Yua$ \,is continuous for all \iinI,
then so is \,$\bigsqcup_\iinI f_i$.
\endpr

\Pf
This follows from the fact that for \,$V\sseq \Yua$,
$$
\big(\bigsqcup_\iinI f_i)^{-1}[V] \ = \ \bigcup_\iinI f_i^{-1}[V].
\tag"$\square$"
$$

\Remarkn{5.1.16}
Note that all the results of this subsection (5.1)
hold for {\it arbitrary\/}
multivalued functions  \,$f:X\to \PPP(Y)$,
\,not necessarily countably-many-valued.
\endpr

\shead{5.2}{Continuity of \WhileCC\ \,computable functions}
Let $A$ be an N-standard topological \Sig-algebra.

In order to prove that \WhileCCx\ \,procedures on $A$ are continuous,
we first state and prove a lemma which says that
such procedures are (almost) equivalent to 
\While\ \,procedures (without `\chooses')
in an extended signature, which includes 
a symbol \,\fs\ \,for an ``oracle function".  Then we apply Lemma 5.1.7.

\Lemman{5.2.1 \,(Oracle equivalence lemma)}
Given a \WhileCCSig\ \,statement $S$, and procedure
$$
P \ \ident \ \funcs\ \ins \ \att\ \outs \ \btt\ \auxs \ \ctt
\ \begins \ S\ \ends,
$$
we can effectively construct a \While(\Sigf) statement \,\Sf\ 
\,and procedure
$$
\Pfs \ \ident \ \funcs\ \ins \ \att\ \outs \ \btt\ \auxs \ \ctt
\ \begins \ \Sf\ \ends
$$
\,in a signature \,\Sigf\ \,which extends \,\Sig\
\,by a function symbol \,$\fs: \nats\to\nats$,
\,such that, putting
$$
\PsqcupA \ \eqdf \ \bigsqcup_{f\in\FFF}\PfA,
$$
where \,$\FFF = \NN^\NN$ \,\ is
the set of all functions \,$f\:\NN\to\NN$
\,and \PfA\ is the interpretation of \,\Pfs\ in $A$ formed by interpreting 
\,\fs\ \,as $f$,
\,we have
$$
\PA \ \sqsseq \ \PsqcupA \ \sqsseq \ (\PA)^\ua.
\tag1
$$
\endpr
\n
(Recall Definitions 5.1.14 and 5.1.5,
and the definition of
\,$\PA\: \Au \,\totop\,\Avua$ \,in \S3.2($g$).)

\Pf
Intuitively, \,\fs\ \,represents a possible implementation 
of the `\chooses' operator:  \,$\fs(n)$
\,is a possible value for the $n$th call of this operator
in any particular implementation of $P$.
We will then take the union of the interpretations 
over all such possible implementations.

In more detail:
the construction of \,\Sf\ \,from $S$ is as follows.
Let \,\ctt\ \,be a new ``counter", \ie,
an auxiliary variable of sort \,\nats\
\,which is not in $S$.
First, it is clear that by ``splitting up" assignments in $S$,
and introducing more auxiliary \,\nats\ \,variables,
we can re-write $S$ in such a way
that every occurrence of the `\chooses' construct
is in the context of an assignment of the form
$$
\ztt' := \chooses \ \ztt:b.
\tag2
$$
where the boolean term $b$ does not contain the `\chooses' construct.
Now replace each assignment of the form (2) by
the pair of assignments
$$
\align
&\ctt := \ctt+1;\\
&\ifs \ \,b\ang{\ztt/\fs(\ctt)} 
  \ \,\thens \ \,\ztt' := \fs(\ctt)\ \,\elses \ \,\divs
\endalign
$$
and initialise the value of \ctt\ (at the beginning of the statement)
to 0.
The result is a \,\Whilex(\Sigf) \,procedure \,\Pfs\
\,with a body \Sf\ \,which, for a given interpretation $f$ of \fs,
``interprets" successive executions of \,`\chooses' 
by successive values of $f$, when this is possible
(\ie, \,$b\ang{\ztt/f(\ctt)}$ 
\,has \,\ttt\ \,as one of its values), and otherwise, 
causes the execution to diverge.

For those $f$ which (for a given input)
always give ``good" values for all the successive executions of \,`\chooses'
assignments (2) in $S$, \,\PfA\ \,will give a possible implementation 
of $P$.
For all other $f$, \,\PfA\ \,will diverge.
Since (for a given input)
each \,\PfA\ \,{\it either\/} simulates one possible implementation 
of successive executions of \,`\chooses' \,in $S$ {\it or\/} diverges,
their ``union" \,\PsqcupA\ \,gives the result of 
{\it all\/} possible implementations
of \,`\chooses', \,plus divergence; hence the conclusion (1).
\endpf

\Thmn{5.2.2}
Let
$$
P \ \ident \ \funcs\ \ins \ \att\ \outs \ \btt\ \auxs \ \ctt
\ \begins \ S\ \ends
\tag3
$$
be a \,\WhileCC\ \,procedure,
where \,$\att:u$ and $\btt:v$.
Then the interpretation 
$$
\PA: \ \Au \ \totop \Avua
$$
is continuous.
\endpr

\Pf
In the notation of the Oracle Equivalence Lemma (5.2.1):
\,\PfA\ \,is continuous for all \,$f\in\FFF$,
\,by the continuity theorem for \,\While\ \,\cite[\S6.5]{tz:hb}.
Hence \,\PsqcupA\ \,is continuous, by Lemma 5.1.15.
Hence, by (1) and Lemma 5.1.7, so is \,\PA.
\endpf

\Remarkn{5.2.3}
In the special case that 
\PA\ is deterministic, \ie, single-valued:
$$
\PA\: \Au \pto \Av,
$$
it follows by Remark 5.1.9 that 
\PA\ is continuous according to our definition (2.5.1)
of continuity for single-valued partial functions.
\endpr

\newpage

\Corn{5.2.4}
A \WhileCCx\ \,computable function on $A$ is continuous.
\endpr

\Pf
Such a function is \WhileCC\ \,computable on \Ax,
hence (by Theorem 5.2.2) continuous on \Ax,
and hence on $A$.
\endpf

\shead{5.3}{Continuity of \WhileCCx\ \,approximable functions}
Recall Definiton 3.5.1 and \S2.7.

\Thmn{5.3.1}
Let $A$ be a metric \Sig-algebra,
and let \,$F\:\Au\pto\Av$.
\sn
($a$) If $F$ is \WhileCCx\ approximable then $F$ is W-continuous.
\sn
($b$) If also \,\dom{F} is open in \Au\ \,then $F$ is continuous.
\endpr

\Pf
\,Suppose $F$ is approximable on $A$ by the 
\WhileCCx\ 
procedure \,$P\:\nats\times\utov$.
We will show that $F$ is W-continuous, using Remark 2.7.2($b$).
Given \,$a\in \dom{F}$ \,and \,$\eps>0$,
\,choose $N$ such that 
$$
2^{-N}<{\eps/3}.
\tag1
$$
Then by Definition 3.5.1,
$$
\nil \ \ne \ P^A_N(a)\ \sseq \ \Bb(F(a),\,2^{-N}).
\tag2
$$
Choose $b\in P^A_N(a)$.
By (2), 
$$
\ds(F(a),b)\ < \ 2^{-N}.
\tag3
$$
By Corollary 5.2.3, $P^A_N$ is continuous on $A$, 
and so by Remark 5.1.4($a$),
there exists $\del>0$ such that 
$$
\all x\in \Bb(a,\del), \ P^A_N(x)\cap \Bb(b,\eps/3)\ne\nil.
\tag4
$$
Take any \,$x\in \Bb(a,\del)\cap\dom{F}$.
By Definition 3.5.1 again,
$$
P^A_N(x)\ \sseq \ \Bb(F(x),\,2^{-N})
\tag5
$$
By (4), choose \,$y\in P^A_N(x)\cap\Bb(b,\eps/3)$.
So
$$
\ds(y,b)\ < \ \eps/3
\tag6
$$
and by (5) 
$$
\ds(F(x),y)\ <\ 2^{-N}.
\tag7
$$
Hence
$$
\align
\ds(F(x),F(a)) \ &\le \ \ds(F(x),y)+\ds(y,b)+\ds(b,F(a))\\
&< \ \eps
\endalign
$$
by (7), (6), (3) and (1).
Part ($a$) follows by Remark 2.7.2($b$).

Part ($b$) then follows by Proposition 2.7.1.
\endpf

\newpage

\itemm{\bbf 6\ \ }
{\bbf Concrete computability and the soundness of \WhileCCxbig
\nl
computation on countable algebras}

\mn
To compute on a metric algebra $A$ using a concrete model
of computation,
we choose a countable subspace $X$ of $A$ and an enumeration
$$
\al: \ \NN\ \to\ X.
$$
From this we build the space \,\CalX\ \,of \al-computable elements 
of $A$, and enumerate it with 
$$
\albar: \ \NN\ \to\ \CalX.
$$
In this section
we step back from topological algebras and consider
computability on {\it arbitrary countable\/} algebras.
We show that if an algebra $A$ is enumerated and
its basic functions are effective, then
functions that are \WhileCCx\ computable on $A$ are also effective.
This result is a key lemma in the ssoundness theorem for
\WhileCCx\ approximation in the next section.

\shead{6.1}{Enumerations and tracking functions for partial functions}
Let 
$$
X \ = \ \ang{\Xs \br \sinSortSig}
$$
be a \SortSig-indexed family of non-empty sets.

\Defn{6.1.1}
An {\it enumeration\/} of $X$ is a family 
$$
\al \ = \ \ang{\als\:\Oms\onto \Xs \br \sinSortSig}
$$
of surjective maps \,$\als\:\Oms\onto \Xs$,
\,for some family 
$$
\Om \ = \ \ang{\Om_s \br \sinSortSig}
$$
of sets \,$\Oms\sseq\NN$.
The family $X$ is said to be
{\it enumerated by\/} \al.
We say that
\,$\al\: \Om\onto X$
\,is an {\it enumeration\/} of $X$, 
and call the pair \,\Xal\ \,an {\it enumerated family of sets\/}.
(The notation `$\onto$' denotes surjections, or onto mappings.)
\endpr

We also write 
\,$\Oms = \Omals$
\,to make explicit the fact that \,$\Oms = \dom{\als}$.

\Defn{6.1.2 \ (Tracking and strict tracking functions)}
We use the notation
\nl
\,$X^u = X_{s_1}\times\dots\times X_{s_m}$
\,and
\ $\Omalu = \Om_{\al,s_1}\times\dots\times\Om_{\al,s_m}$,
\,where \,$u=\tuptimes{s}1m$.
\nl
Let \,$F\:\Xu\pto\Xs$ \,and \,$f:\Omalu\pto\Omals$,
\itemm{($a$)}
$f$ is a {\it tracking function with respect to\/} \al,
\,or \,\al-{\it tracking function\/},
\,for $F$,
\,if the following diagram commutes:
$$
\commdiag{
\Xu& &\mapright^{\dz F}_\bdot& &\Xs\cr
\mapup\lft{\dz{\alu}}\rt\bdot& && &\mapup\lft\bdot\rt{\dz{\ \als}}\cr
\NN^m& &\mapright_{\dz f}^\bdot& &\NN
}
$$
\mn
\itemm{}
in the sense that
\,for all \,$k\in \Omalu$
$$
F(\alu(k))\da \ \ \implies
\ f(k)\da \,\con \,f(k)\in \Omals \,\con 
\,F(\alu(k)) = \als(f(k)).
$$
\itemm{($b$)}
$f$ is a {\it strict \al-tracking function\/} for $F$
if in addition,
for all \,$k\in\Omalu$
$$
f(k)\da \ \ \implies
\ F(\alu(k))\da.
$$

Here we use the notation
\ $\alu(k) = (\al_{s_1}(k_1),\dots,$ $\al_{s_m}(k_m))$,
\ where 
\ $k = (\tup{k}1m)$.
(We will sometimes drop the type super- and subscripts.)

\Defn{6.1.3 \ (\al-computability)}
($a$) \,Suppose $A$ is a \SortSig-family,
and \,\Xal\ \,an enumerated subfamily of $A$,
\,\ie, \,$\Xs \sseq A_s$ \,for all \Sig-sorts $s$.
Suppose 
\,$F\:\Au\pto A_s$ \,and \,$f:\NN^m\pto\NN$,
such that 
$$
\align
F\rest \Xu\:\Xu\ &\pto\ \Xs,\\
f\rest \Omalu\:\Omalu\ &\pto\ \Omals,
\endalign
$$
and \,$f\rest \Omalu$ \,is a ({\it strict\/}) \al-{\it tracking function\/}
for \,$F\rest X$.
We then say that $f$ is a ({\it strict\/}) 
\al-{\it tracking function for\/} $F$.
\sn($b$)
\,Suppose now further that $f$ is a {\it computable\/} (\ie, recursive)
partial function.
Then $F$ is said to be ({\it strictly\/}) \al-{\it computable\/}.
\endpr

\Remarksn{6.1.4}
($a$) In the situation of Definition 6.1.3,
we are not concerned with the behaviour of
$F$ off \Xu, or the behaviour of $f$ off \,\Omalu.
\sn
($b$)
For convenience, we will always assume:
$$
\Om_{\al,\boolss} = \curl{0,1}, \qquad \al_\boolss(0) = \fff, 
\qquad \al_\boolss(1) = \ttt
$$
and also (when \Sig\ is N-standard):
$$
\Om_{\al,\natss} = \NN\ \ \tx{and \,$\al_\natss$ \,is the identity on \NN.}
$$
\endpr

Assume now that
$A$ is a \Sig-{\it algebra\/}
and \,\Xal\ \,is a \SortSig-family of subsets of $A$,
enumerated by \al.
 
\Defn{6.1.5 \,(Enumerated \Sig-subalgebra)}
\Xal\ is said to be an {\it enumerated \Sig-subalgebra of\/} $A$ if
$X$ is a \Sig-subalgebra of $A$.
\endpr
 
\Defn{6.1.6 \,(\Sig-effective subalgebra)}
Suppose $A$ is a \Sig-algebra and \Xal\ is an enumerated \Sig-subalgebra. 
Then \al\ is said to be
\itemm{($a$)}
\Sig-{\it effective\/} if all the basic \Sig-functions on $A$
are \al-computable; \,and
\itemm{($b$)}
{\it strictly\/} \Sig-{\it effective\/} if all the basic \Sig-functions on $A$
are strictly \al-computable.
\endpr
 
\newpage

\shead{6.2}{Soundness Theorem for surjective enumerations}
For the rest of this section we will be considering the special
case of \S6.1 in which the enumerated subalgebra $X$ is $A$ itself,
\ie, we assume the enumeration is {\it onto\/} $A$.
To emphasise this special situation, we will denote the 
enumeration by 
$$
\be\: \Ombe\ \onto\ A,
$$
so that \Abe\ is our {\it enumerated \Sig-algebra\/}.

Given such an enumerated algebra \Abe\ and a function
$$
F\:\Au \ \pto \ A_s,
$$
we have two notions of computability for $F$:
\itemm{($i$)} 
{\it abstract\/}, \ie, \WhileCCx\ computability,
as described in Section 3; \,and
\itemm{($ii$)} 
{\it concrete\/}, \ie, \be-computability,
as in Definition 6.1.3, in the special case 
that $X = A$.

We will prove a {\it soundness theorem\/} (Theorem \Ao),
for these notions of abstract and concrete computability, 
\ie, \,$(i)\tto(ii)$, \,assuming {\it strict effectiveness\/} of \be.

A more general soundness theorem (Theorem A), 
with more general notions of abstract computability
(\WhileCCx\ {\it approximability\/})
and concrete computability
(computability w.r.t\. the {\it computable closure\/}
of an enumeration),
will be proved in Section 7.

\Thmn{$\tx{\bf A}_{\bk0}$ \ (Soundness for countable algebras)}
Let \,\Abe\ \,be an enumerated N-standard \Sig-algebra
such that \be\
is strictly \Sig-effective.
If \,$F:\Au\pto A_s$
\,is \,\WhileCCx\ computable on $A$, then $F$ 
is strictly \be-computable on $A$.
\endpr

\shead{6.3}{Proof of Soundness Theorem $\tx{\bf A}_{\bk0}$}
Assume, then, that \,\Abe\ \,is an enumerated N-standard \Sig-algebra
and \be\
is strictly \Sig-effective.

We will show that each of the semantic functions
listed in \S3.2$(a)$--$(g)$ has a computable 
tracking function.
More precisely, we will work, not with the semantic
functions themselves, but ``localised" functions representing them
(\cf\ \cite[\S4]{tz:hb}).

First we will prove a series of results of the form: 

\pr{Lemma Scheme 6.3.1}\sl
For each semantic representing function
$$
\Phi: \ \Au \ \totop \ \Avua
$$                                                                        
representing one of the semantic functions
listed in \S3.2$(a)$--$(g)$,
\,there is a 
computable tracking function w.r.t. \be,
\,\ie, a function
$$
\ph:\ \Ombeu\ \pto\ \Ombev
$$
which commutes the diagram

$$
\commdiag{
\Au &\ & \bimapright^{\raise1pt\hbox{$\dz\Phi$}} &{}^+& \Avua\cr
\mapup\lft{\dz{\beu}} &\ &&& \mapup\rt{\dz{\,\bev}}\cr
\Ombeu &\ & \mapright_{\dz \ph}^\bdot && \Ombev
}
$$

\mn
in the sense that for all \,$k,l\in\Ombeu$:
$$
\align
\ph(k)\da l\ &\implies \ \bev(l)\,\in \,\Ph(\beu(k)),\\
\ph(k)\,\ua \quad \ &\implies \ \ua\,\in \,\Ph(\beu(k)).
\endalign
$$
\endpr

\Remarksn{6.3.2}
($a$) \,Here \,\ph\ \,is a combination ``strict tracking function"
and ``selection function".
We can think of \,\ph\ \,as giving one possible implementation 
of \,\Ph.
(Compare the representative functions for various semantic functions
in \cite[\S4]{tz:hb}.)
\sn
($b$) \,We are not concerned with the behaviour of \,\ph\ \,on
\,$\NN^m\backslash\Ombeu$
\,(where \,$m = \arity{u}$).
(\Cf\ Remark 6.1.4($a$).)
\endpr

Theorem \Ao\ then follows easily (\S6.5) from this lemma scheme.

\sn
{\bf Proof of Lemma Scheme 6.3.1:}
\ \ We proceed to prove this lemma scheme
by constructing {\it concrete strict tracking functions for the semantic
functions\/} in \S3.2.

Let \xtt\ be a $u$-tuple of variables,
where \,$u = \tuptimes{s}1m$.
Let \,$\PTermx = \PTermx(\Sig)$ \,be the class of all \Sig-terms
with variables among \xtt\ only,
and for all sorts $s$ of \Sig, let \,$\PTermxs = \PTermxs(\Sig)$
\,be the class of such terms of sort $s$.

We consider in turn the semantic functions in \S3.2,
or rather versions of these 
{\it localised to\/} \xtt,
\,\ie, defined only in terms of the state values on \xtt\
(\cf\ \cite[\S4]{tz:hb}).
For example, we localise the set 
\,\StateA\
\,of states on $A$ to the set 
$$
\StatexA \ \eqdf \ \Au
$$
of $u$-tuples of elements of $A$,
where a tuple \,\ainAu\ represents a state \sig\
(relative to \xtt) if \,$\sig[\xtt] = a$.
The set \Au\ is, in turn, represented (relative to \be)
by the set \,\Ombeu.

We assume an effective coding, or G\"odel numbering,
of the syntax of \Sig.
We use the notation
$$
\cnr{\PTerm_s} \eqdf \ \{\cnr{t} \mid t \in \PTerm_s \},
$$
etc.,  for sets of G\"odel numbers of syntactic expressions.

\mn{\bf ({\bi a}) \,Tracking of term evaluation.}
\sn
The function
$$
\PTExsA:\ \PTermxs\times\StatexA \ \totop \ \Asua
$$
defined by
$$
\PTExsA(t,a) \ = \ \tA\sig
$$
for any state \sig\ on $A$ such that \,$\sig[\xtt] = a$,
\,is strictly tracked by a computable function
$$
\ptexsAbe: \ \cnr{\PTermxs}\times \Ombeu \ \pto \ \Ombes
$$
so that the following diagram commutes:

$$
\commdiag{
\PTermxs\times\StatexA&\ & \bimapright^{\raise5pt\hbox{$\dz\PTExsA$}}
&{}^+&\Asua\cr
\mapup\lft{\dz{\ang{\enum,\,\beu}}}&\ &&&\mapup\rt{\dz\,\bes}\cr
\cnr{\PTermxs}\times\Ombeu&\ &\mapright_{\dz{\ptexsAbe}}^\bdot&&\Ombes
}
$$
\mn
(where \,\enum\ \,is the inverse of the G\"odel numbering function),
in the sense that 
$$
\aligned
\ptexsAbe(\cnr{t},k)\da l\ &\implies \ \bes(l)\,\in \,\PTExsA(t,\,\beu(k)),\\
\ptexsAbe(\cnr{t},k)\,\ua \quad\ &\implies \ \ua\,\in \,\PTExsA(t,\,\beu(k)).
\endaligned
\tag1
$$
In order to construct such a representing function,
we first define the {\it state variant representing function\/},
\ie, a (primitive) recursive function 
$$
\vartxbe: \ \Ombeu\times\cnr{\Vars}\times\Ombes\ \to \ \Ombes
$$
such that
$$
\beu(\vartxbe(k,\cnr{\ytt},k_0)) \ = \ \beu(e)\curly{\ytt/\bes(k_0)}.
$$
for \,$k\in\Ombeu$, 
\,$\ytt\in\Vars$
\,and \,$k_0\in\Ombes$
\,(\cf\ Definition 3.2.3($b$)).

We turn to
the definition of \,$\ptexsAbe(\cnr{t},k)$.
This is by induction on \,\cnr{t},
\,or structural induction on \,$t\in\PTermx$,
over all \Sig-sorts $s$.
The cases are:

\bull
$t \ident c$, \ a primitive constant.
Then define
$$
\ptexsAbe(\cnr{t},k) \ = \ k_0
\qquad \tx{where} \qquad \be(k_0) \ = \ c^A.
$$
(Such a $k_0$ exists by the strict \Sig-effectivity of \be).

\bull
$t \ident \xtt_i$ \ for some $i=1,\dots,m$, 
\,where \,$\xtt \ident \tup{\xtt}1m$.
Note that \,$k=(\tup{k}1m)\in \Ombeu$.
\ So define 
$$
\ptexsAbe(\cnr{t},k) \ = \ k_i.
$$

\bull
$t\ident F(\tup{t}1m)$.
\ Let \,$f$ be a computable strict tracking function for $F$,
which exists by the strict \Sig-effectivity of \be.
Then define
$$
\ptexsAbe(\cnr{t},k) \ \sq 
\ f(\pte\,_{\xtt,s_1}^{A,\ssbe}(\cnr{t_1},k),\,\dots,
\,\pte\,_{\xtt,s_m}^{A,\ssbe}(\cnr{t_m},k))).
$$
From the induction hypothesis applied to \,\tup{t}1m, 
\,the definition of \PTE\ (\S3.2($a$))
and the fact that $f$ {\it strictly\/} tracks $F$,
we can infer (1) for $t$.

\bull
$t\ident \ifs(b,t_1,t_2)$. \ Define
$$
\ptexsAbe(t,k)\ \sq\ 
\cases
\ptexsAbe(t_1,k) \tif{$\pte_{\xtt,\boolss}^{A,\ssbe} (b,k) \da 1$}\\
\ptexsAbe(t_2,k) \tif{$\pte_{\xtt,\boolss}^{A,\ssbe} (b,k) \da 0$}\\
\ua           \tif{$\pte_{\xtt,\boolss}^A (b,k) \ \ua$}.\\
\endcases
$$
From the induction hypothesis applied to $b$, $t_0$ and $t_1$,
\,and the definition of \PTE,
\,we can infer (1) for $t$.

\bull
$t \ident (\chooses \ \ztt:t_0)$.
\ We define \,$\ptexsAbe(\cnr{t},\,k)$
\,by specifying its computation:
\,find, by dovetailing (recall the discussion in \S4.1!) some $n$ such that
$$
\ptexsAbe(\cnr{t_0}, \,\vartxbe(k,\cnr{\ztt},n)) \ \da \ 1
$$
(remember, \,$\be(1) = \ttt$, by Remark 6.1.4($b$)),
so that \,$\ptexsAbe(\cnr{t},\,k) =$ some such $n$,
\,if it exists, and \,\ua\ \,otherwise.
From the induction hypothesis applied to $t_0$,
\,and the definition of \PTE,
\,we can infer (1) for $t$.

\mn{\bf ({\bi b}) \,Tracking of atomic statement evaluation.}
\sn
Let \AtStx\ be the class of atomic statements with variables among \xtt\ only.
The atomic statement evaluation function on $A$ localised to \xtt,
$$
\AExA: \ \AtStx \times \StatexA \ \totop \ \StatexAua,
$$
defined by
$$
\AExA(S,a) \ = \ \angg{S}^A\sig
$$
for any state \sig\ such that \,$\sig[\xtt] = a$,
\,is strictly tracked by a computable function
$$
\aexAbe: \ \cnr{\AtStx}\times \Ombeu \ \pto \ \Ombeu
$$
so that the following diagram commutes:

$$
\commdiag{
\AtStx\times\StatexA&\ &\bimaprighturaise5{\AExA}&{}^+ 
&\StatexAua\cr
\mapupl{\ang{\enum,\,\beu}}&\ &&&\mapupr{\,\beu}\cr
\cnr{\AtStx}\times\Ombeu&\ &\maprightd{\aexAbe}^\bdot&&\Ombeu
}
$$
\mn
in the sense that
$$
\aligned
\aexAbe(\cS,k)\da l\ &\implies \ \be(l)\,\in \,\AExA\,(S,\,\be(k)),\\
\aexAbe(\cS,k)\,\ua \quad\ &\implies \ \ua\,\in \,\AExA\,(S,\,\be(k)).
\endaligned
\tag2
$$
The definition of \,$\aexAbe(\cS,k)$ 
\,is given by:
$$
\align
\aexAbe(\cnr{\skips},\,k) \ &\da \ k\\
\aexAbe(\cnr{\divs},\,k) \ &\ \ua\\
\aexAbe(\cnr{\ytt:=t},\,k) \ &\sq 
  \cases
     \vartxbe\,(k,\ytt,l)\tif{$\ptexsAbe(s\cnr{t},k)\da l$}\\
     \ua  \tif{$\ptexsAbe(\cnr{t},k)\ \ua$.}
  \endcases\\
\endalign
$$
Using (1) and the definition of \AExA\ (\S3.2($b$)),
we can infer (2).

\mn{\bf ({\bi c}) \,Tracking of \,\First\ \,and \,\Rest\ \,operations.}
\sn
Let \Stmtx\ be the class of statements with variables
among \,\xtt\ \,only. 
Consider the functions \,\First\ \,and \,\RestA\ \,(\S3.2($c$)).
Then \,\First\ \,is
strictly tracked by a computable function
$$
\first \ : \cnr{\Stmt}\ \to\ \cnr{\AtSt}
$$
defined on G\"odel numbers in the obvious way,
so that the following diagram commutes:

$$
\commdiag{
\Stmt&\ &\maprightu{\First}&\ &\AtSt\cr
\mapupl{\enum}&&&&\mapupr{\enum}\cr
\cnr{\Stmt}&&\maprightd{\first}&&\cnr{\AtSt}
}
$$

\mn
(Note that \,\first, \,unlike most of the other representing functions here,
does not depend on \,\StatexA, \,or, indeed, on $A$ or \,\xtt.)
Next, the localised version of \,\RestA:
$$
\RestxA :
\ \Stmtx\times\StatexA \ \totop \ \Stmtx
$$
defined by
$$
\RestxA(S,a) \ = \ \RestA(S,\sig)
$$
for any state \sig\ such that \,$\sig[\xtt] = a$,
\,is strictly tracked by a computable function
$$
\restxAbe : \ \cnr{\Stmtx}\times\Ombeu\ \pto\ \cnr{\Stmtx}
$$
so that the following diagram commutes:

$$
\commdiag{
\Stmtx\times\StatexA&\ &\bimaprighturaise5{\RestxA}&{}^+&\Stmtx\cr
\mapupl{\ang{\enum,\,\beu}}&&&&\mapupr{\enum}\cr
\cnr{\Stmtx}\times\Ombeu&&\maprightud{\bdot}{\restxAbe}&&\cnr{\Stmtx}
}
$$

\mn
in the sense that 
$$
\aligned
\restxAbe\,(\cS,k)\da \cnr{S'}\ \implies 
\ &S'\,\in \,\RestA\,(S,\,\be(k)),\\
\restxAbe\,(\cS,k)\,\ua \quad\ \implies \ &\divs\,\in \,\RestA\,(S,\,\be(k))
\endaligned
\tag3
$$
The definition of
\,$\restxAbe(\cS,\,k)$, \,as well as the proof of (3),
are by
induction on \,\cS, \, or structural induction on $S$.

\bull
$S$ is atomic. \ Then
$$
\restxAbe (\cS,\,k) \ = \ \cnr{\skips}.
$$
\bull
$S \ \ident \ S_1;S_2$.  \ Then
$$
\restxAbe(\cS,\,k) \ = \ 
\cases
\quad\cnr{S_2} \tif{$S_1$ is atomic}\\
\quad\cnr{\restxAbe(S_1,\,k);S_2} \ow
\endcases
$$
\bull
$S \ \ident \ \ifs \ b \ \thens \ S_1 \ \elses \ S_2 \ \fis$.
\ Then 
$$
\restxAbe (\cS,\,k) \ \sq \
\cases
\cnr{S_1} \tif{$\pteboolsAbe(b,k)\da 1$}\\
\cnr{S_2} \tif{$\pteboolsAbe(b,k)\da 0$}\\
\ua \tif{$\pteboolsAbe(b,k)\ \ua$}.
\endcases
$$
\bull
$S \ \ident \ \whiles\ b \ \dos \ S_0 \ \ods$.
\ Then 
$$
\restxAbe(S,k) \ \sq\
\cases
S_0;S \tif{$\pteboolsAbe(b,k)\da1$},\\
\skips \tif{$\pteboolsAbe(b,k)\da0$},\\
\ua \tif{$\pteboolsAbe(b,k)\ \ua$}.\\
\endcases
$$

\mn{\bf ({\bi d}) \,Tracking of a computation step.}
\sn
The computation step function (\S3.2($d$)) localised to \,\xtt:
$$
\CompStepxA :\  \Stmtx\times\StatexA \ \totop \ \StatexA^\ua
$$
defined by
$$
\CompStepxA(S,a) \ = \ \CompStepA(S,\sig)
$$
for any state \sig\ such that \,$\sig[\xtt] = a$,
\,is represented by the computable function
$$
\compstepxAbe: \ \cnr{\Stmtx}\times\Ombeu \ \pto \ \Ombeu
$$
defined by
$$
\compstepxAbe(\cS,k) \ \sq \ \aexAbe(\first(\cS),\,k).
$$
This makes the following diagram commute:

$$
\commdiag{
\Stmtx\times\StatexA&\ &\bimaprighturaise5{\CompStepxA}&{}^+ &\StatexA^\ua\cr
\mapupl{\ang{\enum,\,\beu}}&\ &&&\mapupr{\,\beu}\cr
\cnr{\Stmtx}\times\Ombeu&\ &\maprightud{\bdot}{\compstepxAbe}&&\Ombeu
}
$$
\mn
in the sense that
$$
\aligned
\compstepxAbe\,(\cS,k)\da l\ &\implies 
  \ \be(l)\,\in \,\CompStepxA\,(S,\,\be(k)),\\
\compstepxAbe\,(\cS,k)\,\ua \quad\ &\implies 
  \ \ua\,\in \,\CompStepxA\,(S,\,\be(k).
\endaligned
\tag4
$$
This is proved easily from the definitions and (2).

\mn{\bf ({\bi e}) \,Tracking of a computation sequence.}
\sn
Now consider localised versions of the computation tree stage
and computation tree of \S3.2.($e$):
$$
\align
\CompTreeStagexA:\ \Stmtx\times\StatexA\times\NN\ 
&\to\ \PPP((\StatexAua)^{<\om})\\
\CompTreexA:\ \Stmtx\times\StatexA \ &\to \ \PPP((\StatexAua)^{\le\om})
\endalign
$$
We will define a function which {\it selects a path
through the computation tree\/}:
$$
\compseqxAbe: \ \cnr{\Stmtx}\times\Ombeu\times\NN\ \pto 
  \Ombeu\cup \curly{\cnr*}
$$
(where `$*$' is a symbol meaning ``already terminated")
by recursion on $n$:
$$
\align
&\compseqxAbe(\cS,\,k,\,0) \ =\ e \\
&\compseqxAbe(\cS,\,k,\,n+1)\ \sq \\
&\qqquad\cases
  \cnr* \ \ \tx{if $S$ is atomic and $n>0$ and 
      \,$\compseqxAbe(\cS,\,k,\,n)\da$}\\
  \ \ua \ \quad\tx{if $S$ is atomic and $n>0$ and 
     \,$\compseqxAbe(\cS,\,k,\,n)\ \ua$}\\
  \compseqxAbe(\restxAbe(\cS,k),\ \compstepxAbe(\cS,k),\ n)\\
  \qquad \ \tx{otherwise.}
\endcases
\endalign
$$
(This is a ``tail recursion": \ compare definition of
\,\CompIA\ \,in \cite[\S3.4]{tz:hb}.)

Writing \,$k_n = \compseqxAbe(\cS,k,n)$,
\,this defines a 
(concrete) {\it computation sequence\/} 
$$
\kbar \ = \ k_0,\ k_1,\ k_2,\ \dots
$$
for $S$ from the 
initial state $k= k_0$.
(Our notation here includes the possibility that 
some of the $k_i$ may be \,\cnr{*} \, or \,\ua.)
As can easily be checked,
there are three possibilities for \,\ebar\
\,(compare the discussion in \S3.2($e$)):

\itemm{($i$)}
For some $n$, \,$k_i \in \Ombeu$ \,for all $i\le n$ 
\,and \,$k_i = *$ \,for all $i>n$.
This represents a computation which {\it terminates\/} at stage $n$,
with {\it final state\/} \,$k_n$.

\itemm{($ii$)}
For some $n$, \,$k_i \in \Ombeu$ \,for all $i< n$ 
\,and \,$k_i = \ua$ \,for all $i\ge n$.
This represents a {\it non-terminating\/} computation, with
{\it local divergence\/} at stage $n$.

\itemm{($iii$)}
For all $i$, \,$k_i\in \Ombeu$.
This represents {\it non-terminating\/} computation, 
with {\it global divergence\/}.

We write 
\,$\kbarn = $ the initial segment \,$k_0,k_1,\dots,k_n$,
\,with length \,$\lgth{\kbarn} = n+1$.
We put \,$\lgth{\kbar}=\infty$.
The \,$k_i$ \,are called {\it components\/} of \,\kbar,
\,and of \,\kbarn, \,for all $i\le n$.

The computation sequence \,\kbar\ \,then has the following connection 
with the computation tree \,\CompTreexA.
Extend (for now) the definition of \,\be\ \,by
\,$\be(\cnr{*}) = *$,
\,$\be(\ua) = \ua$,
\,and
$$
\align
\be(\kbar) \ &\eqdf\ \be(k_0),\ \be(k_1),\ \be(k_2),\ \dots\\
\be(\kbarn) \ &\eqdf\ \be(k_0),\ \be(k_1),\ \be(k_2),
\ \dots, \ \be(k_n).
\endalign
$$
Let \,$\tau = \CompTreexA(S,\,\be(k))$.
Then

\itemm{($i$)}
If the computation sequence \,\kbar\ \,terminates at stage $n$,
then \,\be(\kbarn) \,is a path through \,\ta\
\,from the root to a leaf ($= \be(k_0)$, \,the final state).

\itemm{($ii$)}
If for some (smallest) $n$, \,$k_n=\ua$, 
\,then \,\be(\kbarn) \,is a path through \,\ta\
\,from the root to a leaf ($=\ua$, \,local divergence).

\itemm{($iii$)}
If for all $n$, \,$k_n\in \Ombeu$,
\,then \,\be(\kbar) is an infinite path through \,\ta\
\,(global divergence).

To prove this, we first define
an initial segment of \,\kbar\
\,(including \,\kbar\ \,itself) to be {\it acceptable\/}
if ($i$) no component is equal to `$*$', \,and
\,($ii$) no component, except possibly the last, is equal to \,\ua.
Further, an acceptable initial segment of \,\kbar\ \,is 
{\it maximal (acceptable)\/}
if it has no acceptable extension.
Thus if \,\kbar\ \,is acceptable, it is automatically maximal.
If \,\kbarn\ \,is acceptable,
it is maximal acceptable provided
either \,$k_{n+1}= *$ \,or \,$k_n = \ua$.
We then show:

\Lemman{6.3.3}
Given a computation sequence \,$\kbar = k_0,k_1,\dots$
\,for \,\cS\ \,from $k$,
where \,$k_n = \compseqxAbe(\cS,\,k,\,n)$,
\,let \,$\tau = \CompTreexA(S,\,\be(k))$.
Then
with every acceptable initial segment \,\kbarn\ \,of \,\kbar,
\,$\be(\kbarn)$
\,is a path through \,\ta\
\,from the root.
If \,\kbarn\ \,is maximal,
then \,$\be(k_n)$ \,is a leaf.
\endpr

\Pf
Put \,$\taun = \CompTreeStagexA(S,\,\be(k_0),\,n)$.
The proof is by induction on $n$,
comparing the inductive definitions of \,$k_n$
\,and \,\taun.

\sn
{\it Basis:\/} \ $n=0$.
This is immediate from the definitions:
\,$k_0 = k$, \,and \,$\tau[0] = \curly{\be(k_0)}$.

\sn
{\it Induction step:\/}
Assume the induction hypothesis holds for the initial segment
of length $n$ of the computation sequence for \,\cnr{S'} 
\,from $k_1$, where 
$$
\align
S' \ &= \ \restxAbe(\cS,\,\be(k)),\\
e_1 \ &= \ \compseqxAbe(\cS,\,k,\,1)\\
     &\sq\ \compseqxAbe(\restxAbe(\cS,k),\,\compstepxAbe(\cS,k),\,0)\\
     &\sq\ \compstepxAbe(\cS,e)
\endalign
$$
\ie, assume the induction hypothesis for the segment \,\lbar\
\,of length $n$:
$$
l_0,\ l_1,\ l_2,\ \dots\ ,l_n
$$
where \,$l_i = e_{i+1}$ \,($i=1,\dots,n$).
Now apply the inductive definitions for \,$\compseqxAbe (\cS,$ $\,k, \,n+1)$
\,(above) and \,$\CompTreeStagexA(S,\,\be(k),\,n+1)$
\,(\S3.2($e$)), and use (3) and (4).

\mn{\bf ({\bi f}) \,Tracking of statement evaluation.}
\sn
First we need a 
constructive computation length function 
$$
\complengthxAbe:\ \cnr{\Stmtx}\times\Ombeu\ \pto\ \NN
$$
by (\cf\ \cite[\S3.4]{tz:hb})
$$
\complengthxAbe(\cS,\,k) \ \sq \ \mu n [\compseqxAbe(\cS,\,k,\,n+1)\da *\,]
$$
\ie, the least $n$ (if it exists) such that for all $i\le n$, 
\,$\compseqxAbe(\cS,\,k,\,i)\da\ \ne *$ \,and\
\,$\compseqxAbe(\cS,\,k,\,n+1)\da*$.

Thus \,$\complengthxAbe(\cS,\,k)$ \,is undefined (\ua) 
in the case of local or global divergence
of the computation sequence for \,\cS\ \,from $k$.

Now the statement evaluation function (\S3.2($f$)) localised to \,\xtt:
$$
\SExA: \ \Stmtx\times\StatexA\ \ \totop\ \StatexAua
$$
defined by 
$$
\SExA(S,a) \ = \ \SA(\sig)
$$
for any state \sig\ such that \,$\sig[\xtt] = a$,
\,is strictly tracked by the computable function
$$
\sexAbe: \ \cnr{\Stmtx}\times\Ombeu\ \ \pto\ \Ombeu
$$
defined by
$$
\sexAbe(\cS,\,k) \ \sq\ \compseqxAbe(\cS,\,k,\,\complengthxAbe(\cS,\,k)).
$$
This makes the following diagram commute:

$$
\commdiag{
\Stmtx\times\StatexA&\ &\bimaprighturaise5{\SExA}&{}^+&\StatexA^\ua\cr
\mapupl{\ang{\enum,\,\beu}}&\ &&&\mapupr{\,\beu}\cr
\cnr{\Stmtx}\times\Ombeu&\ &\maprightud{\bdot}{\sexAbe}&&\Ombeu
}
$$
\mn
in the sense that
$$
\aligned
\sexAbe\,(\cS,k)\da l\ &\implies \ \be(l)\,\in \,\SExA\,(S,\,\be(k)),\\
\sexAbe\,(\cS,k)\,\ua \quad\ &\implies \ \ua\,\in \,\SExA\,(S,\,\be(k)).
\endaligned
\tag5
$$
This result is clear from the definition of
\,\complength\ \,and Lemma 6.3.1.

\newpage

\mn{\bf ({\bi g}) \,Tracking of procedure evaluation.}
\sn
For a specific triple of lists of variables 
\,$\att:u, \,\btt:v,\,\ctt:w$,
\,let \,\Procabc\ \,be the class of all \WhileCCx\ \,procedures
of type \,\utov, 
\,with declaration
\ `$\ins \ \att\ \outs \ \btt\ \auxs \ \ctt$'.
The procedure evaluation function (\S3.2($g$))
localised to this declaration:
$$
\PEabcA: \ \Procabc\times\Au \totop\ \Avua
$$
defined by 
$$
\PEabcA(P,\,a) \ = \ \PA(a),
$$
is strictly tracked by the computable function
$$
\peabcAbe: \ \cnr{\Procabc}\times\Ombeu\ \pto\ \Ombev
$$
defined by the following algorithm.
Let \,$P\in \Procabc$; \,say
$$
P \ \ident \ \procs\ \ins \ \att\ \outs \ \btt\ \auxs \ \ctt\
\begins \ S\ \ends
$$
and let \,$k_0\in\Ombeu$.
Take any \,$k_1\in\Ombev$
\,and
\,$k_2 \in \Ombew$.
\,(The choice of $k_1$ and $k_2$ is irrelevant,
by Remark 3.2.4.)
Put
\,$k\ident k_0,k_1,k_2$ \,and put
\,$\xtt \ident \att,\btt,\ctt$.
Compute 
\,$\sexAbe(\cS,k)$.  Suppose this converges to 
\,$l\ident \,l_0,l_1,l_2$, \,where 
\,$l_0\in \Ombeu$,
\,$l_1\in \Ombev$
\,and
\,$l_2\in \Ombew$.
Then we define \,$\peabcAbe(\cP,\,k_0) \da l_1$.
The following diagram then commutes:

$$
\commdiag{
\Procabc\times\Au&\ &\bimaprighturaise5{\PEabcA}&^+&\Avua\cr
\mapupl{\ang{\enum,\,\beu}}&\ &&&\mapupr{\,\bev}\cr
\cnr{\Procabc}\times\Ombeu&\ &\maprightud{\bdot}{\peabcAbe}&&\Ombev
}
$$
\mn
in the sense that 
$$
\aligned
\peabcAbe\,(\cP,k)\da l\ &\implies \ \be(l)\,\in \,\PEabcA\,(P,\,\be(k)),\\
\peabcAbe\,(\cP,k)\,\ua \quad\ &\implies \ \ua\,\in \,\PEabcA\,(P,\,\be(k)).
\endaligned
\tag6
$$
This is proved from (5)
and the definitions of \,\PE\ \,and \,\pe.

This concludes the proof of Lemma Scheme 6.3.1.
\endpf

\mn
{\bf Proof of Theorem $\tx{\bf A}_{\bk 0}$ \,(conclusion):}
\ \ Suppose 
\,$F:\Au\pto A_s$
\,is \WhileCCx\ computable on $A$.
Then there is a deterministic \WhileCCx\ \,procedure (Definitions 3.2.5/6)
$$
P\: u\ \to \ s
$$
such that for all \ainAu,
$$
\align
F(x) \da y \ &\impp \ \PA(x) = \{y\},\\
F(x) \ \ua \,\ \ \ &\impp \ \PA(x) = \{\ua\}.
\endalign
$$
Hence by ($g$) (above) there is a computable (partial) function 
$$
f\: \Ombeu\ \pto\ \Ombes
$$
which strictly tracks $F$, as required.
\endpf

\newpage

\Shead7{\bbf Soundness of \WhileCCxbig\ \,approximation}
In this section we address the general situation 
introduced in \S6.2, of a partial metric \Sig-algebra $A$
with an enumerated subalgebra \Xal,
and prove a more general soundness theorem (Theorem A) for 
\WhileCCx\ approximation.

\shead{7.1}{Enumerated subspace of metric algebra; \,Computational closure}
Let $A$ be an N-standard metric \Sig-algebra,
and  \Xal\ an enumerated \SortSig-family 
\,\ang{\Xsal\mid\sinSortSig}
\,of subsets \,$X_s\sseq A_s$ \,(\sinSortSig).
Each \Xs\ can be viewed as a {\it metric subspace\/}
of the metric space \As.
We call \Xal\ a \SortSig-{\it enumerated (metric) subspace\/} of $A$.

We define from \Xal\ a family
$$
\CalX \ = \ \ang{\CalXs \br \sinSortSig}
$$
of sets \,\CalXs\ \,of \al-{\it computable elements of\/} \,\As,
\,\ie, limits in \As\ of effectively convergent Cauchy sequences 
(to be defined below)
of elements of \,\Xs,
\,so that
$$
\Xs\ \sseq\ \CalXs\ \sseq \ \As,
$$
with corresponding enumerations
$$
\albars:\ \Omalbars \ \onto \ \CalXs.
$$
Writing \,$\albar = \ang{\albars\mid\sinSortSig}$,
we call the enumerated subspace \,\CalXalbar\
\,the {\it computable closure\/} of \Xal\ in $A$.

We will generally be interested in (strictly) \albar-computable
(rather than \al-computable)
functions on $A$ (\cf\ Definition 6.1.3),
as our more general model of {\it concrete computability\/} on $A$.

The sets \,$\Omalbars\sseq\NN$
consist of
{\it codes\/} for \,\CalXs\ \,(w.r.t\. \al),
\,\ie, \,pairs of numbers
\,$c = \ang{e,m}$ \,where
\itemm{($i$)}
$e$ is an index for a total recursive function defining a 
sequence \,$\al\circ\curl{e}$ \,in \Xs, \ie, the sequence
$$
\als(\{e\}(0)),\ \als(\{e\}(1)), \ \als(\{e\}(2)), \ \dots\ ,
\tag1
$$
of elements of \Xs,
\itemm{($ii$)}
$m$ is an index for a modulus of convergence for this sequence:
$$
\all k, l\ge\{m\}(n): 
\ \ds_i(\al(\{e\}(k)),\,\al(\{e\}(l))) < 2^{-n} .
\tag2
$$
\sn
For any such code \,$c=\ang{e,m}\in\Omalbars$,
\,$\albars(c)$ \,is defined as the limit in \As\ of the Cauchy sequence (1),
and \,\CalXs \,is the range of \,\albars:
$$
\commdiag{
\Xs&\quad\sseq\quad&\CalXs&\ \ \sseq\ \ &A\cr
\mapup\lft{\dz{\als}}&&\mapup\lft{\dz{\albars}}\cr
\Omals&\   &\Omalbars
}
$$

\Remarkn{7.1.1}
We may assume, when convenient, 
that the modulus of convergence for a given code is the {\it identity\/},
\ie, replace (2) by the simpler condition
$$
\all k, l\ge n: 
\ \ds_i(\al(\{e\}(k)),\,\al(\{e\}(l))) < 2^{-n} .
$$
or, equivalently,
$$
\all k > n: 
\ \ds_i(\al(\{e\}(k)),\,\al(\{e\}(n))) < 2^{-n} .
\tag3
$$
This is because any code \,$c=\ang{e,m}$ \,satisfying (2)
may be effectively replaced by a code
for the same element of \,\CalXs\ \,satisfying (3),
namely \,$c'=\ang{e',m_1}$,
\,where $m_1$ is a standard code for the identity function on \,\NN,
\,and \,$e'=\comp(e,m)$,
\,where \,$\comp(x,y)$ \,is a primitive recursive function
for ``composition" of (indices of) computable functions, \ie,
\,$\curly{\comp(e,m)}(x) \sq \curly{e}(\curly{m}(x))$.

In case of a code \,$c=\ang{e,m_1}$
\,satisfying (3), the sequence (1)
is called a {\it fast (\al-effective) Cauchy sequence\/}.
In such a case
we will often, for simplicity,
refer to $e$ itself as the ``code",
and the argument of \,\albars.
In this way we will shift between ``$c$-codes" and ``$e$-codes" 
as convenient.
\endpr

\Remarkn{7.1.2}
In the case \,$s = \nats$,
\,we can simply take \,$\Om_{\albar,\natss} = \Om_{\al,\natss} = \NN$,
and \,$\albar_\natss$ \,and \,$\al_\natss$
\,as the identity mappings on \,\NN.
Similarly, in the case \,$s=\bools$, \,we can take
\,$\Om_{\albar,\boolss} = \Om_{\al,\boolss} = \curly{0,1}$,
with \,$\albar(0) = \al(0) = \fff$
\,and
\,$\albar(1) = \al(1) = \ttt$.
(\Cf\ Remark 6.1.3($b$).)
\endpr

\Remarkn{7.1.3 \,(Closure of \alb-computability operation)}
The subspace \,\CalXalbar\ \,is ``computationally closed in $A$",
in the sense that 
the limit of a (fast) \albar-effective Cauchy sequence
of elements of \,\CalX\ \,is again in \CalX,
\,\ie, \,$\Calbar(\CalX) = \CalX$.
({\it Easy exercise.\/})
\endpr

\Remarkn{7.1.4}
We will usually assume that \Omals\ is decidable, 
in fact, that \,$\Omals = \NN$ \,for all sorts $s$, \,which is 
typical in practice, unlike the case for \Omalbar.
(See the following Example.)
\endpr

\Examplen{7.1.5 \,(Constructible reals)}
The best known nontrivial example of an enumerated subspace \Xal,
and its extension to a subspace of \al-computable elements,
is the following.  Let $A$ be the metric algebra \RRR\ of reals 
(Example 2.6.1),
with signature \Sig.
Let \,$X_\realss$ \,be the set of rationals \,$\QQ\sset\RR$, 
\,let \,$\Omalreal = \NN$ 
\,and let
$$
\alreal\: \NN\ \to\ \QQ
$$
be a canonical enumeration of \QQ.
Then \,$\CalQQ \eqdf \Cals(X)_\realss\sset \RR$ 
\,is the subspace of {\it recursive\/} or 
{\it constructible reals\/}.
Note that it is a {\it subfield\/} of \RR, and
hence \,\CalX\ \,is a {\it subalgebra\/} of \RRR.
Further, it is easily verified that \albar\ is strictly \SigRRR-effective.
(\Cf\ Definition 6.1.6.)
Note that \,$\Omalreal = \NN$,
\,unlike \,\Omalbarreal, \,which, by contrast, is non-recursive.
(See the previous Remark.)
\endpr

\Remarkn{7.1.6 \ (Extension of enumeration to \Ax)}
Given an enumeration \al\
of a \Sig-subspace $X$
\,of $A$, we can extend this canonically to an enumeration \alx\
of a \Sigx-subspace \Xx\ \,of \Ax.
({\it Easy exercise\/}.)
This in turn generates an enumeration \,\albarx\
\,of a \Sigx-subspace \CalXx\
\,of \alx-computable elements of \Ax.
It is easy to see that 
\itemm{($i$)}
if \CalX\ is an \Sig-subalgebra of $A$, then \,\CalXx\ \,is a 
\Sigx-subalgebra of \Ax;
\itemm{($ii$)}
if \albar\ is (strictly) \Sig-effective, then \,\albarx\ \,is 
(strictly) \Sigx-effective.
\sn
We will usually use this extension (of \Xal\  and  \,\CalXalbar)
to \Ax\ implicitly,
\ie, writing `\al' instead of `\alx' etc.
\endpr

\shead{7.2}{Soundness Theorem for effective numberings}
We now prove the first main theorem mentioned in the Introduction.

\Thmn{A \ (Soundness)}
Let $A$ be an N-standard metric \Sig-algebra,
and \Xal\ an enumerated \SortSig-subspace.
Suppose the enumerated \SortSig-space \CalXalbar\ of
\al-computable elements of $A$ is a \Sig-subalgebra of $A$,
and \albar\ is strictly \Sig-effective.
If \,$F:\Au\pto \As$
\,is \,\WhileCCx-approximable on $A$, then $F$
is \albar-computable on $A$.
\endpr                                                                          

\Pf
The proof uses the Soundness Theorem $A_0$
(Section 6), or rather the Lemma Scheme 6.4.1
(specifically, part ($g$) of the proof)
applied to the enumerated subalgebra \,\CalXalbar\ 
\,in place of \Abe.

So suppose 
\,$F:\Au\pto \As$
\,is effectively uniformly \WhileCCx\ \,approximable on $A$.
Then there is a \WhileCCxSig\ \,procedure
$$
P:\nats\times u\to s
$$
such that
for all \,\ninNN\ \,and \,all $x\in\dom{F}$:
$$
\ua\,\notin\,\PAn(x)\ \sseq\ \Bb(F(x),\,2^{-n}).
\tag1
$$
(see Definition 3.5.1).
By \S6.4($g$) (applied to \,\CalXalbar\ \,in place of \Abe)
there is a computable function 
$$
f:\ \NN\times\Omalbaru\ \pto\ \Omalbars
$$
which tracks \PA\ strictly,
in the sense that for all \,\ninNN, \,$e\in\Omalbaru$
\,and $e'\in\Omalbars$
\,(and writing $f_n = f(n,\,\cdot\,)$):
$$
\aligned
f_n(e)\da e' \ &\implies\ 
  \albar(e')\in \PAn(\albar(e)),\\
f_n(e)\ \ua \ \ \ \ &\implies\ \ua\in \PAn(\albar(e)).
\endaligned
\tag2
$$
We will show how to define a partial recursive \albar-tracking function 
$$
g: \ \Omalbaru\ \to \ \Omalbars
$$
for $F$ as follows. 

Given any $e\in\Omalbaru$, 
suppose \,$\albar(e)\in\dom{F}$, \,\ie,
$$
F(\albar(e))\da \ \in \As.
\tag3
$$
We must show how to define an \albar-tracking function $g$ for $F$,
\ie, such that
$$
g(e)\in \Omalbars \qquad\tx{and}
\qquad \albar(g(e)) \ = \ F(\albar(e)).
\tag4
$$
By (1), for all $n$
$$
\ua \,\notin \, \PAn(\albar(e)) \ \sseq\ \Bb(F(\albar(e)),\,2^{-n}).
\tag5
$$
Hence by (2), for all $n$
$$
f_n(e)\da \,\in\,\Omalbars
\tag{6$a$}
$$
and
$$
\albar(f_n(e))\,\in\,\PAn(\albar(e)).
\tag{6$b$}
$$
and so by (6$a$) we may assume (by definition of \Omalbar)
that for all $n$
$$
\tx{$\al\circ\curl{f_n(e)}$ \,is a fast Cauchy sequence, 
with limit \,$\albar(f_n(e))$.}
\tag7
$$
Also by (6$b$) and (5), 
$$
\ds\bigl(\albar(f_n(e)), \ F(\albar(e))\bigr) \ <\ 2^{-n}.
\tag8
$$
Now let $e'$ be a ``canonical" index for the (partial) function 
$$
\curl{e'}:\ n\ \mapsto \curl{f_n(e)}(n)
\tag9
$$
obtained uniformly effectively in $e$.
So \curl{e'}
is the ``diagonal" function
formed from the sequence of functions with indices
$f_n(e)$.
Consider
the sequence \,$\als\circ \curl{e'}$, \,\ie,
$$
\als(\{e'\}(0)),\ \als(\{e'\}(1)), \ \als(\{e'\}(2)), \ \dots,
\tag{10}
$$
{\bf Claim:}
\,(10) is a Cauchy sequence in \As, with modulus of convergence 
\,$\lam n(n+2)$.
\endpr

\n{\bf Proof of claim:}
\ For any $n$ and $k>n$:
$$
\align
&\quad\ \ \ds\big(\al(\curl{e'}(k)),\ \al(\curl{e'}(n)\big) \\
&= \ \ds\big(\al(\curl{f_k(e)}(k),\ \al(\curl{f_n(e)}(n)\big) 
  \qquad\tx{by def\. (9) of $e'$}\\
&\le \ \ds\big(\al(\curl{f_k(e)}(k)),\ \albar(f_k(e))\big)
  \ + \ \ds\big(\albar(f_k(e)),\ \albar(f_n(e))\big)
  \ + \ \ds\big(\albar(f_n(e)),\ \al(\curl{f_n(e)}(n))\big)\\
&=\ \ds_1\ + \ \ds_2\ + \ \ds_3 \qquad\tx{(say)}
\endalign
$$
where
$$
\align
&\ds_1 \ \le \ 2^{-k},\\
&\ds_3 \ \le \ 2^{-n},
\endalign
$$
by (7), and
$$
\align
\ds_2 \ &\le \ \ds\big(\albar(f_k(e)),\ F(\albar(e))\big) 
  \ + \ \ds\big(F(\albar(e)),\ \albar(f_n(e))\big) \\
&< \ 2^{-k} \ + \ 2^{-n}
\endalign
$$
by (8).
Therefore 
$$
\align
\ds\big(\al(\curl{e'}(k)),\ \al(\curl{e'}(n)\big) 
  \ &\le \ \ds_1\ + \ \ds_2\ + \ \ds_3\\
&< \ 2\cdot 2^{-k} \ + \ 2\cdot 2^{-n}\\
&< \ 2^{-n+2}.
\endalign
$$
This proves the claim.
\endpf
Further, by the method of Remark 7.1.1
(composing \curl{e'} with the modulus of convergence),
we can replace the index $e'$ by an $e$-code
$e''$ for a fast Cauchy sequence:
$$
\curl{e''}(n) \ \sq \ \curl{e'}(n+2).
\tag{11}
$$
Then we define 
$$
g(e) \ = \ e''.
\tag{12}
$$
We show that $g$ is an \albar-tracking function for $F$,
\ie, (assuming (3)) we show (4).
Since \,$\al\circ \curl{e''}$ \,is a 
fast Cauchy sequence, with the same limit in $A$
(if it exists) as \,$\al\circ \curl{e'}$ \,(by its definition (11)), 
\,to prove (4) it is enough to show (by (12)) that 
$$
\al(\curl{e'}(n)) \ \to \ F(\albar(e)) \qquad \tx{as}\qquad n \ \to\ \infty.
\tag{13}
$$
This follows since
$$
\aligned
\ds\big(\al(\curl{e'}(n),\ F(\albar(e) )\big)
\ &= \ \ds\big(\al(\curl{f_n(e)}(n)),\ F(\albar(e))\big)
\qqquad\tx{by def\. (9) of $e'$}\\
\ &\le\ \ds\big(\al(\curl{f_n(e)}(n)), \ \albar(f_n(e))\big)
\ + \ \ds\big(\albar(f_n(e)), \ F(\albar(e))\big)\\
\ &<\ 2^{-n}\ \ + \ \ 2^{-n} \qquad\qqqquad\tx{by (7) and (8)}\\
\ &=\  2^{-n+1}
\endaligned
$$
proving (13).
\endpf

\Remarkn{7.2.1}
A deterministic version of Theorem A
(\ie, without `\chooses')
was proved in \cite{stewart:thesis}.

\newpage

\itemm{\bbf 8 \ }{\bbf Interpretation of concrete in abstract model:
\,Adequacy of 
\nl
\WhileCCxbig\ \,approximation}

\shead{8.1}{Adequacy Theorem}
In this section we will prove Theorem B, a converse to the result of
the previous section.
Assume that $A$ is an N-standard metric \Sig-algebra,
and \Xal\ an enumerated \Sig-subspace,
with \al-computable closure \,\CalXalbar.

Note that we are not assuming in this section that 
\,\CalX\ \,is a subalgebra of $A$,
\,or even that
\,\albar\ \,is \Sig-effective.

Before stating the theorem,
we need a definition.

\Defn{8.1.1 \,(\al-effective local uniform continuity)}
A partial function
 \,$F:\Au\pto\As$
\,is {\it effectively locally uniformly continuous\/\ } 
(with respect to \al) \,if
there is a recursive sequence 
$$
(k_0,l_0),\ (k_1,l_1),\ (k_2,l_2),\ \dots
$$
of pairs of naturals such that
$$
\dom{F} \ \sseq\ \bigcup_{i=0}^\infty\Bbu\big(\al(k_i),\,2^{-l_i}\big) 
$$
and there is a total recursive function
\ $\LUF: \NN^2\to\NN$
\ (a modulus of local uniform continuity for $F$) such that
for all $i$, all \,$x,y\in\Bbu\big(\al(k_i),\,2^{-l_i}\big)\cap\dom{F}$,
\,and all $n$:
$$
\dsu(x,y)\,< \,2^{-\LUFs(i,n)} \ \implies\ \ \dss(F(x),F(y))\,<\,2^{-n}.
$$
Here \,$\Bbu(a,\del)$ \,is the open ball in \Au\ with centre 
$a$ and radius \del.
(Recall the definition (2.6.3) of the product metric \dsu\ on \Au.).
\endpr

\Examplen{8.1.2}
This phenomenon typically occurs
in the situation where $A$ is a countable union of 
neighbourhoods with compact closure;
for example, in the algebra \,\RRRp\ \,of reals,
\,\RR\ \,is the union of the neighbourhoods 
\,$(-k,\,k)$ \,for $k=1,2,\dots$.
Then a continuous function $F$ on $A$ will be uniformly continuous
on each of these neighbourhoods.
\endpr

We are now ready for the theorem.

\Thmn{B \ (Adequacy)}
Let $A$ be an N-standard metric \Sig-algebra,
\Xal\ an enumerated \SortSig-subspace,
and \,\CalXalbar\ the \SortSig-subspace
\,of \al-computable elements of $A$.  Suppose 
that for all \Sig-sorts $s$:
\sn
($i$) \ \Xs\ is dense in \As, \,and
\sn
($ii$) \ $\als:\NN\to\As$ \,is \WhileCCx-computable on $A$.
\sn
Let \,$F:\Au\pto \As$ \,be a function on $A$ 
such that
\sn
($iii$) \ $F$
is effectively locally uniformly continuous w.r.t\. \al, \,and
\sn
($iv$) \ \dom{F} is open.
\sn
If $F$ is strictly \albar-computable on $A$,
then $F$ is \WhileCCx\ \,approximable on $A$.
\endpr

Note the extra condition in Theorem B
(apart from assumptions ($i$)--($iv$)),
that $F$ be {\it strictly\/} \albar-computable.

\Remarkn{8.1.3}
From the proof of the theorem, it will be apparent that 
only sorts $s$ in the domain of $F$ have to satisfy condition ($i$),
and only sorts $s$ in the domain or range of $F$ have to
satisfy condition ($ii$).
\endpr

The proof uses the following notation.

\Notationn{8.1.4}
For any \,$k\in\NN$, \,let \,\econk\ \,be a canonical index for 
the constant function \,on \,\NN\
\,with constant value $k$,
\ie, for all $n\in\NN$, 
$$
\curl{\econk}(n) \ = \ k.
$$
Note that \,$\econk\in\Omalbar$ \,and
$$
\albar(\econk) \ = \ \al(k).
$$
\endpr

\shead{8.2}{Proof of Theorem B: \,Overview}
As an aid to the reader,
we first give an informal overview 
of the proof of Theorem B.  
(See Figure 3.)

\midinsert
\epsfxsize = 4in
\ce{\epsfbox{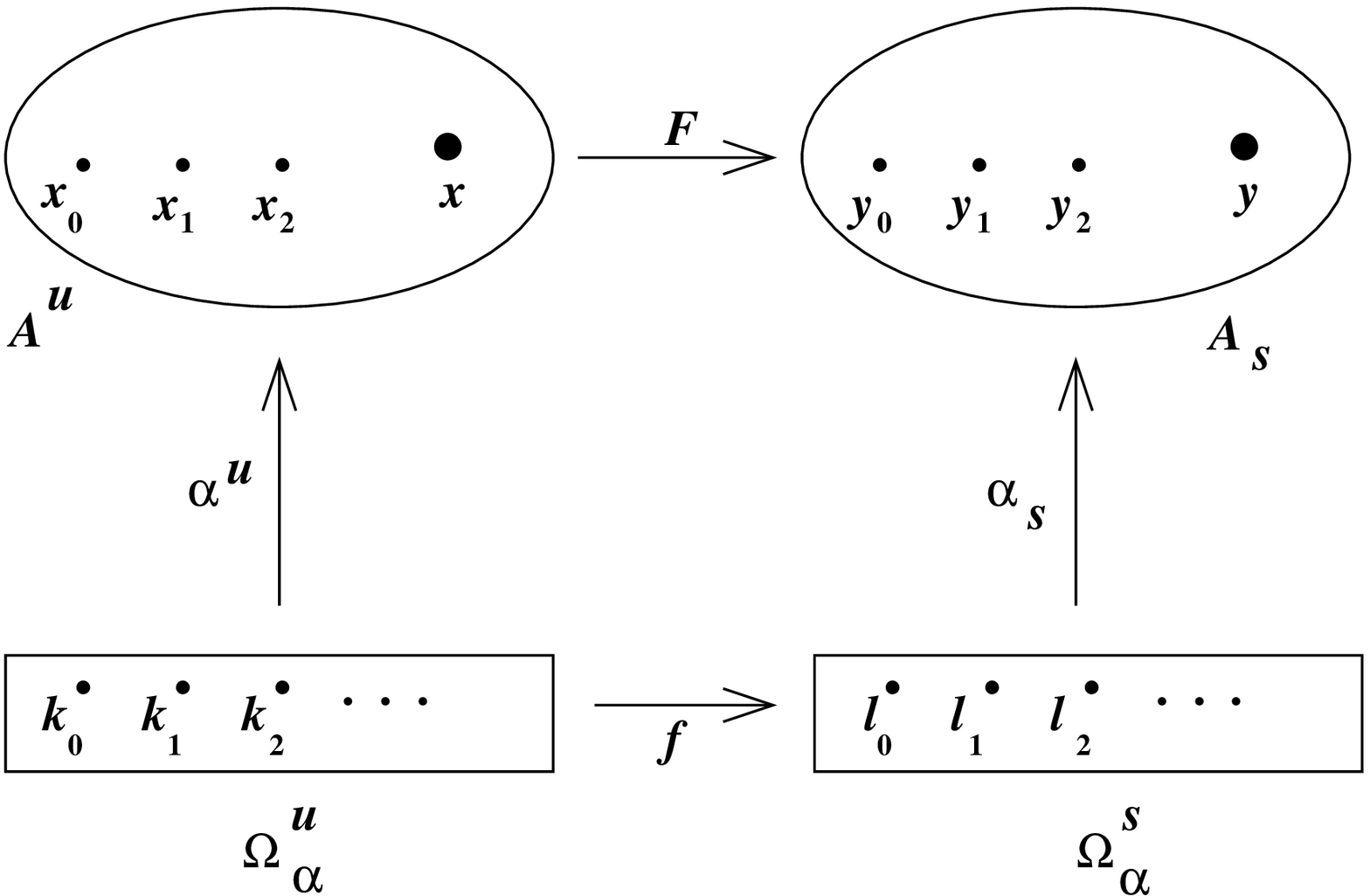}}
\mn
\ce{\sc Figure 3}
\sn
\endinsert

Given the assumptions \,$(i)\to(iv)$
\,of Theorem B,
suppose 
\,$F:\Au\,\pto\,\As$ \,is strictly \albar-computable by 
\,$f:\Omalbaru\pto\Omalbars$.
(In Figure 3, we represent $f$ as mapping 
\,\Omalu\ \,to \,\Omals,
\,rather than mapping
\,\Omalbaru\ \,to \,\Omalbars,
\,as a useful approximation,
as we will see.)
We must describe a \WhileCCx\ \,procedure
which approximates $F$ on $A$.

Let \xinAu.
Suppose \,$F(x)\da y$.
By the {\it density\/} of \,$X=\ran{\alu}$ in \Au,
and by the {\it openness\/} of \dom{F},
for each $n$
we can find
(using the \,`\chooses' \,operator,
as well as the \WhileCCx\ \,computability of \al)
an element \,$k_n$ of \Omalu\ 
\,such that 
\,$x_n \eqdf \alu(k_n) \in\dom{F}$,
\,and also 
\,$\ds(x_n,x) < 2^{-n}$.

Now compute 
an element \,$l_n$ of \Omalu\ 
which is a close approximation to
$f(k_n)$, or rather to $f(e_{\conss[k_n]})$.
More precisely, let
\,$e'_n \eqdf f(e_{\conss[k_n]})$, 
\,and let 
\,$l_n \,\eqdf \,\curl{e'_n}(n)$.
Then 
\,$\ds(\al(l_n), \albar(e'_n)) < 2^{-n}$.
Put \,$y_n = \al(l_n)$.

We must now check that the mapping 
\,$(x,n)\mapsto y_n$
\,defined above is \WhileCCx\
\,computable, and approximates $F$.
By {\it effective local uniform continuity\/} of $F$,
since 
\,$(x_n)_n$
\,is a fast Cauchy sequence with limit $x$,
\,$(y_n)_n$
\,is a Cauchy sequence with computable modulus of convergence and limit $y$.
Note also that \WhileCCx\ \,computability of $y_n$
(as a function of $x$ and $n$)
uses the \WhileCCx\ \,{\it computability\/} of \al.
Hence we can define a \WhileCCx\
\,procedure 
\,$P:\nats\times u \,\to\,s$
\,with \,$P^A(n,x)$
\,equal to
the set of all such $y_n$, obtainable
in this way from all possible implementations of the
\,`\chooses' \,operator.
Hence $F$ is computably approximable by $P$.

We turn to a precise proof of the theorem.

\shead{8.3}{Proof of Theorem B}
First we show, from assumption ($iii$), 
that $F$ has a \WhileCCx\ \,modulus of continuity,
\ie, a function
$$
\MCF:\ \Au\times\NN\ \pto\ \NN
$$
such that \,$\dom{F}\sseq\dom{\MCF}$, \,and
for all \,$x,y\in\dom{F}$ \,and for all $n$,
$$
\ds(x,y)\,<\,2^{-\MCFs(\dz{x,n})}\ \implies \ \ds(F(x),F(y))\,<\,2^{-n}.
\tag1
$$
A \WhileCCx\ \,algorithm for this
is easily constructed as follows
(using the notation of Definition 8.1.1).
With input \xinAu\ \,and $n$:
\,first find $i$ such that 
$$
x\in\Bb(\al(k_i),\ 2^{-l_i}).
\tag2
$$
(If $x\notin\dom{F}$, 
there may be no such $i$, and the algorithm 
for $\MCF(x,n)$ would then diverge, which is fine, from our viewpoint.)
\,Note that 
the sequences ($k_i$) and ($l_i$)
are computable, and also
(by assumption ($ii$))
\,\al\ \,is \WhileCCx\  computable.
We also use 
the primitive operations 
\,\ds\ \,and \,`$<$' (partial!) on \,\RR,
\,as well as the \,`\chooses' \,construct,
in ``finding" a suitable $i$.

Next (by (2)) find a natural number $d_0$ such that 
$$
\ds(x,\,\al(k_i)) \,+ \,2^{-d_0} \ < \ 2^{-l_i}.
\tag3
$$
Here again we use
the \WhileCCx\ \,computability of \al,
and
the primitive operations
\,\ds, \,`$+$' \,and \,`$<$
on \,\RR, \,
as well as the \,`\chooses' \,construct,
to find a suitable $d_0$.

From (2) and (3),
$$
\Bb(x,\ 2^{-d_0}) \ \sseq\ \Bb(\al(k_i),\ 2^{-l_i}).
$$
So define
$$
\MCF(x,n) \ := \ \max(d_0,\,\LUF(i,n))
$$
which is \,\WhileCCx\ \,computable, by the above remarks.

Now we will describe (in pseudo-\WhileCCx\ \,code)
an algorithm for a \WhileCCx-computable function 
$$
G:\ \NN\times\Au\ \totop\ \Asua
$$
which approximates $F$, in the sense that
for all $n$ and all $x\in\dom{F}$,
$$
G_n(x)\ \sseq\ \Bb(F(x),\,2^{-n})\ \sseq\ A_s.
\tag4
$$
With input \,$n,\,x$:
\sn
($1^\circ$)
\ Compute 
$$
M\ :=\ \MCF(x, \,n+1).
\tag5
$$
\ ($2^\circ$)
We want to find some $k$ such that both
$$
\ds(\al(k),\ x) \ < \ 2^{-M}
\tag6
$$
and
$$
\albar(\econk) \ = \ \al(k) \,\in\,\dom{F}.
\tag7
$$
Assume $x\in\dom{F}$.
By the density assumption ($i$)
and \ openness assumption ($iv$),
such a $k$ exists.
Further, by assumption, $F$ has a 
computable strict \albar-tracking function $f$.
Then (7) is equivalent to
$$
f(\econk)\da.
\tag8
$$
So using the \,`\chooses' \,construct again,
search for some $k$ satisfying both (6) and (8).
(Note that in practice 
this `\chooses' operation would be implemented by dovetailing ---
recall the discussion in \S4.1.)
\sn
($3^\circ$)
Compute \,$f(\econk)\da e'$.
By (7), \,$e'\in\Omal$ \,and
$$
F(\al(k)) \ = \ F(\albar(\econk)) \ = \ \albar(f(\econk)) \ = \ \albar(e').
$$
Hence by (1), (5) and (6),
$$
\ds\big(F(x),\ \albar(e')\big)\ 
  = \ \ds\big(F(x),\ F(\al(k))\big) \ < \ 2^{-n-1}.
\tag9
$$
($4^\circ$)
\ Finally compute 
$$
y\ :=\ \al\big(\{e'\}(n+1)\big)
\tag{10}
$$
This is possible by assumption ($ii$) again.
Then, since \,$\al\circ\{e'\}$ \,is a fast Cauchy sequence,
$$
\ds\big(y,\ \albar(e')\big) \ = 
\ \ds\big(\al(\{e'\}(n+1)),\ \albar(e')\big) \ \le\ 2^{-n-1}.
\tag{11}
$$
Hence by (11) and (9),
$$
\align
\ds\big(y,\,F(x)\big) \ &\le
\ \ds\big(y,\,\albar(e')\big)\,+\,\ds\big(\albar(e'),\,F(x)\big)\\
&<\ 2^{-n-1}\,+\,2^{-n-1}\\
&=\ 2^{-n}.
\endalign
$$
Define \,$G_n(x)$ \,to be the set of all possible $y$ computed as in (10),
by all possible implementations of the \,`\chooses' \,construct
as used in the above algorithm.
Then $G$ satisfies (4), and is \WhileCCx\ \,computable, 
by the above discussion.
\endpf

\Shead9{Completeness of \WhileCCxbig\ \,approximation}
Under certain assumptions,
we can combine Theorems A and B into a single equivalence,
Theorem C below.
We will then look at several examples of metric algebras where our
abstract and concrete models are equivalent according to this Theorem.

\shead{9.1}{Effective openness}
Note first the following problem:
Theorem A concludes with \albar-computability of $F$,
whereas Theorem B assumes {\it strong\/} \albar-computability.
To deal with this, we must make an assumption 
of ``effective openness" of \dom{F}.
This is handled by strengthening the ``effective local uniform continuity"
assumption, as follows.

Assume, as before, that $A$ is an N-standard metric
\Sig-algebra,
\Xal\ \,is an enumerated \Sig-subspace of $A$,
and \,\CalXalbar\ \, is its computable closure in $A$.

\Defn{9.1.1 \ (\al-effective openness)}
A subset $U$ of \Au\ ($u$ a \Sig-product type)
is {\it effectively open\/} (with respect to \al)
\,if there is a recursive sequence
$$
(k_0,l_0),\ (k_1,l_1),\ (k_2,l_2),\ \dots
$$
of pairs of naturals such that
$$
U \ =\ \bigcup_{i=0}^\infty\Bbu\big(\al(k_i),\,2^{-l_i}\big).
$$
\endpr

\Defn{9.1.2 \ (Strong \al-effective local uniform continuity)}
A partial function
 \,$F:\Au\pto\As$
\,is {\it strongly effectively locally uniformly continuous\/} 
(with respect to \al) 
\,if there is a recursive sequence
$$
(k_0,l_0),\ (k_1,l_1),\ (k_2,l_2),\ \dots
$$
of pairs of naturals such that
$$
\dom{F} \ =\ \bigcup_{i=0}^\infty\Bbu\big(\al(k_i),\,2^{-l_i}\big)
\tag1
$$
and
there is a total recursive function
\ $\LUF: \NN^2\to\NN$
\ (a modulus of local uniform continuity for $F$) such that
for all $i$, all \,$x,y\in\Bbu\big(\al(k_i),\,2^{-l_i}\big)$,
\,and all $n$:
$$
\ds(x,y)\,< \,2^{-\LUFs(i,n)} \ \implies\ \ \ds(F(x),F(y))\,<\,2^{-n}.
$$
\endpr

\Remarkn{9.1.3}
The only difference between effective local uniform continuity 
(Definition 8.1.1) and the ``strong" version above
is the equality in equation (1).
\endpr

Let
\,$F:\Au\pto \As$
\,be a function on $A$.  Then clearly:

\Lemman{9.1.4}
Strong \al-effective local uniform continuity of $F$ 
implies \al-effective openness of \dom{F}.
\endpr

\Lemman{9.1.5}
Suppose \dom{F} is \al-effectively open,
and \albar\ is strictly \Sig-effective.
Then
$$
\tx{$F$ is \,\albar-computable}
\ \ \ifff
\ \ \tx{$F$ is strictly \,\albar-computable.}
$$
\endpr

\Pf
($\tto$)
Note first that the assumptions imply that 
$$
\domal{F} \ \eqdf\ (\albar)^{-1}(\dom{F})
\ = \ \curly{e\in \Omalbaru\mid \albar(e)\in \dom{F}}
$$
is an r.e. (recursively or computably enumerable) 
subset of \NN, since for all $e\in\NN$
$$
e\in \domal{F} \ \ifff\ 
\ex i \,\big[\ds(\albar(e), \al(k_i)) \,< \,2^{-l_i}\big]
$$
(in the notation of Definition 9.1.2)
which is an r.e\. 
condition, by strict \albar-computability of \,\ds\ \,and \,\lsreal\
\,(implied by strict \Sig-effectiveness of \albar).
Hence, if $f$ is a computable \albar-tracking function for $F$,
it can be replaced by a strict \albar-tracking function $f'$,
defined by
$$
f'(e) \ \simeq\ 
\cases
f(e) \ift{$e\in\domal{F}$}\\
\ua \ow
\endcases
$$
which is easily seen to be computable.
\endpf
\Lemman{9.1.6}
Suppose \dom{F} is \al-effectively open,
and the mappings \,$\als:\NN\to\As$ \,are \WhileCCx\ computable.
Then
$$
\tx{$F$ is \WhileCCx-approximable}
\ \ \ifff
\ \ \tx{$F$ is strictly \WhileCCx-approximable.}
$$
\endpr
\n
(Recall Definition 3.5.1.)
\,The proof is an easy exercise.

\shead{9.2}{Completeness}
We are ready to state the completeness theorem 
for \WhileCCx\ approximability relative to
\albar-computability.

\newpage

\Thmn{C \,(Completeness)}
Let $A$ be an N-standard metric \Sig-algebra,
and \Xal\ an enumerated \SortSig-subspace.
Suppose the enumerated \SortSig-space \CalXalbar\ of
\al-computable elements of $A$ is a \Sig-subalgebra of $A$.
Assume also that for all \Sig-sorts $s$,
\sn
($i$) \ \albar\ \,is strictly \Sig-effective,
\sn
($ii$) \ \Xs\  \,is dense in \As, \,and
\sn
($iii$) \ $\als:\NN\to\As$ \,is \WhileCCx-computable on $A$.
\sn                               
Let \,$F:\Au\pto \As$
\,be a function on $A$,
such that
\sn
($iv$) \ $F$
is strongly effectively locally uniformly continuous w.r.t\. \al.
\sn
Then
$$
\tx{$F$ is (strictly) \WhileCCx\ approximable on $A$}
\ \ \ifff
\ \ \tx{$F$ is (strictly) \,\albar-computable on $A$.}
$$
\endpr
\n
Note that the word ``strictly" in the equivalence
may be omitted or inserted in either side at will.

\Pf
From Theorems A and B, together with Lemmas 9.1.4, 9.1.5 and 9.1.6.
\endpr

\sheads{9.3}{Examples of the application of the Completeness Theorem}
\mn
{\bf ({\bi a}) \,Canonical enumerations\/}
\sn
The purpose of this example is 
to make plausible condition ($iii$) of Theorem C
(and, of course, condition ($ii$) of Theorem B in Section 8),
\ie, the assumption of \WhileCCx\ computability
of the enumeration \al, by describing a commonly occurring
situation which implies it.

Suppose \Xal\ is an enumerated \Sig-subalgebra of $A$.

\Defn{9.3.1}
The enumeration 
\,$\al\:\NN\onto X$
\,is {\it effectively determined by a system of generators\/}
\,$G = \ang{g^s_0,g^s_1,g^s_2,\dots}_\sinSortSig$ 
\,if, and only if,
\sn($i$) \,$G$ generates $X$ as a \Sig-subalgebra of $A$; 
\sn($ii$) \,\al\ \,is defined as the composition of the maps
$$
\commdiag{
\NN\,\maprighturaise3{\enumSig} \,\TermSig\,\maprighturaise3{\evalG}\,X
}
$$
where \,\enumSig\ \,is the inverse of the G\"odel numbering
of \,\TermSig, \,and \,\evalG\
\,is the term evaluation induced by $G$,
\ie,
$$
\evalG(t) \ = \ \bb{t}\sigG,
$$
where \sigG\ is the state defined by
$$
\sigG(\xtt_i^s) \ = \ g_i^s
$$
for some standard enumeration
\,$\xtt_0^s, \,\xtt_1^s, \,\xtt_2^s, \dots$
\,of the \Sig-variables of sort $s$; \,and
\sn($iii$)
if, for any \Sig-sort $s$,
the sequence 
\,\ang{g^s_0,g^s_1,g^s_2,\dots}
\,is finite, then each $g^s_i$ is a \Sig-constant,
\,whereas if this sequence is infinite,
then the map
\,$i \mapsto g_i^s$
\,is a \Sig-function.
\endpr

An enumeration constructed in this way is called {\it canonical\/}
w.r.t\. $G$.

\Remarkn{9.3.2 \,(Totality of \evalG)}
We assume here that \,\evalG\
\,(and hence \al) is total.
This is achieved by assuming that either
\itemm{($i$)}
$A$ is total, \,or
\itemm{($ii$)}
\TermSig\ \,is replaced by some decidable subset 
\,$\Term'(\Sig)$
\,on which \,\evalG\ \,is total
(for example, omitting all terms involving division by 0).
\sn
Either one of these assumptions holds in each of the following examples;
for example, ($i$) holds in example ($b$) below, and ($ii$) in example ($c$),
resulting in the same ``canonical" enumeration \al\ of \QQ\
in both cases (even though the algebras are different).
\endpr

\Propn{9.3.3}
If \,\al\ \,is effectively determined 
by a system of generators, then the canonical
enumerations \,\als\ \,are \Whilex\ computable for all \Sig-sorts $s$.
\endpr

\Pf
This follows from the \Whilex\ computability 
of term evaluation \cite[Cor\. 4.7]{tz:hb}.
\endpf

The significance of the above definition
and proposition is this:
it is quite common for an enumeration to be effectively determined
by a system of generators;
and in such a situation,
condition ($ii$) in Theorem B, and ($iii$) in Theorem C,
will be (more than) satisfied.
This will be the case in the following examples.
\endpr

\mn
{\bf ({\bi b}) \,Partial real algebra\/}
\sn
Recall the example (7.1.5) of the enumeration \al\
of \QQ\ as a subspace of the N-standardised metric 
algebra \,\RRRN\ \,of reals
(Examples 2.5.3($b$) and 2.6.1)
and the corresponding enumeration \albar\
of the set \,\CalQQ\ \,of {\it recursive reals\/}.
Note that \al\ is canonical,
being effectively determined by the generators \curl{0,1},
and is hence \Whilex\ computable over \RRR.
Further, \QQ\ is dense in \RR,
\,\CalQQ\ \,is a subfield of \RR,
and \albar\ is strictly \SigRRR-effective.
We then have,
as a corollary to Theorem C:


\Corn{9.3.4}
Suppose 
\,$F\:\RR^n\ \pto\ \RR$
\,is strongly effectively
locally uniformly continuous.
Then
$$
\multline
\tx{$F$ is (strictly) \WhileCCx-approximable on \RRRN}\\
\llongtofrom
\ \ \tx{$F$ is (strictly) \albar-computable on \RR.}
\endmultline
$$
\endpr

Examples of functions satisfying the assumption
(and also the equivalence)
are all the common (partial) functions
of elementary calculus, such as \ $1/x$, \,$\log x$ \,and \,$\tan x$.

\newpage

\mn
{\bf ({\bi c}) \,Banach spaces with countable bases\/}
\sn
Let $X$ be a Banach space over \RR\ with a countable basis 
\,$e_0,e_1,e_2,\dots $,
which means that any element $x\in X$ can be represented uniquely
as an infinite sum
$$
x \ = \ \sum_{i=0}^\infty r_i e_i
$$
with coefficients \,$r_i\in\RR$
(where the infinite sum is understood as denoting convergence
of the partial sums in the norm of $X$).
(Background on Banach space theory can be found in
any of the standard texts, \eg, \cite{royden,taylor-lay}.)
To program with $X$, we construct a many-sorted algebra \XXX\ of the form
\mn
$$
\boxed{
\matrix \format\l&\quad\l\\
\algebras &\XXX \\
\imports &\RRRN\\
\carrierss &X\\
\functionss
&0\:\ \ \to X,\\
&+\:X^2 \to X,\\
&-\:X \to X,\\
&\odot\:\RR\times X\to X,\\
&\normfn\:X\to \RR,\\
&\es\:\NN\to X,\\
&\ifs_X\:\BB\times X^2\to X\\
\ends&
\endmatrix
}
$$                                                                              
\mn
where \,$\odot$ \,is scalar multiplication,
\,\normfn\ \,is the norm function and
and \,\es\ \,is the enumeration of the basis:
\,$\es(i) = e_i$.
Note that the algebras \BBB\ and \NNN\ are 
implicitly imported, as parts of \RRRN,
so that there are four carriers:
\,$X$, \RR, \BB\ \,and \,\NN,
\,of sorts 
\,\vectors, \,\scalars, \,\bools\ \,and \,\nats\
\,respectively.

Let \,$\Sig = \Sig(\XXX)$.  
Let \Sigo\ be \Sig\ without the norm function \,\normfn,
\,and let \XXXo\ be the reduct of \XXX\ to \Sigo.
Then \Sigo\ is the signature of an N-standardised vector space
over \RR, with explicit countable basis.

This can be turned into a metric algebra
in the standard way,
by defining a distance function on $X$
in terms of the norm:
$$
\ds(x,y) \ \eqdf\ \norm{x-y}.
$$

Let \,$\LQQe\sset X$ \,be the set of all finite 
linear combinations of basis elements from \es\ with 
coefficients in \QQ.
The following are easily shown:
\bull
\LQQe\ \,is countable; in fact it has a canonical enumeration
$$
\al\: \NN\ \onto\ \LQQe
$$
w.r.t\. the generators \,\es,
\,which (by ($a$) above)
is \Whilex\ computable;
\bull
\LQQe\ is dense in $X$;
\bull
\LQQe, with scalar field \QQ\
(together with carriers \NN\ and \BB)
is a \Sigo-subalgebra of \XXXo.

\sn
Now let \,(\CalLQQe,\ \albar) \,be the enumerated subspace of 
\al-computable vectors.  Then we can see that

\bull
\CalLQQe, with scalar field \,\CalQQ\
(together with carriers \NN\ and \BB)
is also a \Sigo-subalgebra of \XXXo; 
\,and moreover,

\bull
\albar\ \,is strictly \Sigo-effective.

\sn
However \,\CalLQQe\ \,is {\it not\/} necessarily a {\it normed\/} 
subspace of \XXX, since it may not be
closed under \,\normfn,
\,\ie, \,\norm{x} may not be in \,\CalQQ\ \,for all $x\in \CalLQQe$;
for example, if \XXX\ is the space \,\lp\ \,or
\,$\Lp[0,1]$
\,where $p$ is a nonrecursive real (see Examples 9.3.8 below).
We must therefore make an explicit assumption
for the Banach space \,\Xnorm\ \,with respect to 
both the {\it closure\/} of \,\CalLQQe\ \,under \,\normfn,
\,and the \,\albar-{\it computability\/} of \,\normfn.

\Assumptionn{9.3.5 \,(\albar-computable norm assumption for \,\Xnorm)}
\nl
For all $x\in \CalLQQe$, \,$\norm{x}\in\CalQQ$.
Furthermore, the norm function \,\normfn\
\,is strictly \,\albar-computable.
\endpr

\n
As we will see, many common examples of Banach spaces satisfy 
this assumption.

Note that ssumption 9.3.5 is equivalent to the 
following (apparently weaker) assumption,
which is often easier to prove:
\endpr

\Assumptionn{9.3.6 \,(\alalbar-computable norm assumption for \,\Xnorm)}
For all $x\in \LQQe$, \,$\norm{x}\in\CalQQ$.
Further, \,\normfn\ \,has a computable \alalbar-tracking function,
\ie, a computable function \,$f\:\NN\to\NN$ 
\,such that the following diagram commutes:

\bn
$$
\commdiag{
\LQQe& &\maprighturaise3{\normfn}& &\CalQQ\cr
\mapupl{\al}& && &\mapupr{\ \albar}\cr
\NN& &\maprightd{f}& &\Omalbar
}
$$
\endpr

\mn
Suppose now that \,\Xnorm\ \,satisfies the \albar-computable norm assumption.
Then the \Sigo-subalgebra \,\CalLQQe\ \,of \XXXo\
can be expanded to a \Sig-subalgebra of \XXX\ 
(which we will also write as \,\CalLQQe), enumerated by \albar,
which is strictly \Sig-effective.

Now let \,$F\:X\to \RR$ \,be a (total) {\it linear functional on\/} $X$.
$F$ is said to be {\it bounded\/} if for some real $M$,
$$
|F(x)| \ \le \ M\norm{x} \ \ \ \tx{for all}  \ \,x\in X.
\tag1
$$
Write \norm{F} for the least $M$ for which (1) holds.
Then if $F$ is bounded,
$$
|F(x)-F(y)| \ \le\ \norm{F}\cdot\norm{x-y} \ \ \ \tx{for all}\  \,x,y\in X,
$$
and so $F$ is uniformly continuous, in fact it is clearly 
{\it effectively locally uniformly continuous\/},
and strongly so (since it is total).
We may therefore apply Theorem C to $F$.

\Corn{9.3.7 \,(Completeness for computation on Banach spaces)}
Let $X$ be a Banach space over \RR\ with countable basis, and
let \,\CalLQQe\ \,be the enumerated subspace of \al-computable vectors,
where \al\ is a canonical enumeration of the subspace \,\LQQe.
Suppose \Xnorm\ satisfies the \,\alalbar-computable norm assumption.
Then for any bounded linear functional $F$ on $X$,
$$
\tx{$F$ is (strictly) \WhileCCx\ approximable on \XXX}
\ \ifff
\ \tx{$F$ is (strictly) \,\albar-computable on $X$,}
$$
where \XXX\ is the N-standard algebra formed from $X$ as above.
\endpr

Finally we give examples of Bananch spaces which satisfy this
\albar-computable norm assumption.

\Examplesn{9.3.8 \,(Banach spaces with computable norms)}
\itemm{($i$)}
For $1\le p<\infty$, we have the space \,\lp\ 
\,of all sequences $x = \ang{x_n}_{n=0}^\infty$
\,of reals such that \,$\sum_{n=0}^\infty |x_n|^p < \infty$,
with norm defined by
$$
\norm{x}_p \ = \ \big(\sum_{n=0}^\infty |x_n|^p\big)^{1/p},
$$
and a countable basis given by \,$e_i = \ang{e_{i,n}}_{n=0}^\infty$, 
\,where 
$$
e_{i,n} \ = \
\cases
1 \ift{$i=n$},\\
0 \ow.
\endcases
$$
It is not hard to see that 
$$
\tx{\sl if $p$ is a recursive real, then 
\,\lp\ \,satisfies the computable norm assumption,}
$$
and hence Corollary 9.3.7 can be applied to it.

\itemm{($ii$)}
For $1\le p<\infty$, we have the space 
\,$\Lp[0,1]$
\,of all Lebesgue measurable functions $f$ on 
the unit interval $[0,1]$ 
such that \,$\int_0^1 |f|^p<\infty$,
\,with norm defined by
$$
\norm{f}_p \ = \ \big(\int_0^1 |f|^p\big)^{1/p},
$$
and a countable basis given by (\eg)
some standard enumeration of all 
step functions on $[0,1]$ with rational values
and (finitely many) rational points of discontinuity,
or of all
polynomial functions on $[0,1]$ with rational coefficients.
Again, it is not hard to see that 
$$
\tx{\sl if $p$ is a recursive real, then \,$\Lp[0,1]$
\,satisfies the computable norm assumption,}
$$
and hence Corollary 9.3.7 can be applied to it.

\itemm{($iii$)}
The space \,$C[0,1]$ 
\,of all continuous functions $f$ on $[0,1]$,
with norm defined by
$$
\norm{f}_{\tx{sup}} \ = \ \sup_{t\in I}|f(t)|
$$
and a countable basis given by a standard enumeration
of all 
zig-zag functions on $[0,1]$ with (finitely many) turning points
with rational coordinates,
or of all
polynomial functions on $[0,1]$
with rational coefficients.
Again, we see that 
$$
\tx{\sl $C[0,1]$ \,satisfies the computable norm assumption.}
$$

\Shead{10}{Conclusion}
We have compared two theories of computable functions on topological
algebras, one based on an abstract, high level model of programming and
another based on a concrete, low-level implementation model. Our
examples and results here, combined with our earlier results 
\cite{tz:top,tz:hb}
and those of Brattka \cite{brattka96,brattka:thesis}, 
show that the following are surprisingly necessary features
of a comprehensive theory of computation on topological algebras:

\item{1.}
The algebras have partial operations.
\item{2.}
Functions are both continuous and multivalued. 
\item{3.}
Classical algorithms in analysis require nondeterministic
constructs for their proper expression
in programming languages.
\item{4.}
Indeed, multivalued subfunctions are needed to compute even single-valued
functions, and abstract models must be nondeterministic even to compute
deterministic problems.
\item{5.}
Abstract models and effective approximations by abstract models
are generally sound for concrete models. 
\item{6.}
Abstract models even with approximation or limit operators are
adequate to capture concrete models only in special circumstances.
\item{7.}
Nevertheless there are interesting examples where equivalence holds.
\item{8.}
The classical computable functions of analysis can be characterised
by abstract models of computation.

Specifically, we examined abstract computation by the basic
imperative model of \qwhiles-array programs. Many
algorithms in practical computation are presented in
pseudo-code based on the \qwhiles\ language.  To meet the
requirement of feature 2 above we added the simplest form of
countable choice to the assignments of the language, and we
defined the \WhileCCx\ approximable computations. We proved
a Soundness Theorem (Theorem A) and an Adequacy Theorem
(Theorem B), and combined these into a Completeness Theorem (Theorem C), 
in the case of metric algebras with partial
operations. We considered algebras 
of real numbers and Banach spaces where equivalence
theorems hold.

There are, of course, interesting technical questions to
answer in working out the details of the computability
theory for the \WhileCCx\ model (\cf\ the theory
for single-valued functions on total algebras in
\cite{tz:hb}). There are several other important
abstract models of computation that may be extended
with nondeterminsitic constructs in order to establish
equivalence with concrete models. The abstract model of
schemes in \cite{brattka:thesis} is quite general in a number of
ways. The topological properties of many valued functions
are also in need of investigation.

However, returning to the general problem posed in the
Introduction, the features 1--8 above suggest that new
research directions are needed to develop a comprehensive
theory of specification, computation and reasoning
with infinite data. What are the appropriate programming
constructs for working with topological computations? What
specification techniques are appropriate for continuous
systems? What logics are needed to support verification
of programs that approximate functions?  Our work on
computation suggests that some advanced semantic features
are necessary. It suggests that the nondeterminism that
played an important role in programming methodologies of
the late 1970s  (\eg, \cite{dijkstra:book} seems to be needed in
the proper development of topological programming. There
are plenty of algorithms in scientific modelling, numerical
analysis and graphics to investigate, using such new theoretical tools.

\bigskip
\bigskip
\goodbreak

\cbb{References}\bigskip

\bibliographystyle{alpha}
\bibliography{abbrev,bib}

\enddocument